\documentclass{jfm}
\usepackage{graphicx}
\usepackage{epstopdf,epsfig,siunitx,caption,subcaption,multirow,color}

\shorttitle{Levitation of non-magnetizable droplet inside ferrofluid}
\shortauthor{C. Singh, A. K. Das and P. K. Das}

\title{Levitation of non-magnetizable droplet inside ferrofluid}

\author{Chamkor Singh\aff{1}\footnote{\centering Current address: Max-Planck Institute for Dynamics and Self-Organization, Goettingen, 37077, Germany.},
Arup K. Das\aff{2}
\and Prasanta K. Das\aff{1}\corresp{\email{pkd@mech.iitkgp.ernet.in}}}

\affiliation{\aff{1}Department of Mechanical Engineering, Indian Institute of Technology,
Kharagpur, India
\aff{2}Department of Mechanical and Industrial Engineering, Indian Institute of Technology,
Roorkee, India}

\begin{document}

\maketitle
\begin{abstract}
The central theme of this work is that a \emph{stable} levitation of a denser \emph{non-magnetizable} liquid droplet, against gravity, inside a relatively lighter ferrofluid -- a system barely considered in ferrohydrodynamics -- is possible, and exhibits unique interfacial features; the stability of the levitation trajectory, however, is subject to an appropriate magnetic field modulation. We explore the shapes and the temporal dynamics of a plane {non-magnetizable} droplet levitating inside ferrofluid against gravity due to a spatially complex, but systematically generated, magnetic field in two dimensions. The coupled set of Maxwell's magnetostatic equations and the flow dynamic equations is integrated computationally, utilizing a conservative finite-volume based second order pressure projection algorithm combined with the front-tracking algorithm for the advection of the interface of the droplet. The dynamics of the droplet is studied under both the {constant ferrofluid magnetic permeability} assumption as well as for more realistic field dependent permeability described by the Langevin's non-linear magnetization model. Due to the non-homogeneous nature of the magnetic field, unique shapes of the droplet during its levitation, and at its steady state, are realized. The complete spatio-temporal response of the droplet is a function of the Laplace number $La$, the magnetic Laplace number $La_m$, and the Galilei number $Ga$; through detailed simulations we separate out individual roles played by these non-dimensional parameters. The effect of the viscosity ratio, the stability of the levitation path and the possibility of existence of multiple-stable equilibrium states is investigated. We find, for certain conditions on the viscosity ratio, that there can be developments of \emph{cusps} and \emph{singularities} at the droplet surface; this phenomenon we also observe experimentally and compare with the simulations. Our simulations closely replicate the singular projection on the surface of the levitating droplet. Finally, we present an dynamical model for the vertical trajectory of the droplet. This model reveals a condition for the onset of levitation and the relation for the equilibrium levitation height. The linearization of the model around the steady state captures that the nature of the equilibrium point goes under a transition from being a \emph{spiral} to a \emph{node} depending upon the control parameters, which essentially means that the temporal route to the equilibrium can be either monotonic or undulating. The analytical model for the droplet trajectory is in close agreement with the detailed simulations.    
\end{abstract}
\section{Introduction}\label{secIntro}
The buoyancy due to the gravitational field is the oldest known mechanism of levitation of matter inside fluids. It is only the past few decades that the researchers have used other gravity compensation techniques for levitating objects in liquids or gases. For example using acoustics \citep{trinh1985compact}, optical technique by utilizing photon momentum transfer \citep{price2015vacuo}, using magnetic fields to levitate objects inside paramagnetic substances \citep{ikezoe1998making}, inside magnetic nanofluids \citep{rosensweig1966buoyancy}, inside air \citep{geim1999magnet}, and more recently studying the effect of lasers \citep{limbach2016toward}. 

In the context of \emph{magnetic nanofluids} or ferrofluids\footnote{Ferrofluid is a colloidal suspension of surfactant coated magnetic nanoparticles (characteristic size $\sim 10$ nm) inside a suitable carrier liquid \citep{rosensweig2013ferrohydrodynamics}.}, one of the earliest observation of solid levitation inside ferrofluid was made by \citet{rosensweig1966buoyancy}. It was observed that a solid magnet dispersed inside a ferrofluid levitates itself against the gravity. Thereafter, the levitation of solid objects inside ferrofluids has found numerous technological and research applications in recent years. The principle has been investigated for non-magnetic solid particle separation from a continuous stream of ferrofluid \citep{pamme2006magnetism,vojtivsek2012microfluidic}. In a similar manner, it has also been used in biological cell sorting at micro-scale utilizing a magnetic fluid as the outer phase liquid \citep{zhu2013microfluidic}. The transport of diamagnetic particles is another application \citep{dunne2007levitation,liu2014horizontal,zhu2011analytical}.  As sometimes required in biology, the gravity compensating environment can also be achieved through the levitation of non-magnetizable objects inside ferrofluids \citep{beysens2015generation}. For example, the research has shown that the magnetically created microgravity environment through magnetic levitation is technically applicable in the control of crystallization \citep{huber1996technology}. Recently, the magnetic levitation has been successfully utilized to measure the density of solid and immiscible liquids \citep{mirica2009measuring}. As the magnetic fields can be generated using electromagnets integrable to electronics, the process has a tremendous potential for smart sensor applications. Such efforts have already been made, for example the development of magnetic actuators \citep{olaru2013maximizing}.

Though the experimental evidence of solid phase levitation inside ferrofluid came right after the invention of ferrofluids, very little efforts have been made till date to investigate the same for a liquid phase levitation. A distinct hydrodynamics is certainly expected in the latter case due to the presence of a deformable liquid-liquid interface instead of a rigid liquid-solid interface. It is physically tempting to study this system and to look for, if any, the distinguishing behaviors. Similar to the practical applications of solid object levitation inside a ferrofluid, the liquid phase levitation might also have useful applications, especially in the modern small scale devices.

Many of the facets of the interface between a magnetizable and a non-magnetizable liquid are exhibited by the ferrofluid droplets. Numerous intriguing features of the ferrofluid interfacial phenomena, such as deformations, appearance of peculiar shapes, small scale instabilities at the interface, wetting, hysteresis, merging and break-up, stretching and pinning, and many more can be fundamentally studied using the droplet systems. The motivation also comes from looking at the in-use and future possible applications of ferrofluid droplets such as micro scale mixing \citep{mugele2006microfluidic}, inkjet-printing \citep{verkouteren2011inkjet}, transport of surfactant \citep{wojciechowski2009interfacial,kovalchuk2001numerical}, transport of drugs in biological systems, vibrating interfaces \citep{kim2015mode,whitehill2011droplet} and so on.

The investigations focusing ferrofluid droplets seem to start accelerating from $1980s$, although seminal works had been already performed on relevant droplet systems, \emph{e.g.} by \cite{taylor1964disintegration} on the disintegration of water droplets due to electric field and \cite{rosenkilde1969dielectric} on dielectric droplet in an electric field. \citet{bacri1983bistability} found that magnetic field over a threshold value can destabilize a ferrofluid droplet. The researchers found that the shape of the ferrofluid droplet can change from an elongated one to a slender one. They used anionic ferrofluid in their experiments to get high agglomerate concentrations to achieve this transition regime. \citet{sherwood1988breakup} studied the breakup dynamics of droplets from a more general point of view as he investigated the effect of both electric and magnetic field. A more rigorous analysis of the equilibrium shapes of the ferrofluid droplet seems to be first performed by \citet{sero1992shape}. The researchers used an energy minimization principle and studied both partially and totally free ferrofluid droplets. They found interesting bifurcating solutions and hysteresis mechanisms. \cite{wohlhuter1993effects} investigated polarizable drops and their stability in external fields. Near the same time, \citet{bacri1994behavior} reported, for the first time, the magnetic fluid micro droplet behavior under a rotating magnetic field. A starfish like shape instability was observed by the investigators. Later \citet{sandre1999assembly} studied the behavior of highly magnetic droplet under rotating and modulated fields. They observed the rotations of the droplet to be synchronous with the applied field and also the breakup of the droplet at increased vorticity. Recently, the researchers have shown interest in the ferrofluid droplet patterns \citep{jackson2005theory,timonen2013switchable}, formation processes \citep{chen2010ordered,liu2011numerical2,liu2011numerical} and instabilities at its interface \citep{bashtovoi1999dynamics,chen2008experimental}. 

In the last decade and near the start of the present one, efforts have been made to investigate the deformation dynamics, motion and manipulation of the ferrofluid droplets under different field configurations and using different tools such as numerical programs \citep{afkhami2008field,afkhami2010deformation}, image processing \citep{koh2013digital} and other experimental techniques \citep{nguyen2013deformation}. \citet{jackson2007confined} observed unique regular and irregular ferrofluid droplet shapes inside a Hele-Shaw cell under cross magnetic fields applied normal to the cell plane. \citet{afkhami2008field} investigated the motion of a ferrofluid droplet through a viscous medium and subsequently \citep{afkhami2010deformation} did a numerical and experimental study to predict its deformation and shape. \citet{chen2008experimental} experimentally studied the Rosensweig instability of a ferrofluid droplet. \citet{zhu2011nonlinear} simulated the experimentally observed droplet shapes and found its deformation to be non-linearly related to the magnetic Bond number. From microfluidic application point of view, \citet{wu2013ferrofluid} investigated the formation and breakup of a ferrofluid droplet in a microfluidic flow focusing device. Before that, \citet{tan2010formation} studied similar aspects in a microfluidic T-junction. New applications have also emerged in recent for example energy harvesting using ferrofluid droplets \citep{kim2015liquid,kim2015energy}, micro structure printing \citep{fattah2016printing} and optofluidic devices \citep{gu2015ferrofluid}. Recently in an experimental work, \citet{gu2016motion} have observed the non-linear behavior of a ferrofluid droplet under the application of a periodic field. \citet{lira2016ferrofluid} found a very interesting family of stable polygonal shapes of ferrofluid droplet for quasi-two dimensional conditions using a vortex-sheet formulation. Most recently \citet{rowghanian2016dynamics} have obtained further important insights into the ferrofluid deformation dynamics.

It is notable that though a decent amount of research has been conducted on the ferrofluid droplet behavior inside non-magnetizable environment, the inverse system has not been explored up to its fundamental dynamical and interfacial details. { Relevant exceptions are the studies by \citet{duplat2013bubble}, where the researchers have studied a bubble shape in magnetically compensated gravity environment inside liquid oxygen, by \cite{ueno1995study,ueno1999numerical} and \cite{korlie2008modeling} for bubbles in simplified \emph{uniform} field conditions. Again a heavy settled droplet rise \emph{against} the gravity in a \emph{non-uniform} field condition is not tackled.} Though globally the droplet will respond along the applied field gradient in the inverse system (repel away from the magnetic source), the difference lies in the fact, as will be shown in our study, that the local flipping of the direction between the curvature and the magnetic force at the interface in the inverse system can cause intriguing interplay between the magnetism and the fluid flow, especially under the non-uniformity of the magnetic field, and results in unique droplet shapes and its interfacial features. Furthermore, the inverse system is technically equally relevant to the droplet manipulation technologies, creating gravity compensation environments for the small scale droplets, and in sensing applications based on such principles, though it has not received due attention. 

{ In free space, it is well known that a stable levitation of a permanent magnet is proven to be not possible by the Earnshaw's theorem --- a consequence of the fact that the Maxwell's equations do not permit a magnetic field maximum in free space \citep{geim1999magnet}. Although the opposite is possible for a diamagnetic substance, which requires a filed minimum rather than a maximum. However, if the surrounding medium is not a free space and is fluid, and at the same time, is itself magnetizable, the situation is then different for a dispersed diamagnetic substance. In addition this system even permits for the levitation of a non-magnetic object. }

In this work, the temporal dynamics and the spatial shapes of a plane non-magnetizable liquid droplet levitating inside a ferrofluid against gravity due to a spatially complex, but systematically generated, magnetic field have been studied in a two dimensional environment, primarily through numerical computations. The coupled Maxwell's magnetostatic equations and the flow dynamic equations are integrated computationally, utilizing a finite-volume based conservative second order pressure projection algorithm combined with the front-tracking algorithm for the advection of the interface of the droplet. To support our simulations, we demonstrate the non-magnetizable droplet levitation experimentally and compare the interfacial singular projections obtained from the simulations with the experimental observations. Finally, we present a non-linear analytical model for the droplet trajectory in the vertical direction. 

The mathematical formulation of the problem, and the physical basis for it, is presented in \S\ref{secPhenomenon} and \S\ref{secFormulation} while the numerical solution methodology is described in \S\ref{secNumerical}. The discussion and analysis of the levitation phenomenon is broken into several parts --- { \S\ref{secResultsConstantPermeability} describes basic characteristics of the droplet shape, time-dependent levitation height, and the effect of non-dimensional parameters on the same;} the effects due to viscosity ratio between the two phases are explored in \S \ref{secResultsViscosityRatio}; the effects of the non-linear magnetization of the ferrofluid are presented in \S \ref{secResultsVariablePermeability}. We study the stability conditions for the levitation path and the final equilibrium location in \S \ref{secResultsStability} and compare the experimentally observed phenomenon with the simulation in \S \ref{secResultsExperiments}. In \S \ref{secOneDimensional}, we describe the analytical model for the levitation height, the necessary condition for the onset of the levitation as well as the transitions in the solution behavior near the equilibrium levitation point. Finally we summarize our findings in \S \ref{secConclusions}.
\section{Phenomenon}\label{secPhenomenon}
\subsection{Setup and visualization}
\begin{figure}
	\centering
	\begin{subfigure}[]{0.287\textwidth}
		\centerline{\includegraphics[width=0.99\textwidth]{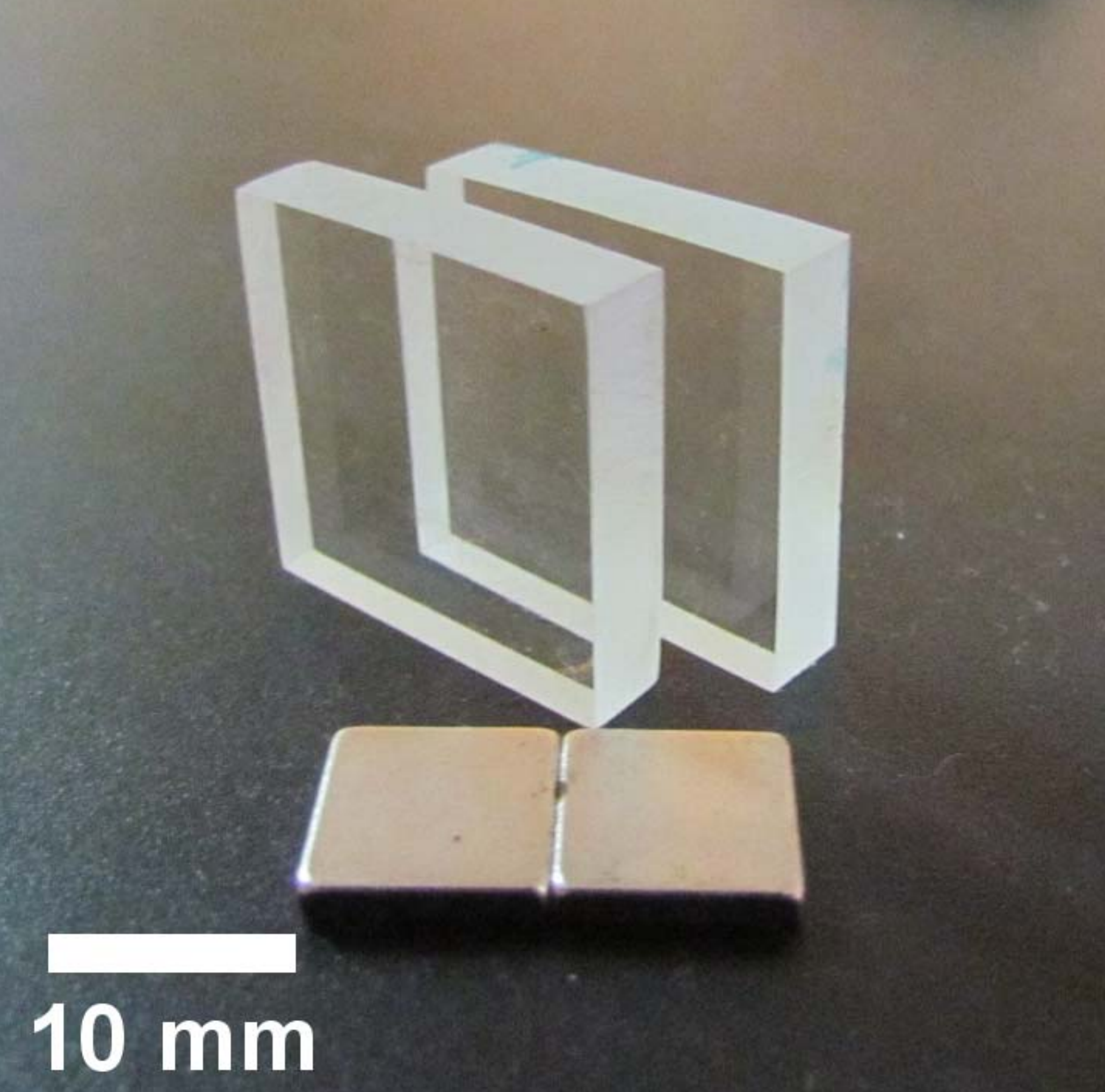}}
	\end{subfigure}
	\begin{subfigure}[]{0.70\textwidth}
		\centerline{\includegraphics[width=0.99\textwidth]{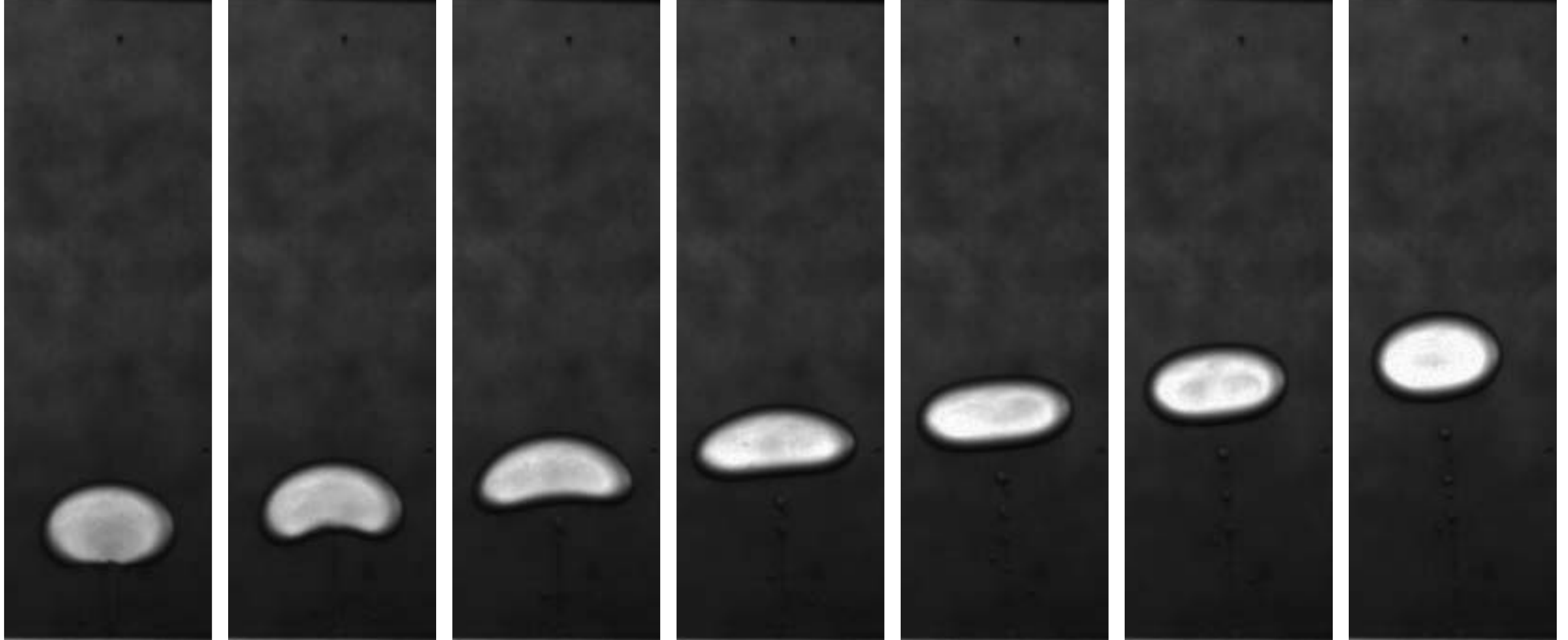}}
	\end{subfigure}	
	\caption{(Left) The PMMA sheets used to manufacture the Hele-Shaw cell together with a pair of neodymium magnets. (Right) Recorded image sequence showing a water droplet levitating inside a ferrofluid sample contained in the Hele-Shaw cell. Only a part of the cell is shown. {The droplet radius is $\sim 2\;\si{mm}$.} The levitation is initiated when the cell is placed on a magnet surface. The gravity acts vertically downward in the figure.}
	\label{figExpIntroduction}
\end{figure}
To lay down the physical basis for the numerical computations, first we experimentally demonstrate the phenomenon of a non-magnetizable droplet levitation inside a ferrofluid. This approach has actually helped us to first understand about a realistic set of magnetic field boundary conditions that can lead to a stable levitation of the droplet.  

The levitation of the droplet is visualized in a Hele-Shaw cell arrangement. The cell is made up of two closely spaced transparent poly-methyl-methacrylate (PMMA) sheets of size $20\:\si{mm}\times 20\:\si{mm}$, shown in figure \ref{figExpIntroduction} (left). The figure shows the PMMA sheets prior to the manufacture of the cell. The four sides of the cell are closed by inserting a $1.0\pm0.01\:\si{mm}$ thin polymer sheet cuttings in between the PMMA sheets. The final effective volume of the cell is measured to be $17.7\:\si{mm}\times 17.7\:\si{mm}\times 1.0\:\si{mm}$. 

The cell is filled with a sample of kerosene based ferrofluid (a precise description of the magnetic characterization of the ferrofluid sample, apparatus and image analysis is given later in \S \ref{secResultsExperiments} while comparing with the simulations). A small droplet of water with volume $\pi \delta_H R^2$ is placed into the cell before its closure, where $R$ is the droplet radius and $\delta_H$ is the gap between the walls of the Hele-Shaw cell. The size of the droplet is predicted using image processing after it is dispensed into the cell {($R\sim 2\;\si{mm}$)}. The plane of the cell is kept parallel to the direction of the gravity, and thus in the absence of the field, the droplet remains settled at the bottom. 

The levitation against the gravity is now initiated by placing the cell on the surface of a permanent magnet; the magnets used are of Neodymium and a pair is shown in figure \ref{figExpIntroduction} (left) together with the cell. Figure \ref{figExpIntroduction} shows the phenomenon recorded during one of the experiments under the above stated conditions. Seven different time instants during the evolution of the droplet are shown in the figure. Initially, the droplet rests on the bottom wall. After the application of the field at the bottom of the cell, the droplet begins to levitate and the shape of the droplet alters -- it elongates laterally and attains a concavity at the bottom. Eventually it attains almost an elliptic shape. 
\subsection{Stable levitation}
\begin{figure}
	\centerline{\includegraphics[width=0.999\textwidth]{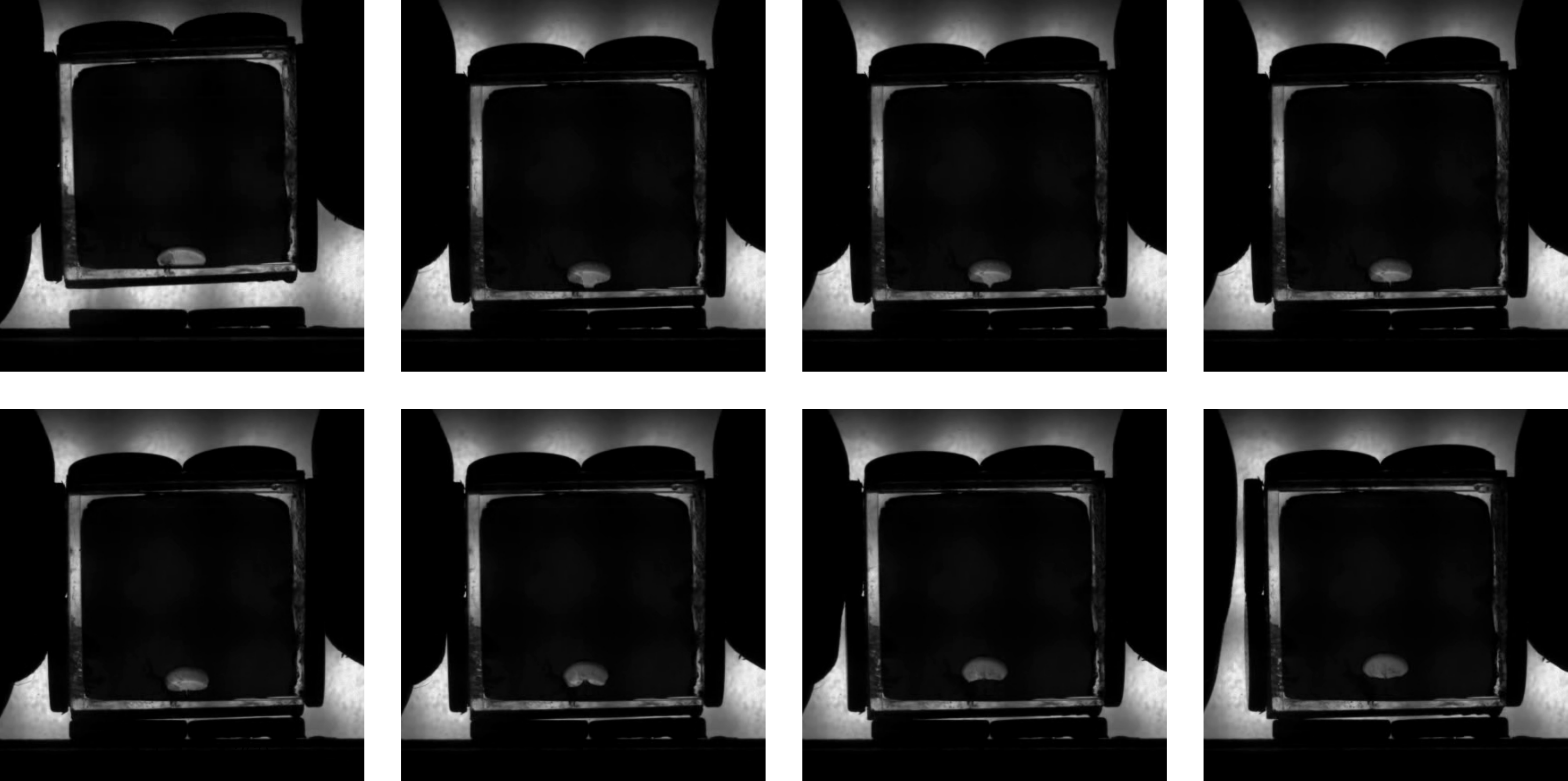}}
	\caption{The recording of one demonstrative experiment with a weakly magnetizable ferrofluid sample. The bottom pair of magnets remains fixed at the base and the cell is placed onto the magnets, initiating the levitation of the non-magnetizable droplet.}
	\label{figExpTest3}
\end{figure}
Under the influence of a single magnet at the base, it is noticed that the location of the levitated droplet is not exactly stable; there is a side wise movement of the droplet in addition to its vertical rise and the levitation path is not exactly a straight trajectory. In other words a stable levitation is not achieved using a single magnet at the bottom; after some initial period the droplet path tilts towards the side walls. To achieve a stable levitation and to restrict the side wise drift of { the droplet, an additional pair} of neodymium magnets, one at each side wall of the cell, is attached to provide a force on the droplet directing from the side walls towards the center of the cell. The single magnet at the bottom is now also replaced with a pair so that the north and south ends of the magnets make contact to each other in alternative fashion and thus remain attached to the cell. That is, if the north pole of a magnet is in contact with the wall, then its adjacent magnets will have their south poles in contact with the respective walls --- an arrangement very similar to the Halbach array of magnets {\citep{halbach1985application}}. One pair of magnets at the top wall closes the loop. The presence of this complimentary pair at the top wall is expected to reduce the levitation height of the droplet, but its absence gives rise to a field configuration which affect the stability of the levitation negatively and in an interesting way (here briefly, the absence of this pair of magnets at the upper wall gives rise to more than two possible equilibrium locations for the droplet; we elaborate on it in \S \ref{secResultsStability}). In figure \ref{figExpIntroduction} (left), one such pair of magnets is shown. The complete schematic for the magnet arrangement is shown in figure \ref{fig:ka}, and is further explained below in section \ref{secFormulation}.
 
One of the experimental demonstration conducted with this arrangement of magnets is shown in figure \ref{figExpTest3}. Eight different time instants are shown in the figure. Initially the water droplet remains in the vicinity of the bottom wall. Under this arrangement of magnets, the droplet eventually finds an stable equilibrium position depending upon the balance between the gravitational, buoyancy and the magnetic forces. Certainly, other magnet arrangements for a stable levitation of the water droplet may also be possible; the current alternating arrangement has proven specifically helpful in incorporating the magnetic field boundary conditions conveniently in terms of sine functions multiplied by the maximum strength of the magnet (\S \ref{secBcs}). This Halbach array arrangement is also practically helpful in keeping the magnet edges in contact with each other without any external forcing. 
\subsection{Physical explanation}
A possible physical explanation of the phenomena can be given after \cite{rosensweig2013ferrohydrodynamics}. First let us assume that there is no droplet. When the cell is brought in contact with the magnet, and then held stationary, the ferrofluid is attracted towards the magnet. For an initial period of time, some transients appear inside the ferrofluid due to this attractive force. However, as the cell is then kept stationary and there is not a continuous supply of work, the stationary magnetic force cannot make the ferrofluid to move continuously due to the thermodynamic constraints. The stresses inside the ferrofluid reorient themselves to counter this perpetual motion, much like the pressure redistribution in the gravitational field. The net response from the ferrofluid in this static condition is the redistribution of the mechanical pressure to balance the magnetic and gravitational forces; the pressure being higher in higher magnetic field regions. Now if a non-magnetizable object is placed inside the ferrofluid, it experiences the developed pressure gradient and starts moving away from the magnet. In the next section we describe the mathematical basis of our computations.
\section{Mathematical formulation}\label{secFormulation}
\begin{figure}
	\centerline{\includegraphics[width=0.5\textwidth]{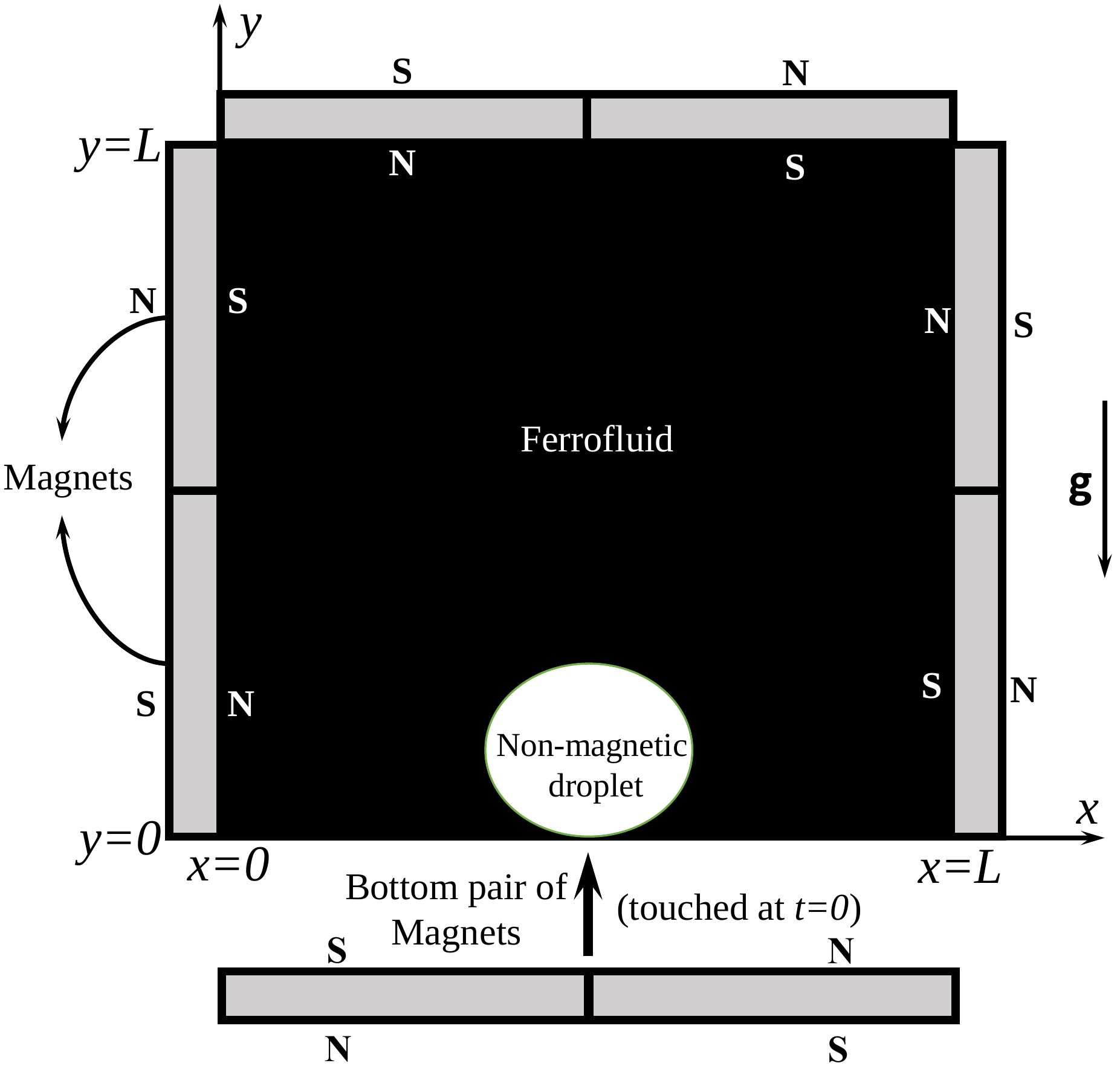}}
	\caption{The initial description of the problem domain, together with the arrangement of external permanent magnets. The flow domain $(0\leq x \leq L, 0\leq y \leq L)$ consists of a non-magnetizable droplet $(\Omega_d)$ immersed in an immiscible ferrofluid $(\Omega_f)$. The magnets are arranged in an alternate arrangement and the bottom pair of magnets is brought to contact at $t=0$, which initiates the levitation of the droplet. Here N and S represent north and south pole of the magnets respectively.}
	\label{fig:ka}
\end{figure}
The kerosene based ferrofluid-water combination inside the cell in our experiments represents an immiscible, incompressible two-phase system in two dimensions. We mimic this flow environment inside the cell by considering a square domain $\Omega\subset\mathbb{R}^2$. The domain consists of two fluid phases. The ferrofluid makes the bulk phase $(\Omega_f)$, while a non-magnetizable droplet of an immiscible liquid $(\Omega_d)$ is considered inside it (figure \ref{fig:ka}). The cell is covered with four magnet pairs in a Halbach array arrangement, as shown. The bottom pair of magnets initially remains detached from the cell and the droplet remains settled at the bottom. The levitation is initiated when this pair is brought in contact to the bottom wall. { In other words, both the fluid phases initially remain quiescent under the gravity and the magnetic field (applied at the top and the side walls). The flow is then initiated at some time instant (marked as $t=0$) when the magnetic field is applied at the bottom wall.} For computational simplicity, we utilize a two-dimensional idealization of the actual cell in our simulations. Considering the fact that the PPMA sheets were coated against the adhesion of the liquids, any effect arising due to the contact line pinning with the parallel walls are neglected. We also neglect the effects of the interfacial curvature along the third dimension perpendicular to the plane of the cell. Though, in general this can have an effect on the stability of the interface, we shall show, while comparing the simulations in section \S \ref{secResultsExperiments}, that these idealized simplifications have not introduced any considerable and qualitative change in the dynamics and { geometry} of the droplet interface during its rise.  

\subsection{Dimensional form of the governing equations\label{secGoverningEquationsDimensional}}
The motion of the non-magnetic droplet inside ferrofluid medium, and the continuous local flow and the magnetic field, are described by a coupling between the Navier-Stokes equations and the Maxwell's equations of electromagnetism. { If the fluids are considered electrically non-conducting, and assuming that the relaxation time of the magnetic nano-particles in the ferrofluid is much smaller than the relevant hydrodynamic time scales, the Maxwell's equations reduce to the magnetostatic form},
\begin{equation}
\left. \begin{array}{ll}  
\bnabla\bcdot\boldsymbol{B} = 0,\quad \bnabla\times\boldsymbol{H} = 0,\\[3pt]
\boldsymbol{B}=\mu_o (\boldsymbol{M+H}),
\end{array}
\right\}\mbox{ in }\Omega,
\label{eqMaxwell}
\end{equation}
where $\boldsymbol{B}$, $\boldsymbol{H}$ and $\boldsymbol{M}$ are the magnetic flux density, the magnetic field and the magnetization respectively while $\mu_o$ is the free space magnetic permeability. The irrotationality of $\boldsymbol{H}$ permits the relation
\begin{equation}
\boldsymbol{H}=\nabla\phi,
\label{eqGradPhi}
\end{equation}
where $\phi$ is a scalar magnetic potential. Also $\boldsymbol{M}$ is constitutively related to $\boldsymbol{H}$ through 
\begin{equation}
\boldsymbol{M}=\chi\boldsymbol{H},
\label{eqM} 
\end{equation}
where $\chi$ is the magnetic susceptibility. { For ferrofluids in general, the magnetization is itself governed by a differential equation involving magnetization relaxation time, however here we assume that the relaxation time is small and the magnetization relaxes in infinitesimally small time, so called quasi-equilibrium ferrohydrodynamic hypothesis}. A single scalar equation can be obtained using equation \ref{eqGradPhi} and \ref{eqM} in equation \ref{eqMaxwell}, expressed as
\begin{equation}
\bnabla\bcdot\mu_o(1+\chi)\nabla\phi=0,\mbox{ in }\Omega,
\label{eqPhi} 
\end{equation}
where the quantity $\mu_o(1+\chi)$ is equal to the magnetic permeability. 

{ The ferrofluid susceptibility varies with the magnetic field strength and other thermodynamic variables. In this study we consider it either a constant or a function of the local magnetic field. The simplifying assumption of constant permeability of ferrofluid is used in the first two set of simulations to obtain the basic characteristics of the droplet shape and levitation height, while being computationally efficient. It has also served as a reference for further refinement of our simulations using a more realistic field-dependent permeability model, when the experimentally observed shape is compared with the simulations.} In the latter case, the magnetization relation \ref{eqM} can be written in its non-linear form, i.e. $\boldsymbol{M}=\chi(H)\boldsymbol{H}$. Then the equation for magnetic potential is different for $\Omega_f$ and $\Omega_d$ and is expressed as (using Langevin's non-linear magnetization equation) 
\begin{equation}
\left. \begin{array}{ll}  
\displaystyle\bnabla\bcdot\mu_o\left(1+\frac{M_s}{|\nabla\phi|}\left[\mbox{coth}\;\gamma |\nabla\phi|-\frac{1}{\gamma |\nabla\phi|}\right]\right)\nabla\phi=0,\quad \mbox{in }\Omega_f\mbox{ if }\chi_f=\chi_f(H),\\[8pt]
\bnabla\bcdot\mu_o(1+\chi_f)\nabla\phi=0,\quad \mbox{in }\Omega_f\mbox{ if }\chi_f \mbox{ is constant},\\[8pt]
\bnabla\bcdot\mu_o(1+\chi_d)\nabla\phi=0,\quad\mbox{in }\Omega_d,
\end{array}
\right\}
\label{eqPhiBothPhases}
\end{equation}
where $M_s$ is the saturation magnetization of ferrofluid and $\gamma=3\chi_o/M_s$. { It should be noted that this specific non-linear ferrofluid magnetization model assumes that the ferrofluid exhibits nearly a paramagnetic behavior and also does not reflect the hysteresis of magnetization. It further assumes nearly mono-disperse size distribution of the magnetic nano-particles and negligible dipole-dipole interactions.} For non-magnetic liquids, the magnetic susceptibility is usually negligible, or in other words, their permeability can be considered equal to the free space permeability $\mu_o$. The equation \ref{eqPhiBothPhases} serves as the key to obtain $\boldsymbol{H}$ over the whole domain $\Omega$.

The equations for the isothermal and incompressible flow field are 
\begin{equation}
\left. \begin{array}{ll}  
\bnabla\bcdot\boldsymbol{v} = 0,\\[3pt]
\displaystyle\rho\frac{\boldsymbol{Dv}}{\boldsymbol{D}t}=\bnabla\bcdot(-p\mathsfbi{I}+\mathsfbi{S}+\mathsfbi{S}_m)+\rho \boldsymbol{g}+\boldsymbol{f}_s,
\end{array}
\right\}\quad \mbox{in}\quad \Omega,
\label{eqFlow}
\end{equation}
where $p, \mathsfbi{S}, \mathsfbi{S}_m, \boldsymbol{g}$ and $\boldsymbol{f}_s$ are the mechanical pressure, viscous stress tensor, magnetic stress tensor, gravitational acceleration and the interfacial force respectively. The Newtonian viscous stress tensor is $\eta(\bnabla\boldsymbol{v}+(\bnabla\boldsymbol{v})^T)$. The force $\boldsymbol{f}_s$ is expressed as $\sigma\kappa\boldsymbol{n}\delta_s$ and acts singularly at the interface. Here $\kappa$ is the local curvature of the interface, $\boldsymbol{n}$ is the unit outward normal at the interface and $\delta_s$ is the delta function at the interface.

The driving force in the present study is due to the magnetic stresses. There are number of different expressions for $\mathsfbi{S}_m$ that exist in the ferrohydrodynamic literature; interestingly, under incompressible and isothermal conditions and for isotropic permeability $\mu=\mu(H)$, the existing expressions for magnetic stress tensor reduce to the form 
\begin{equation}
\mathsfbi{S}_m=-a\mathsfbi{I}+\mu\boldsymbol{HH},
\label{eqMgStress}
\end{equation}
where the first part $-a\mathsfbi{I}$, which contains the isotropic magnetic pressure, can safely be lumped with $-p\mathsfbi{I}$ \citep{rosensweig2013ferrohydrodynamics, afkhami2008field}. This form of the magnetic stress tensor is particularly well adaptable to a conservative finite-volume formulation in comparison to the magnetic body force density expressions such as the Kelvin force density or the Korteweg-Helmholtz force density which are derivable by taking the divergence of $\mathsfbi{S}_m$. We directly discretize the divergence of \ref{eqMgStress} on finite volume cells utilizing $\int_V \bnabla\cdot\mathsfbi{S}_m dV=\int_S \mathbf{n}_s\cdot\mathsfbi{S}_m dS$ (where $V,S$ and $\mathbf{n}_s$ denote the computation cell volume, cell surface and the outward normal at the cell surface respectively), which conserves the magnetic force fluxes for each individual computational cell.
\subsection{Initial, boundary and interfacial conditions\label{secBcs}}
The no-slip and the no-penetration flow boundary conditions are considered at all the four walls while sinusoidal boundary conditions are considered for the gradient of magnetic potential to replicate the Halbach array of magnets, described as
\begin{equation}
\left.
\begin{array}{ll}
\boldsymbol{v} = \boldsymbol{0}, \mbox{ on }\p\Omega,\\[3pt]
\nabla\phi\bcdot\boldsymbol{n}_b=
\left\{
\begin{array}{ll}
H_o\mbox{sin}\left(2\upi x/L\right), \mbox{ on }\p\Omega_T,\\[3pt]
H_o\mbox{sin}\left(2\upi y/L\right), \mbox{ on }\p\Omega_L,\\[3pt]
-H_o\mbox{sin}\left(2\upi y/L\right), \mbox{ on }\p\Omega_R,\\[3pt]
-H_o\mbox{sin}\left(2\upi x/L\right), \mbox{ on }\p\Omega_B,
\end{array}
\right.
\end{array}
\right\}
\label{eqBcs}
\end{equation}
where $\boldsymbol{n}_b$ is the unit normal at the walls pointing into $\Omega$ and subscripts $T,L,R$ and $B$ represent top, left, right and bottom wall respectively.

Although the conditions at the interface are not explicitly needed in the numerical treatment of the two-phase, one-fluid formulation \citep{tryggvason2011direct}, we state them below for the sake of completeness. At the interface, the normal component of the magnetic flux density, the tangential component of the magnetic field, the tangential component of the total stress, and the velocity are continuous, while there is discontinuity in the normal component of the total stress,
\begin{equation}
\left.
\begin{array}{ll}
\displaystyle\boldsymbol{n}\bcdot[\boldsymbol{B}]=0,\quad
\boldsymbol{n}\times[\boldsymbol{H}]=0,\quad
[\boldsymbol{t}^T\mathsfbi{T}\boldsymbol{n}]=0,\quad
[\boldsymbol{v}]=0,\quad\\[3pt]
[\boldsymbol{n}^T\mathsfbi{T}\boldsymbol{n}]=\sigma\kappa,
\end{array}
\right\}
\label{eqIcs}
\end{equation}
where $\mathsfbi{T}=-p\mathsfbi{I}+\mathsfbi{S}+\mathsfbi{S}_m$ is the total stress tensor and $[x]$ denotes the difference of a quantity, $x$, right across the interface.
\subsection{Non-dimensionalization}
Looking at the governing model, the flow solution can be considered as dependent on the following dimensional parameters
\begin{equation}
\boldsymbol{v}=\boldsymbol{v}(\boldsymbol{x},t;\rho,\eta,R,g,\sigma,\mu,\gamma,H_o,M_s),
\end{equation}
The number of independent variables are reduced by properly identifying the non-dimensional groups which influence the flow solution. Taking $\Omega_d$ as the reference for all the properties except the magnetic permeability, for which FF has the permeability higher than the droplet medium, 
and considering the following reference scales,
\begin{equation}
\left.
\begin{array}{ll}
\nabla\sim R^{-1},\;\rho\sim\rho_d,\;\eta\sim\eta_d,\;\mu\sim \mu_f,\\[3pt]
\boldsymbol{v}\sim \eta_d/\rho_d R,\;t\sim \rho_d R^2/\eta_d,\;p\sim \eta^2_d/\rho_d R^2, \\[3pt]
\boldsymbol{H}\sim H_o,\;\boldsymbol{B}\sim \mu_oH_o,\;\boldsymbol{f}_s\sim\sigma/R^2,\\
\end{array}
\right\}
\label{eqScales}
\end{equation}
the equations of fluid motion are normalized to the following form
\begin{equation}
\left. \begin{array}{ll}  
\bnabla^*\bcdot\boldsymbol{v}^* = 0,\\[3pt]
\displaystyle\rho^*\frac{\boldsymbol{Dv^*}}{\boldsymbol{D}t^*}=\bnabla^*\bcdot(-p^*\mathsfbi{I}+\mathsfbi{S}^*)+La_m\;\bnabla^*\bcdot\mathsfbi{S}_m^*+Ga\;\rho^*\boldsymbol{g}^*+La\;\boldsymbol{f}_s^*,
\end{array}
\right\}\quad \mbox{in}\quad \Omega,
\label{eqFlowNormalized}
\end{equation}
where the non-dimensional group
\begin{equation}
La=\frac{\sigma\rho_d R}{\eta_d^2}
\end{equation}
is the Laplace number signifying {the ratio of the interfacial force} to the the viscous force,
\begin{equation}
La_m=\frac{\mu_f\rho_d H_o^2 R^2}{2\eta_d^2}
\end{equation}
is the magnetic Laplace number signifying the ratio of the magnetic force to the the viscous force, and
\begin{equation}
Ga=\frac{g\rho_d^2 R^3}{\eta_d^2}
\end{equation}
is the Galilei number signifying the ratio of the gravitational force to the viscous force. 

The Maxwell's equations transform to the following non-dimensional form,
\begin{equation}
\bnabla^*\bcdot\boldsymbol{B}^*=0,\quad
\bnabla^*\times\boldsymbol{H}^*=0,\quad
\boldsymbol{B}^*=\boldsymbol{H}^*+\left[\frac{1}{\xi_o}\right]\boldsymbol{M}^*
\end{equation}
where $\xi_o=H_o/M_s$. The magnetization relation takes the following normalized form
\begin{equation}
\boldsymbol{M}^*(\boldsymbol{H}^*)=\displaystyle\left[\mbox{coth}\;\gamma_o H^*-\frac{1}{\gamma_o H^*}\right]\frac{\boldsymbol{H}^*}{H^*}=\chi^*(\gamma_o,H^*)\boldsymbol{H}^*,
\end{equation}
where 
\begin{equation}
\gamma_o=\frac{3\chi_oH_o}{Ms}=3\chi_o\xi_o.
\end{equation}
Using this relation for magnetization, the equation for magnetic potential takes the following non-dimensional form
\begin{equation} 
\displaystyle \bnabla^*\bcdot\left(1+\frac{1}{\xi_o|\nabla^*\phi^*|}\left[\mbox{coth}\;\gamma_o |\nabla^*\phi^*|-\frac{1}{\gamma_o |\nabla^*\phi^*|}\right]\right)\nabla^*\phi^*=0,
\label{eqPhiBothPhasesNormalized}
\end{equation}
or
\begin{equation} 
\displaystyle \bnabla^*\bcdot\left(1+\frac{1}{\xi_o}\chi^*(\gamma_o,H^*)\right)\nabla^*\phi^*=0.
\label{eqPhiBothPhasesNormalized2}
\end{equation}
Therefore, four non-dimensional parameters, $La,La_m,Ga$ and $\gamma_o$ are significant and thus the flow solution in case of field dependent ferrofluid susceptibility can be expressed by the functional form
\begin{equation}
\boldsymbol{v}^*=\boldsymbol{v}^*(\boldsymbol{x}^*,t^*;La,La_m,Ga,\gamma_o).
\label{equFunctionalFormV}
\end{equation}
In case of constant ferrofluid susceptibility, where there is no bound on $\boldsymbol{M}$, the reference scale for the magnetization is chosen equal to $H_o$. In such case the functional form of the flow solution reads
\begin{equation}
\boldsymbol{v}^*=\boldsymbol{v}^*(\boldsymbol{x}^*,t^*;La,La_m,Ga).
\end{equation}
The parameter $\gamma_o$ depicts the effect of non-linearity of the magnetization on the flow solution. In the following sections we drop out the star symbols on the non-dimensional variables for convenience.
\section{Numerical method}\label{secNumerical}
We solve the flow dynamic equations \ref{eqFlow} numerically utilizing the one-fluid approach \citep{tryggvason2011direct} in which the set of equations over the whole domain $\Omega$ can be solved if the spatial distributions of the fluid properties are known. We perform the space discretizations of the flow dynamic equations \ref{eqFlow} as well as the magnetic potential equation \ref{eqPhiBothPhases} using the finite volume method over a standard staggered rectangular grid. To march in time, we utilize a second order pressure projection algorithm. The two phases are recognized using a marker function 
\begin{equation}
C(\boldsymbol{x},t)=
\left\{
\begin{array}{ll}
1, \mbox{ in }\Omega_f,\\[3pt]
0, \mbox{ in }\Omega_d.\\
\end{array}
\right.
\label{eqMarker}
\end{equation}
To advect the marker function, we use the front-tracking scheme of \citet{unverdi1992front}, in which the interface location is first updated using the velocity field solution and then the marker function is reconstructed from the known interface location. The detailed description of the projection algorithm, the front-tracking algorithm and the principles of finite-volume discretization are presented in \citet{tryggvason2011direct}, and here we only describe the incorporation of equation for magnetic potential (\ref{eqPhi}), the handling of the magnetic permeability field $\mu$ and the incorporation of the magnetic stresses in the equation of motion \ref{eqFlow} in the overall algorithm.
\begin{table}
	\begin{center}
		\def~{\hphantom{1}}
		\scalebox{1.0}{
			\begin{tabular}{lcccccc}
				Phase&$\rho\:(\si{kg.m^{-3}})$& $\eta\:(\si{Pa.s})$ & $\mu\:(\si{N.A^{-2}})$ &$M_s\:(\si{G})$ &$\chi_o$& $\sigma\:(\si{N.m^{-1}})$\\[8pt]
				Ferrofluid ($\Omega_f$)  &$868.0$& $0.025$ & $^{*}$  & 57.7 &$ 0.0819$ & \multirow{2}{*}{$0.97\times10^{-3}$}\\
				Water ($\Omega_d$)       &$1000.0$& $0.001$ & $\mu_o$ & - & 0.0 &\\
			\end{tabular}}
			\caption{Physical properties of both the phases. The saturation magnetization of the ferrofluid sample is measured by EverCool SQUID VSM DC magnetometer and the interfacial tension is determined by a technique used in \cite{zhu2011nonlinear}.\\
				$^*$Computed using Langevin's function for ferrofluid magnetization.}
			\label{tableExpVsSim}
		\end{center}
	\end{table}	
	
\subsection{Description of the algorithm\label{secDescriptionOfAlgorithm}}
Initially at $t=0$, both the fluids are considered quiescent $(\boldsymbol{v}(\boldsymbol{x},0)=\boldsymbol{0})$. The initial location of the interface between the two phases is considered to be known, or in other words, the initial discrete distribution of the fluid property fields - $\rho_{i,j}, \eta_{i,j}, \mu_{i,j}$ - on the grid is considered to be known; $i,j$ being the indices associated with concerned grid point. Knowing $\mu_{i,j}$ at $t=0$, the equation for the magnetic potential $\phi$ (\ref{eqPhi}) is first solved. Obtaining $\phi$, $\boldsymbol{H}$ is obtained from $\boldsymbol{H}=\nabla\phi$. Once $\boldsymbol{H}$ is known, the magnetic body force is computed. The equation of motion \ref{eqFlow} is then solved for the velocity field $\boldsymbol{v}$ using the pressure projection algorithm \citep{tryggvason2011direct}. Using $\boldsymbol{v}$, the location of the interface is then advected, and from this updated location of the interface, the discrete marker function $C_{i,j}$ is reconstructed using the front-tracking algorithm \citep{tryggvason2011direct,unverdi1992front}. The discrete density and viscosity fields are then interpolated from $C_{i,j}$ as
\begin{equation}
\rho_{i,j}=\rho_f C_{i,j}+\rho_d(1-C_{i,j}),\quad \eta_{i,j}=\eta_f C_{i,j}+\eta_d(1-C_{i,j}).
\end{equation}
The permeability $\mu_{i,j}$ is obtained in a similar fashion if it is assumed constant in both the phases. However, when the non-linear magnetization model is considered, the permeability is considered constant only in the non-magnetizable phase. In the magnetizable phase, it is predicted using the Langevin's relation for the ferroflud magnetization. Both the above cases are summarized as
\begin{equation}
\left. \begin{array}{ll}  
\displaystyle\mu_{i,j}=\mu_f C_{i,j}+\mu_d(1-C_{i,j}),\quad \mbox{if }\mu_f \mbox{ is assumed constant},\\[8pt]
\mu_{i,j}=\left(\mu_o\left(1+\frac{M_s}{|\nabla\phi|}\left[\mbox{coth}\;\gamma |\nabla\phi|-\frac{1}{\gamma |\nabla\phi|}\right]\right)\right)C_{i,j}+\mu_d(1-C_{i,j}),\quad \mbox{if }\mu_f=\mu_f(H),
\end{array}
\right\}
\label{eqPhiBothPhasesMuij}
\end{equation}
Once the property fields are advected, the algorithm is advanced to the next time step and the numerical cycle is repeated till desired time.

{ The use of constant magnetic permeability is a greater computationally simplification and results in lower computational times. Therefore it is useful for obtaining very basic features of the flow, and also when parameters other than magnetic permeability (or susceptibility) were varied. An example is the variation of viscosity ratio, which is not related to the permeability or magnetization of the ferrofluid in our formulation. However, instead of varying the permeability ratio, and for comparison with experiments, we adopt the non-linear magnetization model.}

The spatial as well as the temporal discretizations in our scheme are second order accurate. The advection term in the momentum equation is handled using a second order essentially non-oscillatory (ENO) scheme while the standard second order centered in space discretization is applied to the diffusive viscous terms. The interpolations of the viscosity at the computational cell faces are performed harmonically while the density and the permeability are interpolated arithmetically. A highly optimized V-cycle MULTIGRID method is implemented for the solution of the pressure Poisson equation and the equation for the magnetic potential.  
\subsection{Setup of simulations}

\begin{table}
	\begin{center}
		\def~{\hphantom{1}}
		\scalebox{0.85}{
		\begin{tabular}{lccccc}
			Section  & \S\ref{secResultsConstantPermeability}   &   \S\ref{secResultsViscosityRatio} &   \S\ref{secResultsVariablePermeability} &   \S\ref{secResultsStability}&   \S\ref{secResultsExperiments}\\[8pt]
			
			{Focus}  & \begin{tabular}{@{}c@{}}Droplet shapes and\\ levitation height\end{tabular} &  \begin{tabular}{@{}c@{}}Effect of \\viscosity ratio \end{tabular} &   \begin{tabular}{@{}c@{}}Effect of non-linear\\ magnetization \end{tabular} &   \begin{tabular}{@{}c@{}}Stability of\\ levitation \end{tabular} & \begin{tabular}{@{}c@{}}Comparison \\with experiments\end{tabular}\\[16pt]
			
			$La$  & 0.1,1.0,10.0    &   0.1,1.0,10.0 &   0.1 & 10.0 & 2.09\\[3pt]
			
			$La_m$  & 40,160,360,640,1000    &   40,160,360,640,1000 &   1000$^{*}$  & 360,1000 & 5862.79$^{*}$\\[3pt]
			
			$Ga$  & 0.1,1.0,10.0    &   0.1,1.0,10.0 &   0.1  & 1.0 & 38.6\\[3pt]
			
			$\rho_d/\rho_f$  & 2 &  2 &  2  & 2 &  1.152\\[3pt]
			
			$\eta_d/\eta_f$  & 2 &  0.5 &  2  & 2 & 0.04\\[3pt]
			
			$\mu_d/\mu_f$  & 0.25 &  0.25 &  $^{**}$  & 0.25 & $^{**}$\\[3pt]
			
			$\gamma_o^1$  & $^{***}$   &   $^{***}$ & \begin{tabular}{@{}c@{}}0.111,0.278,0.389,0.556,\\0.833,1.111,1.389,1.667 \end{tabular} & $^{***}$ &  2.06\\[3pt]
			
		\end{tabular}}
		\caption{Breakup of simulation results and corresponding values of input parameters.\\
			$^{1}\gamma_o=3\chi_oH_o/M_s=3\chi_o\xi_o$.\\
			$^{*}$The vacuum permeability is used to compute characteristic $La_m$, as the permeability of the ferrofluid in this case is a variable determined by Langevin's function.\\
			$^{**}$The quantity is computed using Langevin's function for the field dependent ferrfluid magnetization.\\
			$^{***}$The quantity is not uniquely defined, or do not appears in the formulation.}
		\label{tableSim}
	\end{center}
\end{table}

 The properties of the two fluids are presented in table \ref{tableExpVsSim}, where $\rho_d\sim 10^{3}\:\si{kg.m^{-3}}$, $\eta_f\sim 10^{-2}\:\si{Pa.s}$ and $\sigma\sim 10^{-3}\:\si{N.m^{-1}}$. We consider a water droplet of radius of $\sim1\:\si{mm}$. The magnetic fields used in our experiments are of $\sim10^2\:\si{kA.m^{-1}}$ while the order of permeability of ferrofluid is considered a multiple of $\mu_o$. For these orders of magnitude of the properties, the $La\sim10$ and $Ga$ can be as high as $\sim10^{2}$, while a realistic $La_m$ turns out to be $\sim10^{4}$. Considering these, we simulate for $La$ and $Ga$ in the range $0.1-100$ while for $La_m$ in the range $10-1000$. Being conservative, the $La_m$ is not increased further to avoid any numerical instability arising due to the higher order of the magnetic source term in the momentum equation. 
 
 The value of the parameter $\gamma_o$ changes due to change in $H_o$, as $\chi_o$ or $M_s$ do not vary for a given ferrofluid sample. Here, the initial susceptibility of ferrofluids is measured to be $\sim 0.0819$, while $M_s\sim 10^{3}\:\si{A.m^{-1}}(57.7\:\si{G})$. Thus for applied field between $10^3-10^4\:\si{A.m^{-1}}$, the order of $\gamma_o=3\chi_oH_o/M_s$ is expected to vary between $0.2-2.0$. We simulate for $\gamma_o$ in the range $0.1-2$. 
 
 The breakup of simulations and all the corresponding simulation parameters are summarized in table \ref{tableSim}. The simulations are divided into six different sets with different foci. In the first set, we simulate for constant permeability in both the phases. Next, the effect of viscosity ratio is investigated. In the third set of simulations, the effects of field dependent ferrofluid permeability, or equivalently the effects of non-linear magnetization, are simulated. The stability of levitation is the focus of the subsequent set, while the appearance of interfacial singularity and comparison with experiment is addressed through the last set of simulations. The same hierarchy, as listed in table \ref{tableSim}, is utilized for sectioning the results.

\subsection{Output fields and variables \label{secFormulationHeightAndSpeed}}
Besides the magnetic and velocity field solutions, we study the shapes of the levitating droplet, the final levitation height, the time dependent displacement and velocity of the droplet tip/nose, the droplet deformation (defined below), and the time averaged change in the droplet deformation. In certain cases, we also utilize the absolute magnetic field contours and the vorticity contours.

The discrete information about the interface is described by two vectors - $\{x_k\}=(x_1,x_2,...,x_N)^{T}$ and $\{y_k\}=(y_1,y_2,...,y_N)^{T}$ - storing $x$ and $y$ co-ordinates of $N$ discrete interface points. This structure is compatible with the implementation of front-tracking algorithm. The total number of interface points $N$, however, can be different for each time step due to their dynamic addition or deletion, a part of the front-tracking algorithm. From this information about the interface, the vertical distance of the tip of the droplet from the bottom wall is extracted using
\begin{equation}
h_{tip}(t)=\mbox{max}(\{y_k\}),
\end{equation} 
and the speed of the tip is computed using 
\begin{equation}
v_{tip}(t)=\left[h_{tip}(t+\Delta t)-h_{tip}(t)\right]/\Delta t,
\end{equation} 
where $\Delta t$ is the time step. 
 
To study the time dependent deformation of the levitating droplet, a global droplet deformation parameter or shape factor, $\mathcal{D}$, is computed from the interface information using
\begin{equation}
\left.
\begin{array}{ll}
\displaystyle \mathcal{D}(t)=\frac{A_x(t)-A_y(t)}{A_x(t)+A_y(t)},\\[3pt]
\displaystyle A_y(t)=\mbox{max}(\{y_k\})-\mbox{min}(\{y_k\}),\\[3pt]
\displaystyle A_x(t)=\mbox{max}(\{x_k\})-\mbox{min}(\{x_k\}).
\end{array}
\right\}
\end{equation} 
Notice that $\mathcal{D}(t)$ can be negative if $A_x(t)<A_y(t)$, which implies that the vertical span of the droplet is more than its horizontal span.

To show the complete $\mathcal{D}(t)$ curve for all the simulations is a cumbersome task. To mark the average extent of the droplet deformation over all times, a time averaged shape factor, $\langle \mathcal{D}\rangle$, is computed. It reduces the results for the shape factor to one value for one simulation. The expression for $\langle \mathcal{D}\rangle$ is
\begin{equation}
\displaystyle \langle \mathcal{D}\rangle=\frac{1}{t_{end}}\sum_{t=0}^{t_{end}} (\mathcal{D}(t+\Delta t)-\mathcal{D}(t)),
\label{equDeformationParameter}
\end{equation}
where $t_{end}$ is the time up to which the flow is simulated. 
\subsection{Code Validation and Grid and time step independence}
The computational code is developed and tested for single phase as well as two phase viscous incompressible flow, with and without the magnetic effects, and has been successfully applied in problems by the authors to predict the interfacial \citep{singh2016single} and relaxation mechanisms \citep{singh2016flow} in ferrofluids. We incorporate the non-linear magnetization/field dependent permeability model and the conservative discretization of the full magnetic body stress tensor, and a highly optimized multigrid technique for the solution of the pressure Poisson and the magnetic potential equation. For the problem at hand, the grid as well as the time step independence of the code is carefully studied. The details of these consistency checks have been presented in appendix \ref{appA}. Besides these quantitative consistency checks, the physical correctness also confirmed in our study through its ability to predict the experiments, especially the fine features at the interface (\S\ref{secResultsExperiments}). 
\section{ Droplet shapes and levitation height: Effect of $La,\;La_m$ and $Ga$}\label{secResultsConstantPermeability}
In the first set of simulations, the magnetic permeabilities ($\mu_f,\;\mu_d$) are considered constant. As discussed before, this theoretical assumption served as a reference for further refining of our simulations using a more realistic field-dependent permeability model. The simulation parameters are given in the second column of table \ref{tableSim}. The intermediate case of $La, Ga$ and $La_m$ -- respectively $1.0,1.0$ and $360$ -- is first discussed, while the variations in $La, Ga$ and $La_m$ are discussed afterwards. 

\begin{figure}
	\centerline{\includegraphics[width=0.7\textwidth]{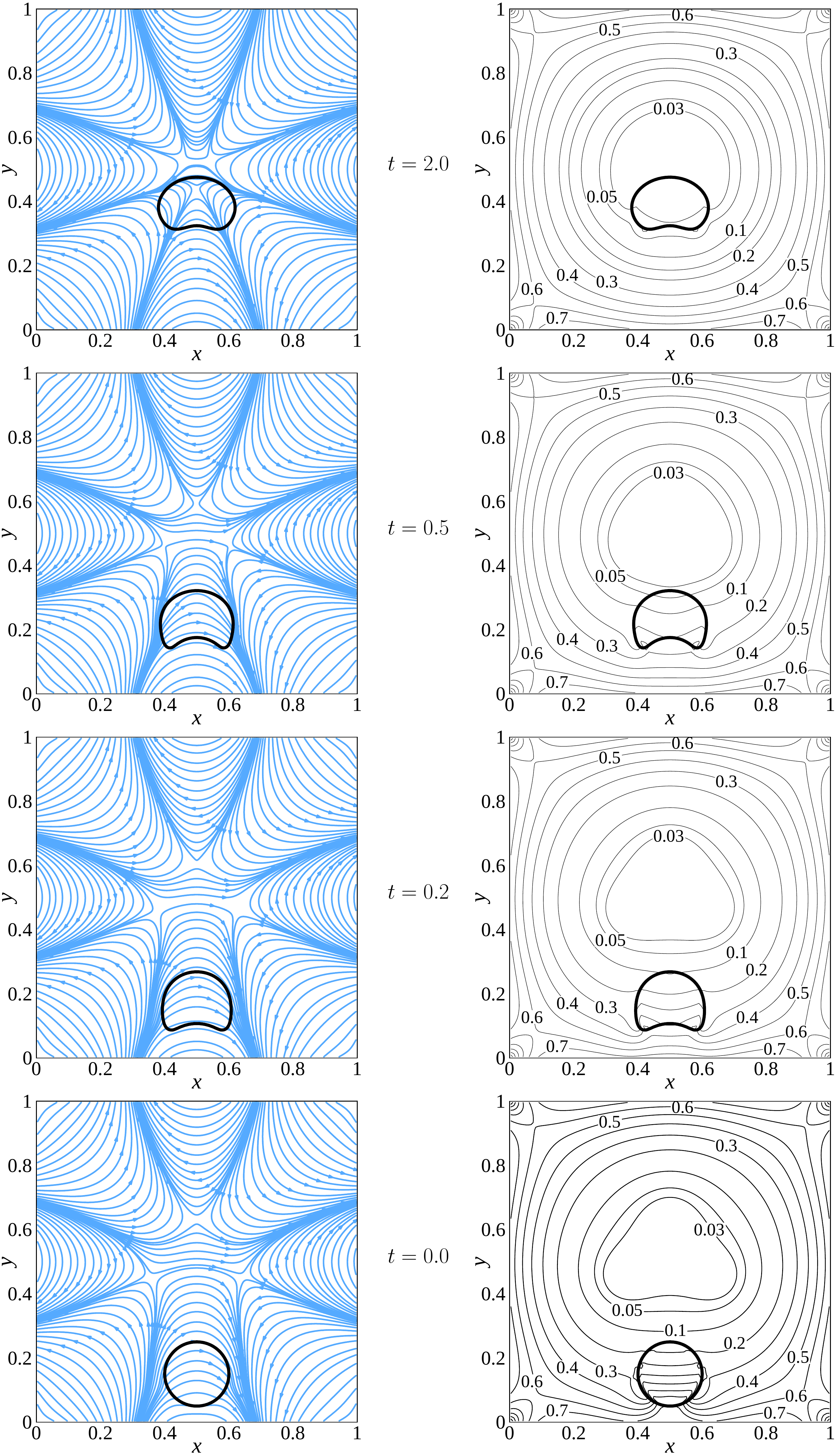}}
	\caption{The interface of the levitating droplet, the magnetic field lines (left column) and the absolute magnetic field contours (right column) at different instants of time. $La=1.0,Ga=1.0,La_m=360$. Here gravity points downwards.}
	\label{figPhenomena}
\end{figure}

The numerically predicted movement of the droplet for the above values of the non-dimensional numbers is shown in figure \ref{figPhenomena}. The gravity is pointing vertically downward and the droplet is levitating upward, defying the gravitational field.
Both the magnetic filed lines (left column) and the absolute magnetic field contours (right column) around the droplet are shown. The patterns of the magnetic field lines due to the Halbach array of magnets at the walls are apparently quite complex but symmetric in configuration. The lines are originating and are terminating at the alternate magnet surfaces; the originating lines are more concentrated near the center of the magnets. 

The mechanism behind the droplet levitation against the gravity is better understood by looking at the absolute field contours in the right column of figure \ref{figPhenomena}. The absolute magnetic field contours are normalized with $H_o$ and thus their magnitude vary from $0$ to $1$. The absolute $H$ field is higher near the walls of the domain while approaches zero near the center of the domain, implying that the field gradient is acting from the walls towards the center of the domain. Near the bottom part of the droplet at the initial condition, the value of $H/H_o$ is nearly $0.7$ while it is in between $0.1$ to $0.2$ near the top of the droplet. Thus in a global sense, the droplet experiences a net upward magnetic force proportional to this gradient, which in this case, has turned out to be sufficient for the levitation of the droplet against the downward gravitational force. A minimum magnetic field region has established at the center of the domain, and the droplet (as a bulk) seeks this region of minimum magnetic field strength. As the droplet moves upward, the field gradient across its poles reduces. Observing the time instants $t=0.2,0.5,1.0$ and $2.0$ in figure \ref{figPhenomena}, it comes out that the difference between the field strength at the bottom and the top of the droplet reduces, respectively as $0.3,0.16,0.09$ to approximately $0.03$. As the droplet approaches the center of the domain, it retards and finally reaches an equilibrium location and shape. A concavity at the tail of the droplet develops during its rise, its extent reaches a maximal, and then it reduces until an equilibrium configuration of the tail is attained. 

\begin{figure}
	\centerline{\includegraphics[width=0.8\textwidth]{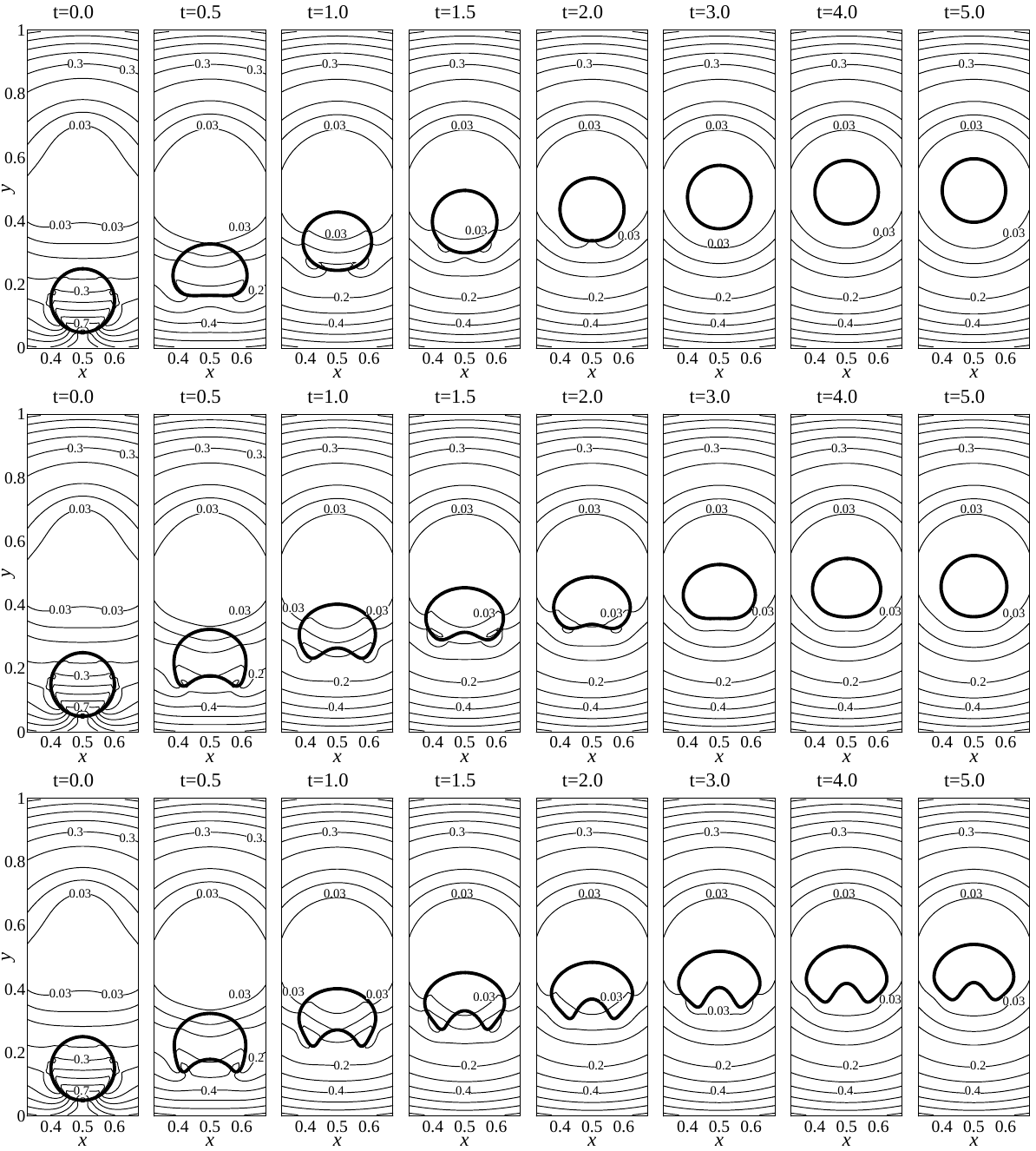}}
	\caption{The interface of the levitating droplet and the absolute magnetic field lines for $La=10.0$ (top row), $1.0$ (middle row) and $0.1$ (bottom row). $Ga=0.1,La_m=360$.}
	\label{figEffectOfLa}
\end{figure}
\subsection{The effect of change in $La$}
The non-dimensional number $La$ is varied for fixed $Ga$ and $La_m$. The interface of the levitating droplet for $Ga=1.0$ and $La_m=360$ is shown in figure \ref{figEffectOfLa}. The flow is simulated for three distinct $La$ -- $0.1,1.0$ and $10.0$. Note that only a part of the domain ($0.325\leq x\leq 0.675, 0.0\leq y\leq 1.0$) is focused. The absolute field strength contours are also presented. Whereas the steady state levitation height of the droplet seems nearly independent of the $La$, the difference lies in the extent of the deformation of the interface. For $La=10.0$, it is only during the very start of the phenomenon (instant $t=0.5$ and $1.0$) {when} the droplet shape deviates from its initial round configuration. The droplet regains nearly the round shape at around $t=1.5$. As the $La$ is decreased, the deformation of the interface becomes more prominent. The tail of the levitating droplet deforms to have a concavity. As time progresses, this feature at the bottom of the droplet disappears in case of $La=1.0$. However, this feature at the tail of the droplet stays in case of $La=0.1$. The shape in case of $La=0.1$ seems to resemble a \emph{crescent}, except the fact that round projections are appearing instead of sharp cusps at the tail. The increase in the droplet deformation due to a decrease in the $La$ number is expected, as a smaller $La$ signifies a lower interfacial energy. However, the transition specifically towards the \emph{crescent} like shapes is worth noticeable. The droplet retains its symmetry about the vertical axis and is asymmetric about the horizontal axis. Its shape is nearly an \emph{oval} in case of $La=10.0$ and at low time $t=0.5$. Lowering the interfacial energy further by reducing $La=1.0$ results in a transition from \emph{oval} towards \emph{crescent}. At longer time, the droplet regains its round shape at high $La$ while the deformed shapes persist for low $La$. The change in the final average levitation height of the droplet due to change in $La$ is meager.    

\begin{figure}
	\centerline{\includegraphics[width=0.8\textwidth]{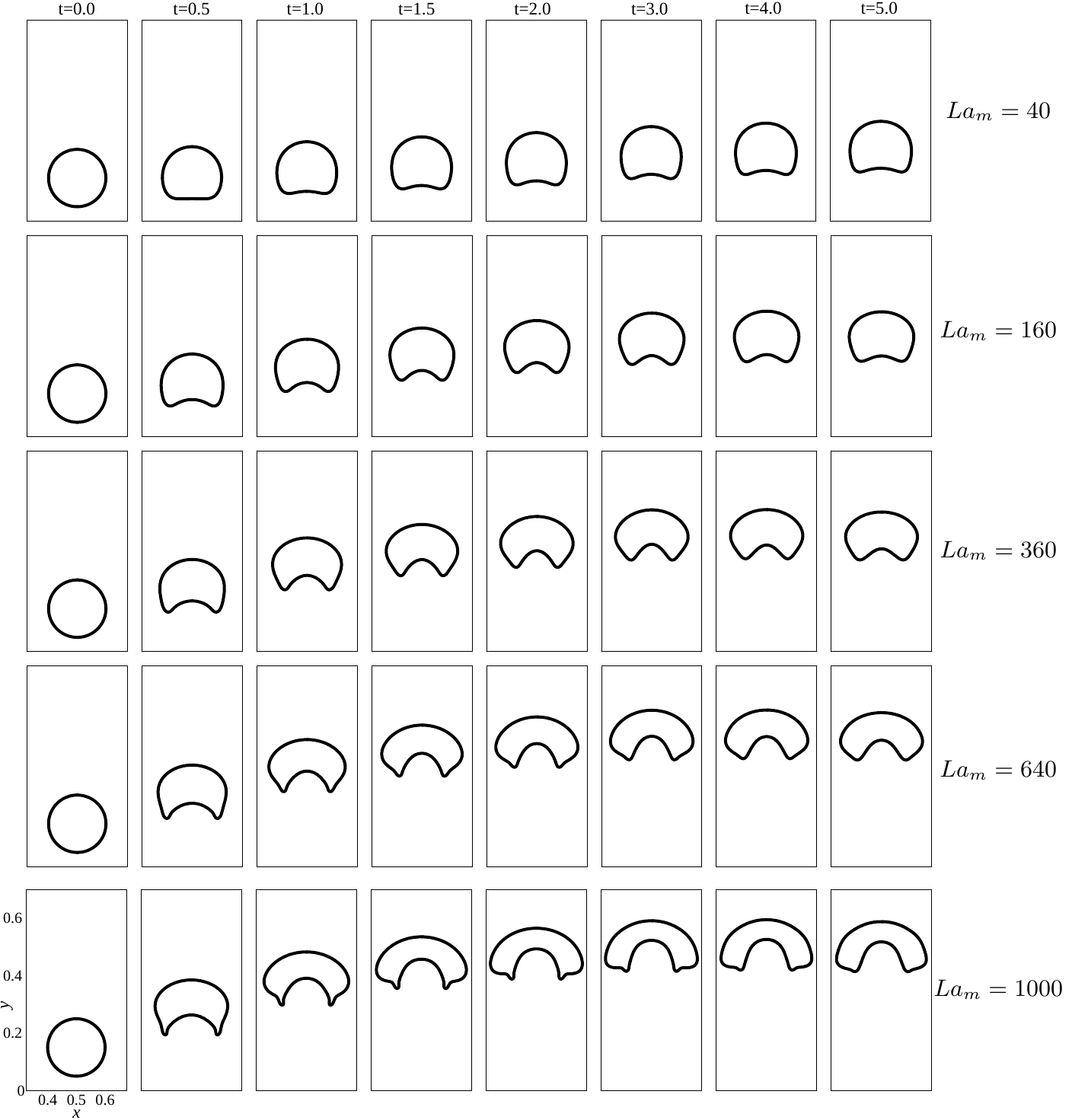}}
	\caption{The interface of the levitating droplet for different $La_m$. $Ga=1.0,La=0.1$.}
	\label{figEffectOfLam}
\end{figure}
\subsection{The effect of change in $La_m$}
The deformation of {the droplet is not solely sensitive} to $La$. In this subset of simulations, the $La$ and $Ga$ numbers are fixed while the deformation is studied with changing $La_m$. These results are depicted in figure \ref{figEffectOfLam}. The droplet rises to a relatively greater height and its deformation is also enhanced at increased $La_m$. A further difference is noted for $La_m=640$ and $1000$. In these two cases, the shape of the rising droplet is more \emph{skirted} for $t\leq 2.0$. For $La_m=1000$, the interfacial deformation is more pronounced and the droplet eventually gains a \emph{segmented ring} like shape for $t\geq1.0$. The increase of $La_m$ from $40$ to $1000$ leads to a further shape transition at late times -- first from \emph{oval} to nearly \emph{crescent}, {and then to a nearly \emph{segmented-ring} }. 

The levitation height increases with $La_m$ due to the stronger field gradients caused by the increase in $La_m$. The reasons behind the transitions in the shape of the droplet, however, seem more involved. For $La_m=1000$, the initial transition from the round towards a skirted configuration, and eventually to the segmented-ring shape, is particularly intriguing. A possible physical explanation is as follows. In figure \ref{figEffectOfLam}, the droplet at $t=5.0$, for all $La_m$, is close to its steady state. It is clear that as $La_m$ increases, the droplet is attaining its equilibrium more and more closer to the center of the domain ($y=0.5$), where the magnetic field lines (not absolute field contours) are more distorted. In case of lower interfacial energy for lower $La$, the field patterns near the center of the domain tend to deform the droplet to a {segmented ring like shape}. On the other hand, lower $La_m$ magnitudes fail to raise the droplet to such field line regions, and this ends up with less alteration in the interface configuration. The magnetic field is primarily responsible for the deformation of the droplet at steady state when flow velocity vanishes. The hydrodynamic flow field is thus at play primarily during the rise of the droplet before the steady state. As the maximal extent of the concavity at the tail of the droplet is observed during the rise of the droplet, the hydrodynamic flow helps in enhancing the deformation caused by the magnetic interfacial force. Once the steady state is reached, the flow field die out and only the balance between magnetic, interfacial tension and gravity forces govern the deformed shape. This insight is also quantitatively supported by the time curves of the droplet deformation $\langle\mathcal{D}\rangle)$ (appendix A), where a peak in $\langle\mathcal{D}\rangle$ is observed before the steady state. The deformation relatively decreases after this hydrodynamically dominated regime.

\begin{figure}
	\centerline{\includegraphics[width=0.7\textwidth]{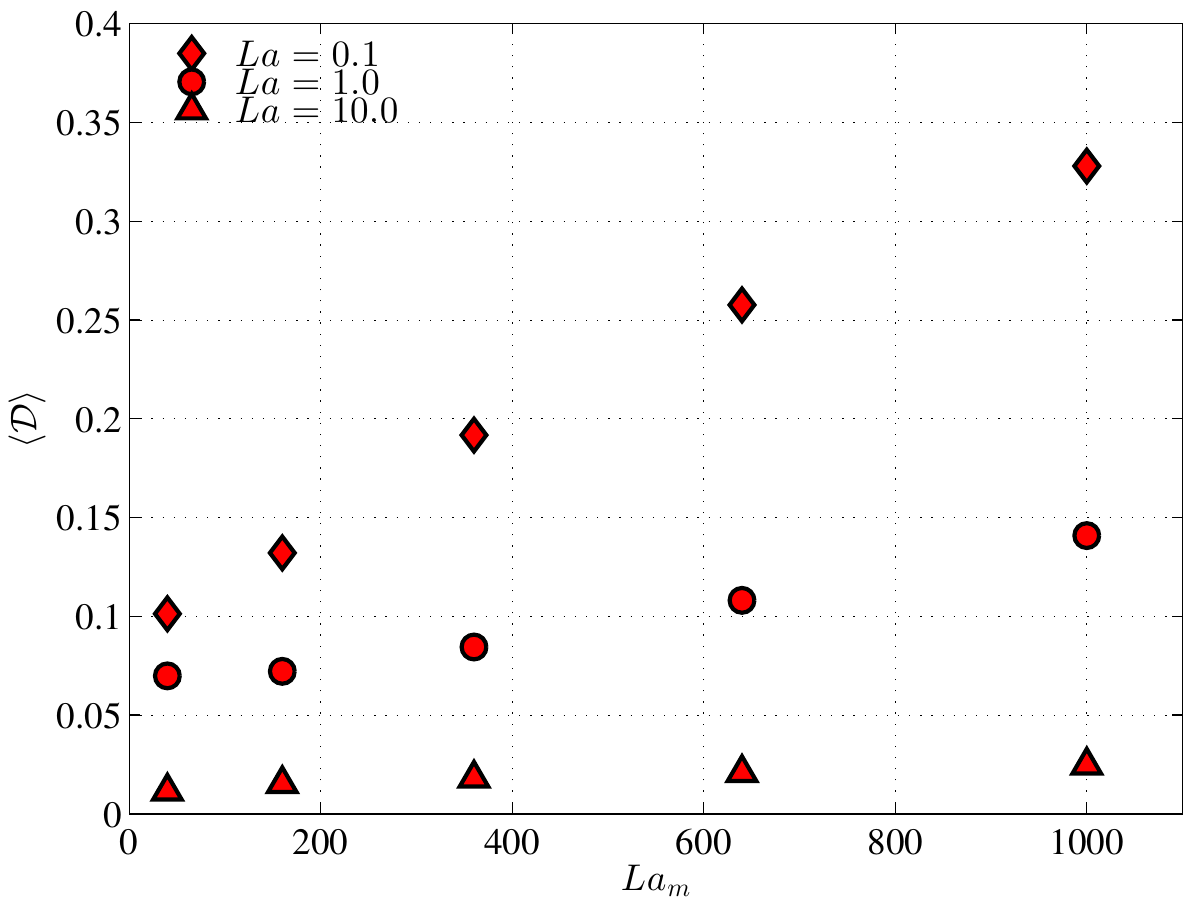}}
	\caption{The change in the global time averaged deformation parameter $\langle \mathcal{D}\rangle$ (equation \ref{equDeformationParameter}) with respect to $La_m$ and $La$, quantitatively signifying that the deformation of the droplet increases with increasing $La_m$ but decreasing $La$. $Ga=1.0$.}
	\label{figDeformationParameter}
\end{figure}

The extent of deformation of the droplet with respect to the $La$ and $La_m$ is quantified in figure \ref{figDeformationParameter} with the help of the time averaged droplet deformation parameter. It supports that the deformation significantly increases with increasing $La_m$, provided $La$ is low enough. For example at $La=0.1$, the deformation parameter increases from $0.1$ to $0.33$ as $La_m$ is increased from $40$ to $1000$. On the other hand, at three orders of magnitude higher $La=10$, the value of the deformation parameter is negligible, and stays close to zero even for high $La_m$. This clearly suggest that the deformation of the droplet is maximized at low $La$ and higher $La_m$. In addition, the time averaged deformation increases nearly linearly with $La_m$. The slope of the linear dependence is dependent on $La$. At high $La$, the slope is close to $0$. A decrease in $La$ amplifies the effect of $La_m$ on the global deformation of the droplet. 

\subsection{The effect of change in $Ga$\label{secEffectGa}}
\begin{figure}
	\centerline{\includegraphics[width=0.8\textwidth]{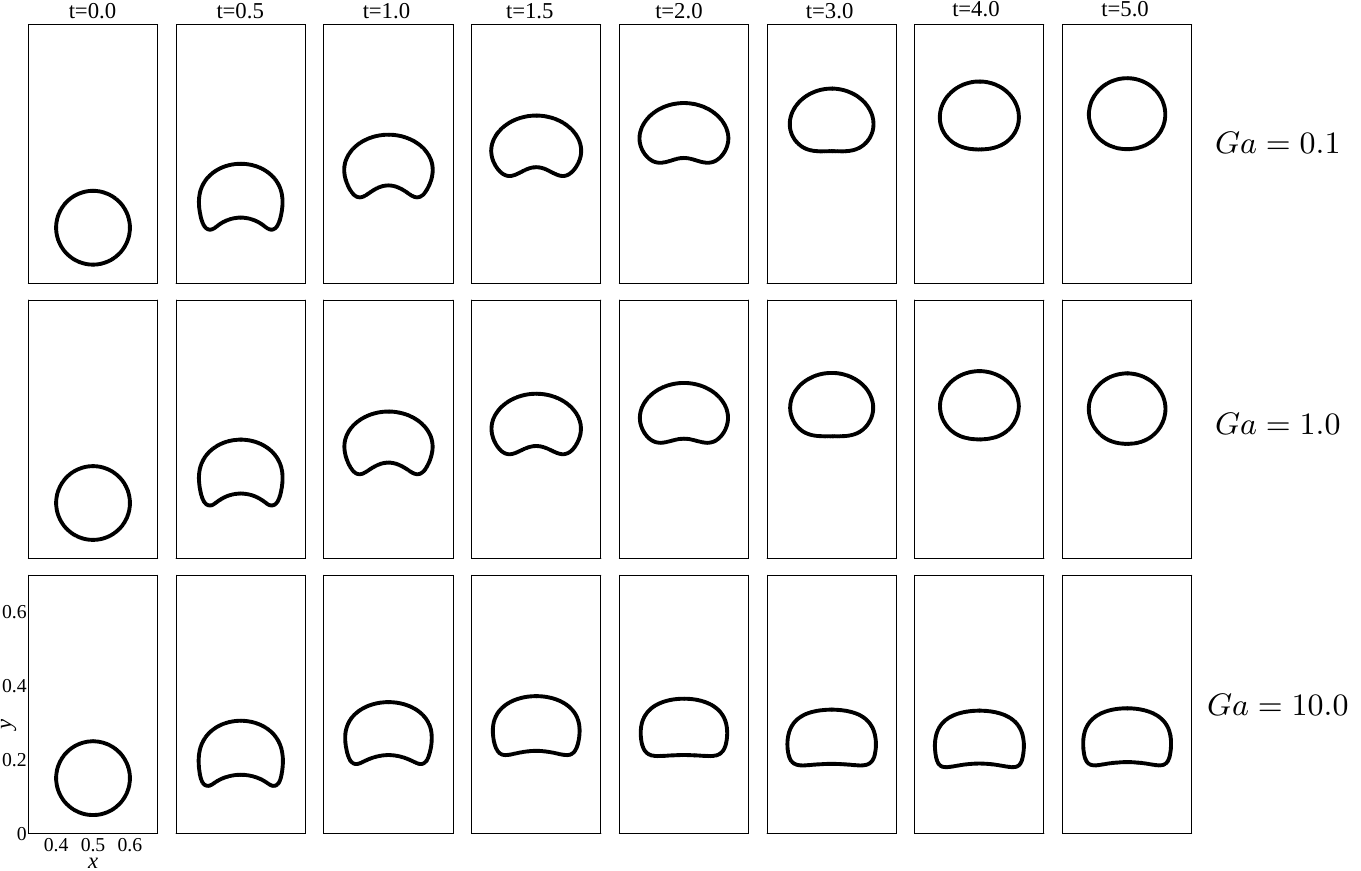}}
	\caption{The interface of the levitating droplet for different $Ga$. $La=1.0,La_m=360$.}
	\label{figEffectOfGa}
\end{figure}

The $Ga$ number signifies the extent of the gravitational force on the droplet relative to the viscous force. We vary $Ga$ for fixed $La$ and $La_m$. Figure \ref{figEffectOfGa} depicts the results when $Ga$ is varied from $0.1$ to $10.0$ at intermediate values of $La=1.0$ and $La_m=360$. As expected, increased $Ga$ decreases the levitation height of the droplet. In addition, the droplet is more suppressed in case of $Ga=10.0$ as compared to $Ga=0.1$ and $1.0$. Observing this, it is certainly possible that if $Ga$ is increased further, the droplet may settle down instead of levitating. Although there is not a considerable deviation of the droplet shape for $Ga=0.1$ and $1.0$ from the previously discussed simulations, a new transition to tooth-like shape occurs near the steady state in case of higher $Ga\:(=10)$. In this case, if the top portion of the droplet interface becomes concave upward, the shape will nearly be a tooth surface. We show in \S \ref{secSummaryConstMu} that for certain combinations of $La,La_m$ and $Ga$, the droplet closely resembles such a shape. 

\subsection{Levitation height and speed characteristics\label{secDisplacementAndVelocity}}
\begin{figure}
	\centerline{\includegraphics[width=0.99\textwidth]{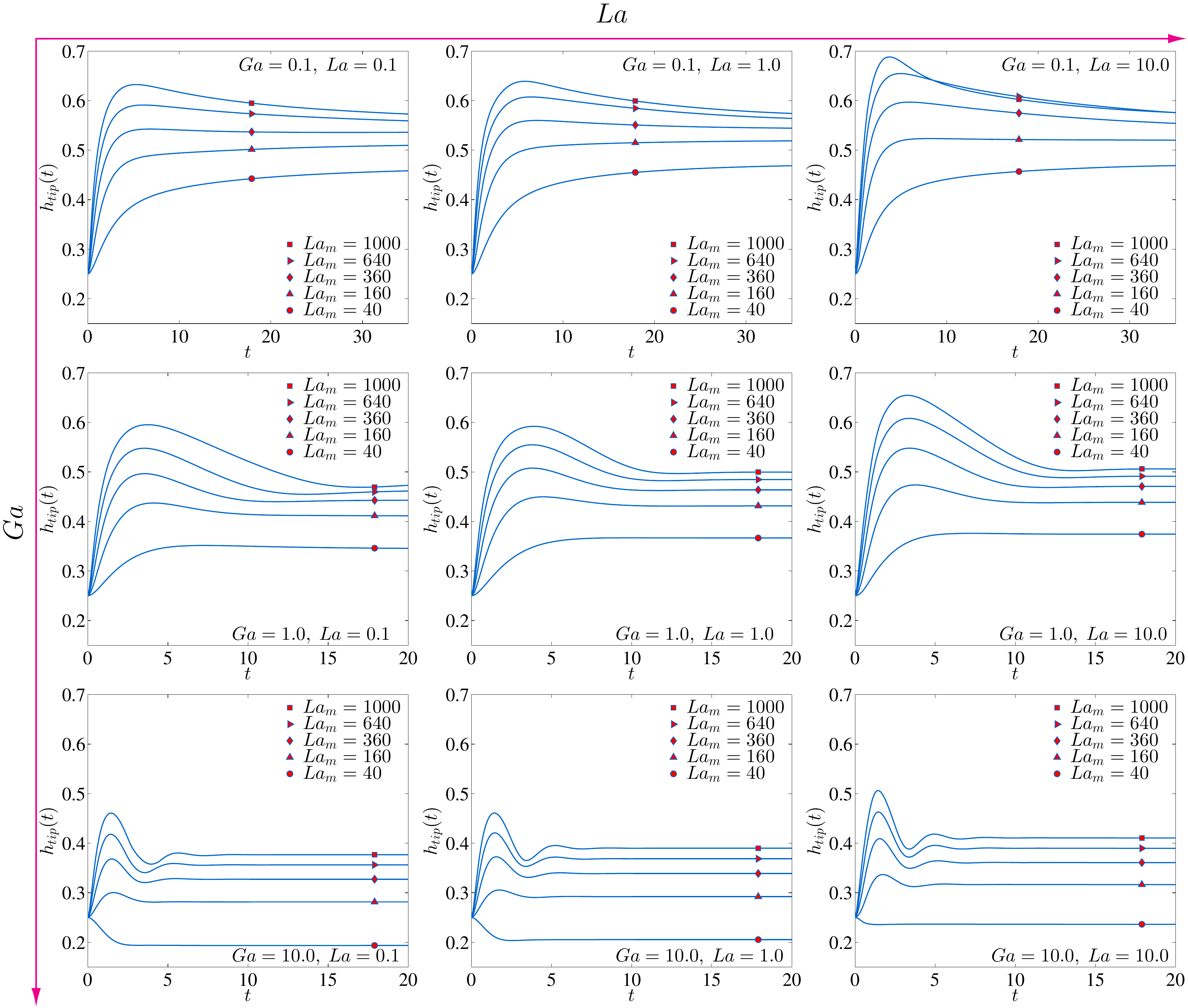}}
	\caption{The time dependent vertical location of the tip of the droplet for different $La,Ga$ and $La_m$.}
	\label{figDisplacement}
\end{figure}

In this subsection, we study the levitation height and the speed of the droplet as a function of time and look for different temporal modes to reach the equilibrium shape. For the time dependent levitation height, the vertical location of the tip/nose of the droplet, denoted as $h_{tip}(t)$ (\S\ref{secFormulationHeightAndSpeed}), is tracked with time. The results are depicted in figure \ref{figDisplacement}. The $La$ number increases from left to right while the $Ga$ number increases from top to bottom. In each of the nine plots, five different $La_m$ cases are considered.

It is observed that the $La$ number has the least effect on the $h_{tip}(t)$ curves. Though the droplet shape has turned to be quite sensitive to $La$ (as discussed previously), its evolution in time shows an opposite nature. 

The effect of $La_m$ on the $h_{tip}(t)$ curves is quite intriguing. First we take the case when both $La$ and $Ga$ are $0.01$ (top-left plot in figure \ref{figDisplacement}). For $La_m$ values of $40$ and $160$, the droplet approaches an equilibrium location, without crossing the steady state levitation height. However, at further increase of $La_m$ to $360$ and $640$, there starts appearing a crossover/overshoot in the curve. At $La_m=1000$, the overshoot is quite apparent; the value of $h_{tip}$ starts increasing, reaches a maximum and then begin to decrease while approaching towards the stationary state. Physically which implies that if the order of magnitude of the magnetic levitation force on the non-magnetic droplet, relative to the viscous and gravitational forces on it, is considerably high then the droplet can cross the equilibrium location before finally attaining it. This also suggest that the effect of the hydrodynamic flow during the rise of the droplet is not always the same; the viscous resistance is opposed more efficiently at increased $La_m$ suggested by the overshoot. This same behavior is depicted for all the three $La$ and $Ga$ between $0.1$ and $10$. The system behavior experiences a transition in the nature of its temporal response; here the transition is due to the control parameter $La_m$. In the analytical model in \S \ref{secOneDimensional} it is shown that this transition, is indeed, a standard transition between two fixed points of type \emph{node} and \emph{spiral}.

At a given $La_m$, the increase in the $Ga$ number alters the time response in three ways. One of them has already been discussed that as the $Ga$ is increased, the equilibrium height for the droplet tip decreases. The second effect is that the time to reach the equilibrium state considerably reduces with increasing $Ga$. Third observation is that {at $Ga=10$ and $La_m=40$,} the droplet tip moves downward instead of moving upward, as expected in \S \ref{secEffectGa}. Also, provided that the value of $La_m$ is such that it causes an overshoot, the number of crossovers around the steady state location increase with increasing $Ga$.

Whereas the droplet reaches a steady state relatively early in case of higher $Ga$, { reach to equilibrium} is apparently asymptotic in case of $Ga=0.1$. For $Ga=10.0$ and $1.0$, the steady state is reached exactly well before $t=20$. However when $Ga=0.1$, the $h_{tip}$ value has not exactly reached its steady state even at $t=35$. The droplet essentially creeps towards the equilibrium. The viscosity is thus more dominant at low $Ga$ and $La_m$. This also holds true from the fact that $\mathcal{O}(Ga[\rho^*g^*])<\mathcal{O}(\bnabla^*\cdot\mathsfbi{S}^*)$ as $Ga\rightarrow 0$ for $Ga<1$ and 
$\mathcal{O}(La_m[\bnabla^*\cdot\mathsfbi{S}^*_m])\rightarrow \mathcal{O}(\bnabla^*\cdot\mathsfbi{S}^*)$ as $La_m\rightarrow 1$ for $La_m>1$.

\begin{figure}
	\centerline{\includegraphics[width=0.99\textwidth]{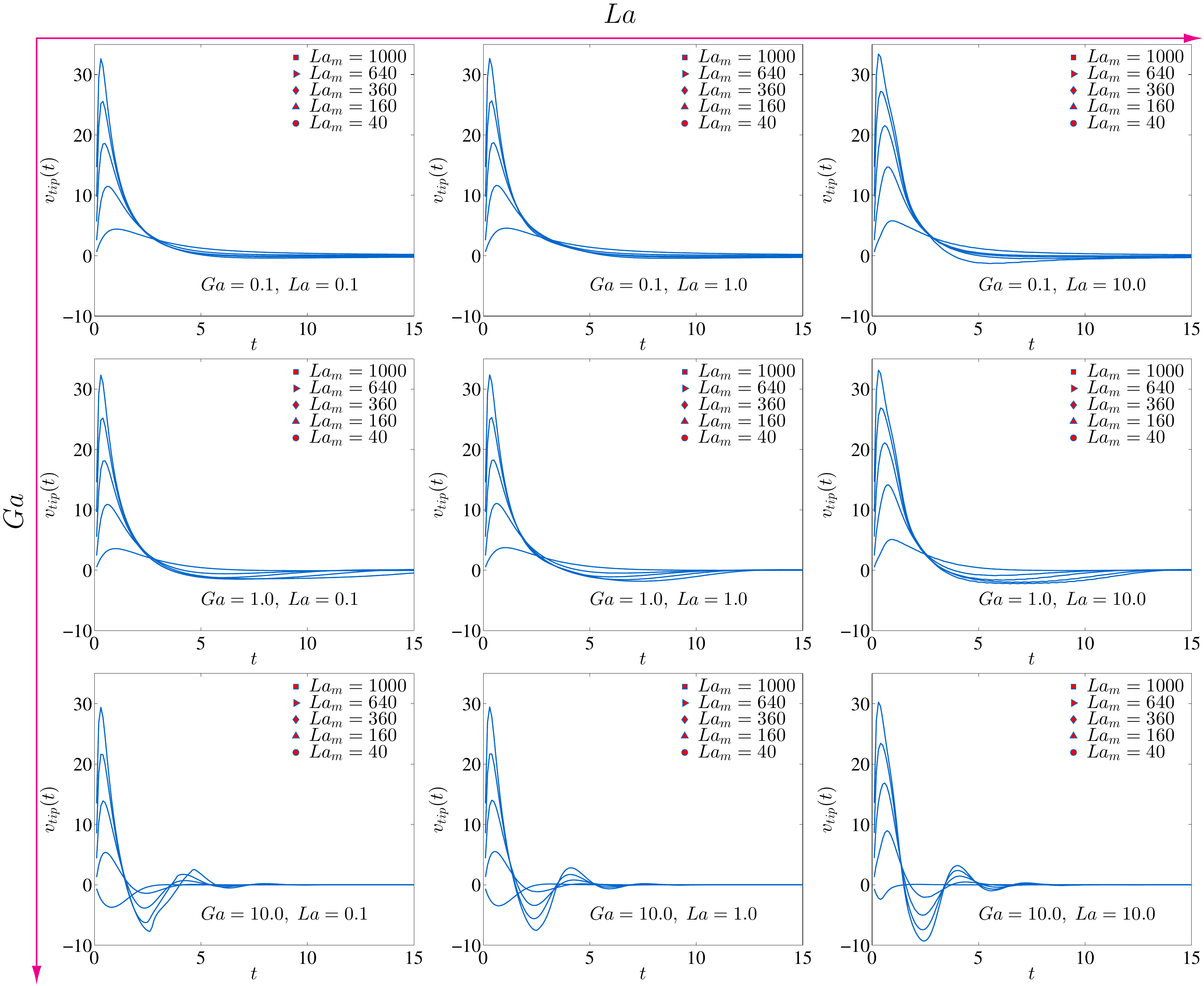}}
	\caption{The time dependent vertical velocity of the tip of the droplet for different $La,Ga$ and $La_m$.}
	\label{figVelocity}
\end{figure}

We further elaborate on the different characteristics of the time dependent evolution of the droplet by analyzing the vertical velocity of the tip of the droplet. The results are presented in figure \ref{figVelocity}. First, the maximum rise velocity is encountered only during the very initial transience. In all the cases, starting from the rest, the droplet gains a maximum velocity which again becomes negligible within non-dimensional time approximately $t=5.0$. This further refines our intuition that the hydrodynamic flow is dominant during the initial rise of the droplet. It is again observable that $La$ has a negligible effect on the time dependent evolution of the droplet and it primarily influences only the shape of the droplet. The $Ga$ and the $La_m$ numbers, on the other hand, considerably change the levitation velocities. The oscillations of the droplet around the final equilibrium position at increased $Ga$ is appreciably resolved via the velocity characteristics. This feature was less appreciable in the displacement plots. Now at least two periods of oscillations around the equilibrium are clearly identifiable for $Ga=10.0$. For this value of $Ga$, it is depicted that the levitation velocity can also be negative after reaching its positive maximum, or even at the very beginning of the phenomenon if $La_m$ is low, indicating the downward motion of the droplet after the overshoot.

The temporal route to the final equilibrium state thus can be either monotonic, or through undulations. Further, the monotonic levitation can be asymptotic in nature where the droplet takes g a long time to reach its final equilibrium (creep). The displacement plots have revealed about the monotonicity and the asymptotic nature, while the velocity plots have resolved the undulations around the equilibrium more apparently.

\begin{figure}
	\centerline{\includegraphics[width=0.9\textwidth]{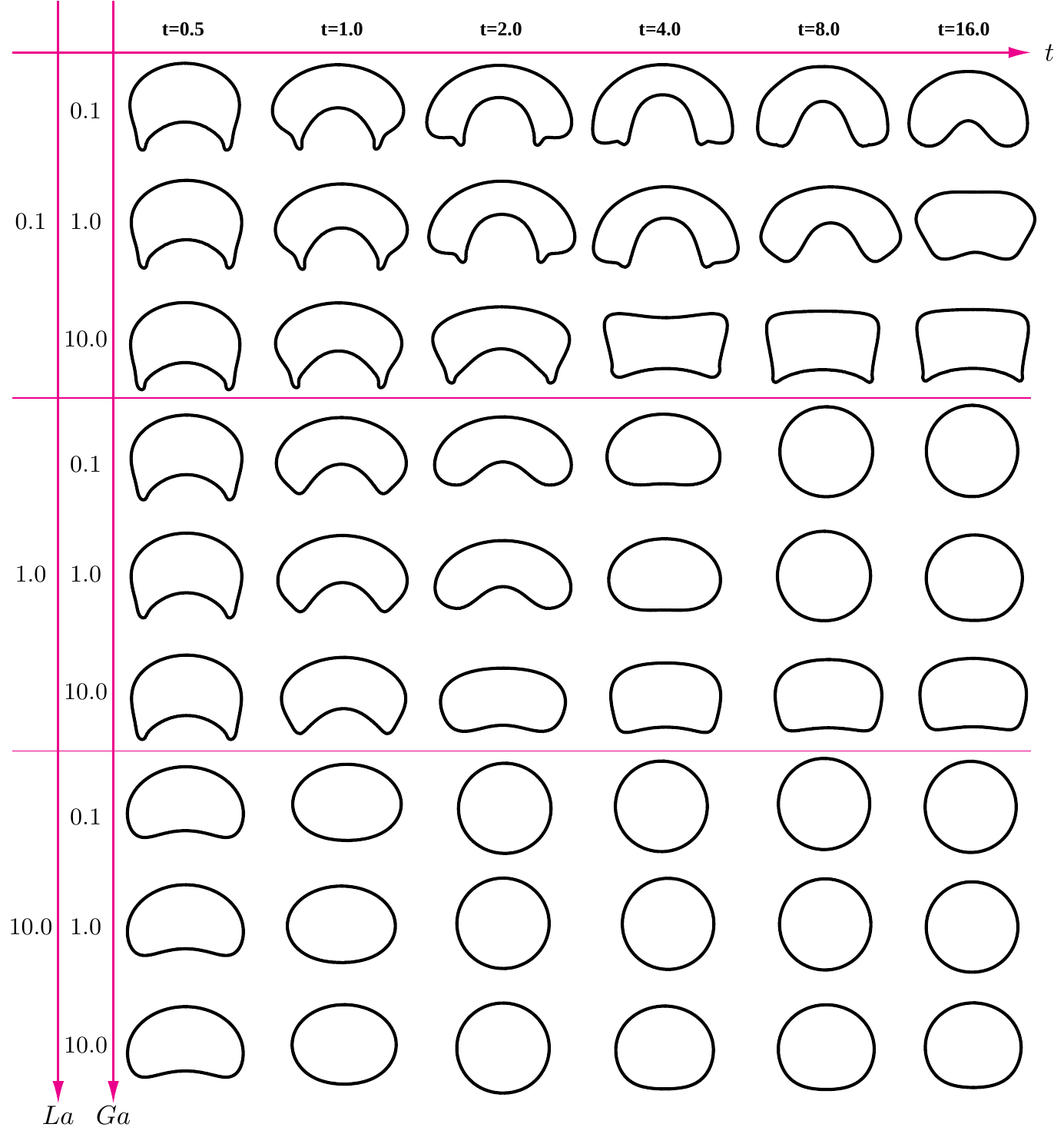}}
	\caption{The summary of the shapes of the droplet for different $La$ and $Ga$ at various time instants. The $La_m=1000$ in this case.}
	\label{figShapes}
\end{figure}

\begin{figure}
	\centerline{\includegraphics[width=0.9\textwidth]{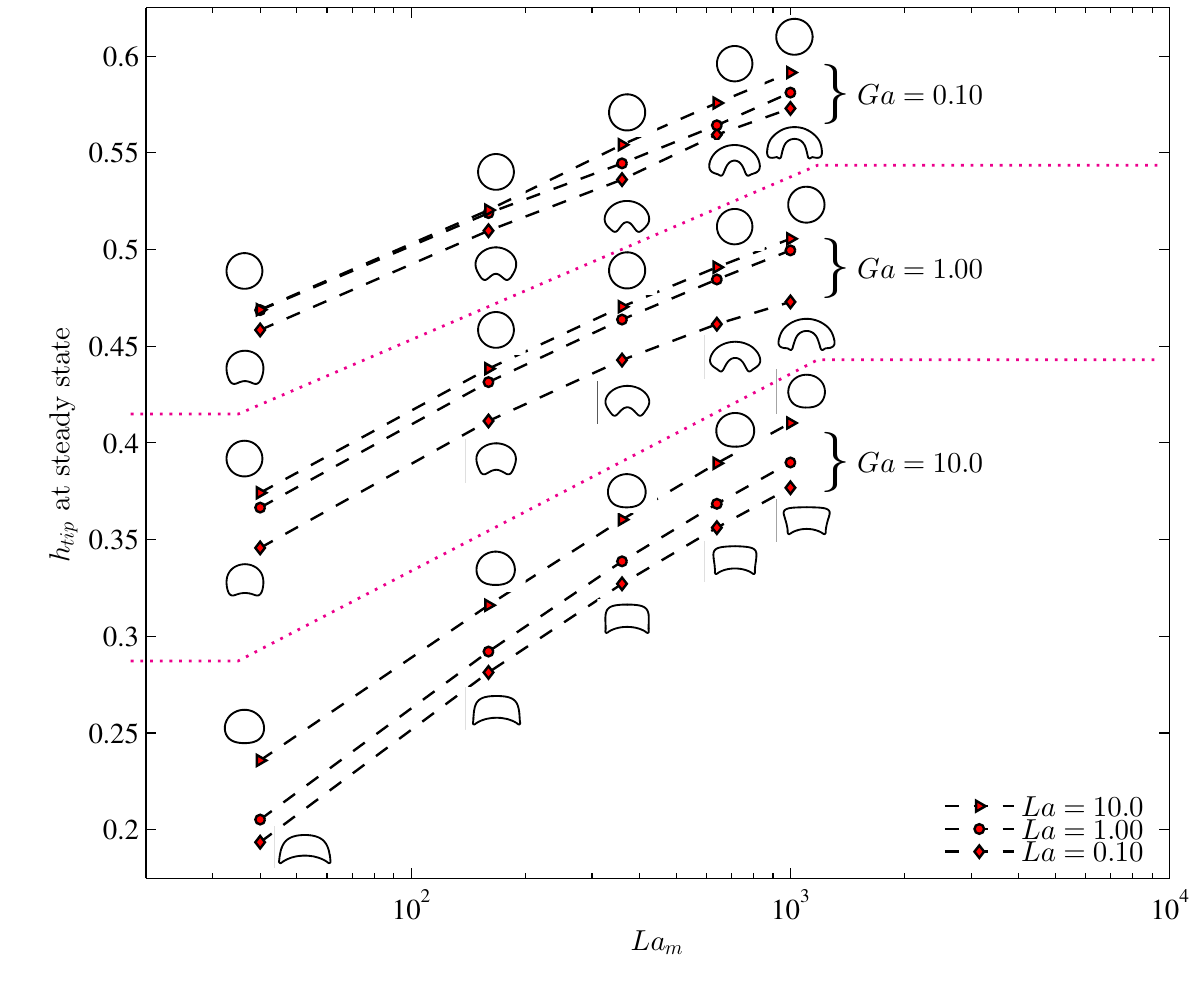}}
	\caption{The summary for the steady state height of the tip of the droplet for different $La,Ga$ and $La_m$. The steady state droplet shapes are also presented.}
	\label{figLevitationHeight}
\end{figure}
\subsection{Summary of droplet shapes and levitation height for constant permeability\label{secSummaryConstMu}}
The observations under the constant ferrofluid permeability assumption are summarized in terms of the droplet shapes and the corresponding levitation height. First we discuss the shapes.

Different interfacial configurations of the droplet emerge when the combination of $La, La_m$ and $Ga$ is changed. These shapes also vary with time for given $La, La_m$ and $Ga$. A summary of the levitating droplet shapes with changing time, $La$ and $Ga$ is presented in figure \ref{figShapes}. The $La_m$ number is $1000$. The droplet shapes are more stable for higher magnitude of $La$. On the other hand at low $La=0.1$, highly deformed droplet shapes are observed. In this regime, the shape of the droplet is \emph{skirted} near the initiation of the phenomena, at $t=0.5$ and $1.0$. Afterwards a transition to a \emph{tooth} shaped or \emph{segmented-ring} shaped configuration takes place, depending on the value of $Ga$. For example, for $La=0.1$ and $Ga=0.1$, the transition is from \emph{skirted} towards \emph{ring-segment}, while for $La=0.1$ and $Ga=10.0$, the droplet transforms from a \emph{skirted} to a \emph{tooth} like shape. These shapes are suppressed for higher $Ga\;(=10.0)$. An increase in $La$ transforms the droplet shape towards an \emph{oval}.

It is interesting to note that, for $La=0.1$ and $1.0$, the maximum deformation of the droplet occurs between $t=1.0$ and $t=4.0$. From the velocity plots, this is the period when the droplet is retarding after attaining a maximum velocity. However, for $La=10.0$, the maximum deformation occurs before $t=1.0$, when the droplet is accelerating. Thus a stiffer interface (higher $La$) deforms maximum while in acceleration while a flexible interface (lower $La$) exhibits the maximum deformation while in retardation.  

Figure \ref{figLevitationHeight} presents the vertical height of the droplet tip as $La_m$ is varied at different $La$ and $Ga$. The levitation height is along the abscissa and the $La_m$ increases along the ordinate logarithmically. The three zones, separated by two dashed lines, are for different $Ga$ {while in each zone}, the line plots are for different $La$. The corresponding droplet shapes are also shown, except for the intermediate case of $La=1.0$. A significant outcome from this semi-log graph is that the equilibrium state levitation height of the droplet increases nearly exponentially with $La_m$, for all the simulated $La$ and $Ga$. The roles of the non-dimensional numbers - $La,Ga$ and $La_m$ - are now more understandable from this plot. Whereas $Ga$ has the least effect on the droplet shape, $La$ has the least effect on the levitation height. The $La_m$ number influences the phenomenon significantly in both ways. { The shapes in figure \ref{figLevitationHeight} are taken at t=30. At this time the shapes are at steady state, except for the cases where the droplet \emph{creeps} towards the equilibrium. However, even for the \emph{creeping} cases, the shapes at $t=30$ are expected to be quite close to the equilibrium, as indicated from the displacement plots in figure \ref{figDisplacement}.}
\section{The effect of viscosity ratio}\label{secResultsViscosityRatio}
In the previous discussions the density, as well as viscosity, of the non-magnetizable droplet was considered to be greater than the ferrofluid. As the levitation against gravity makes sense only if the droplet density is greater than the ferrofluid, the density ratio $\rho_d/\rho_f$ is kept greater then $1.0$ for all the simulations in this study. However, the viscosity ratio $\eta_d/\eta_f$ for an arbitrary combination of a ferrofluid sample and a non-magnetizable droplet can be either $>1.0$ or $<1.0$. In the last section, the value of $\eta_d/\eta_f$ was $2.0$. In this section, the ratio is reduced below $1.0$ and the changes in the dynamics of the levitation of the droplet and its shapes are discussed. 

\begin{figure}
	\centerline{\includegraphics[width=0.45\textwidth]{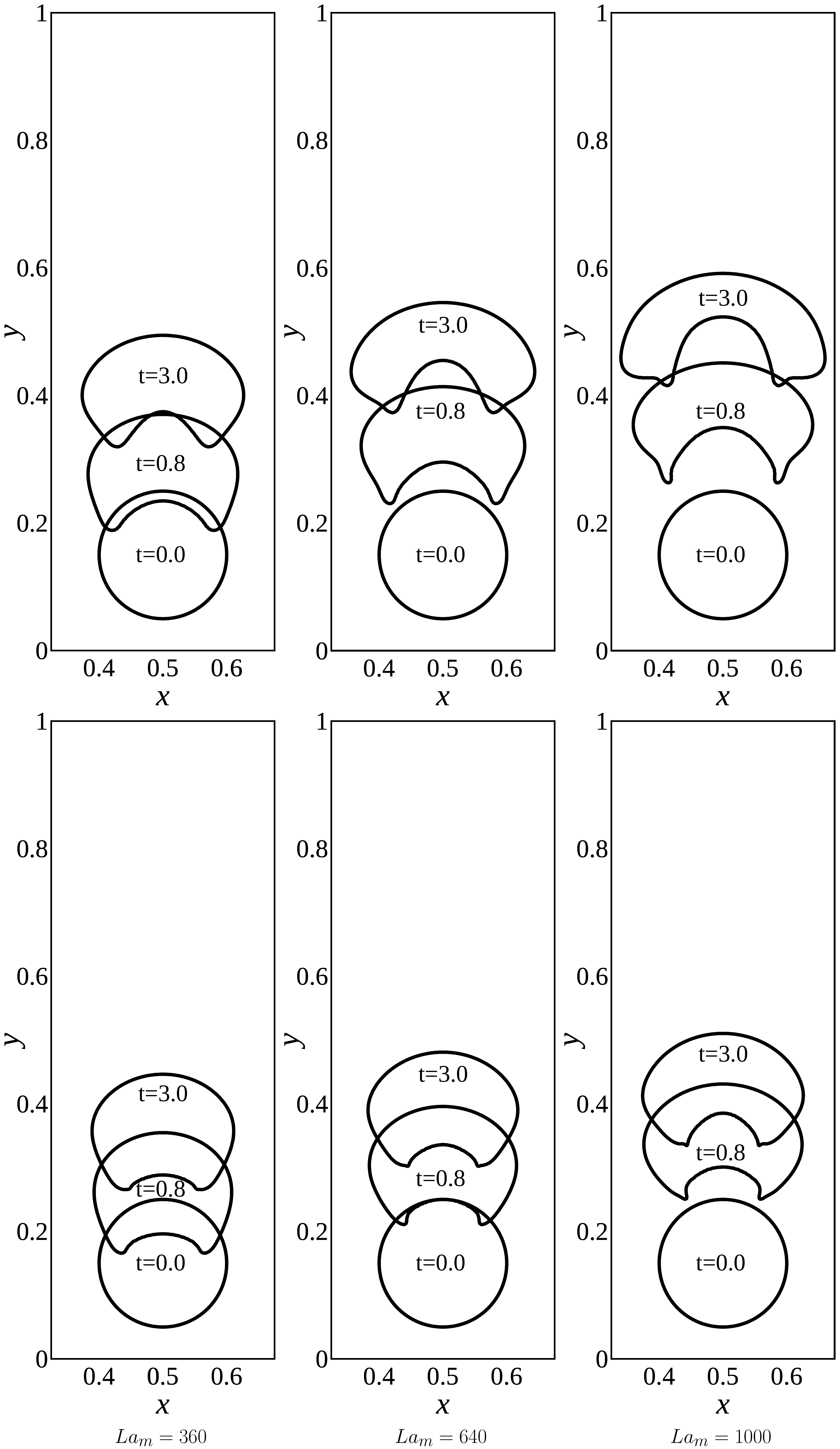}}
	\caption{The effect of viscosity ratio on the levitating droplet - $\eta_d/\eta_f=2.0$ for the top row while $\eta_d/\eta_f=0.5$ for the bottom row. $Ga=1.0,La=0.1$.}
	\label{figEffectOfViscosity}
\end{figure}

For the same set of non-dimensional and other parameters, the viscosity ratio $\eta_d/\eta_f$ is now changed to $0.5$ (table \ref{tableSim}, second row). The ferrofluid is now more viscous than the non-magnetic fluid and its viscosity is now taken as the reference viscosity. The comparison between these two cases of $\eta_d/\eta_f>1.0$ and $\eta_d/\eta_f<1.0$ is shown in figure \ref{figEffectOfViscosity}.
A significant difference appears at the tail of the droplet where the geometry of the two projections has changed due to the change in the viscosity ratio. The two projections are now more cusped. The droplet is no more skirted during its rise. For the viscosity ratio $0.5\:(\eta_d/\eta_f<1)$, it resembles a crescent shape more closely than the case $\eta_d/\eta_f=2.0$. 

The cusped nature of the projections at the tail gives rise to a curiosity that whether any singularity is possible at the interface of a non-magnetizable droplet levitating in a ferrofluid, and as it seems that it is, then what might be the physical mechanism behind such a behavior. Considering the unit normal pointing outwards at the interface, the signed curvature at the top of the droplet is positive while at the middle of the tail, it is negative. Thus it must change its sign, either smoothly or abruptly, at some point along the interface in between these two locations.  The abruptness or smoothness must depend on the capability of the interfacial tension against other forces. It is observed that the curvature changes its sign smoothly at the projections when $\eta_d/\eta_f=2.0\:(>1)$. However, the surface tension fails to maintain the smoothness of the interface if the fluidity of the outer liquid is lesser than the liquid of the droplet, or in other words, if the outer phase (ferrofluid) is relatively more viscous. Also the height of levitation has reduced, in other words, the phenomenon is now slower than the previous case of viscosity ratio $>1$. It indicates that the net downward component of the viscous forces at the surface of the droplet has essentially increased due to increased outer viscosity, retarding the droplet motion. It is rather intriguing that the droplet travels to a greater height in a given time when $\eta_d/\eta_f=2.0$ rather than when $\eta_d/\eta_f=0.5$, even though it is comparatively more blunt in the former case. This indicates that the increase in net viscous drag on the rising droplet is less influenced by the geometric configuration of the droplet, specifically the frontal area, and the increased viscosity of the outer phase is dominating the change in the net drag.

In addition to the fact that the droplet levitates to a lesser height in a given time in case of $\eta_d/\eta_f<1$, the time dependent change in its shape is also swift. The droplet do not continue to deform for a longer period of time. This fact becomes apparent when the positions at the time instants $t=0.8$ and $t=3.0$ are compared for the two cases of the viscosity ratio (figure \ref{figEffectOfViscosity}). Whereas the droplet is still under deformation after $t=0.8$ in case of $\eta_d/\eta_f=2.0$, it gains nearly a steady shape in case of $\eta_d/\eta_f=0.5$. The higher outer viscosity helps the droplet to attain its equilibrium shape relatively early in time, and at the same time, it reduces the levitation height. {It should also be noted that at the equilibrium, the flow field and the viscous force vanish, and it is expected that the final droplet shape and height do not depend on the viscosity ratio, if other parameters are fixed. During the later times, when the flow field vanishes and the droplet approaches steady state, we do observe similar equilibrium shapes for different viscosity ratios, provided that the simulation parameters do not belong to \emph{creeping} regime. Of course it is expected that in the \emph{creeping} regime the droplet shape can take comparably long time to relax, but eventually should also attain shape independent of the viscosity ratio. For creeping case, $t\sim30$ is found to be sufficient to obtain similar equilibrium shapes for different viscosity ratios. This order of the time to obtain steady shapes in the creeping scenario was also indicated in figure \ref{figDisplacement}.}

\section{The effect of non-linear magnetization}\label{secResultsVariablePermeability}
\begin{figure}
	\centerline{\includegraphics[width=0.9999\textwidth]{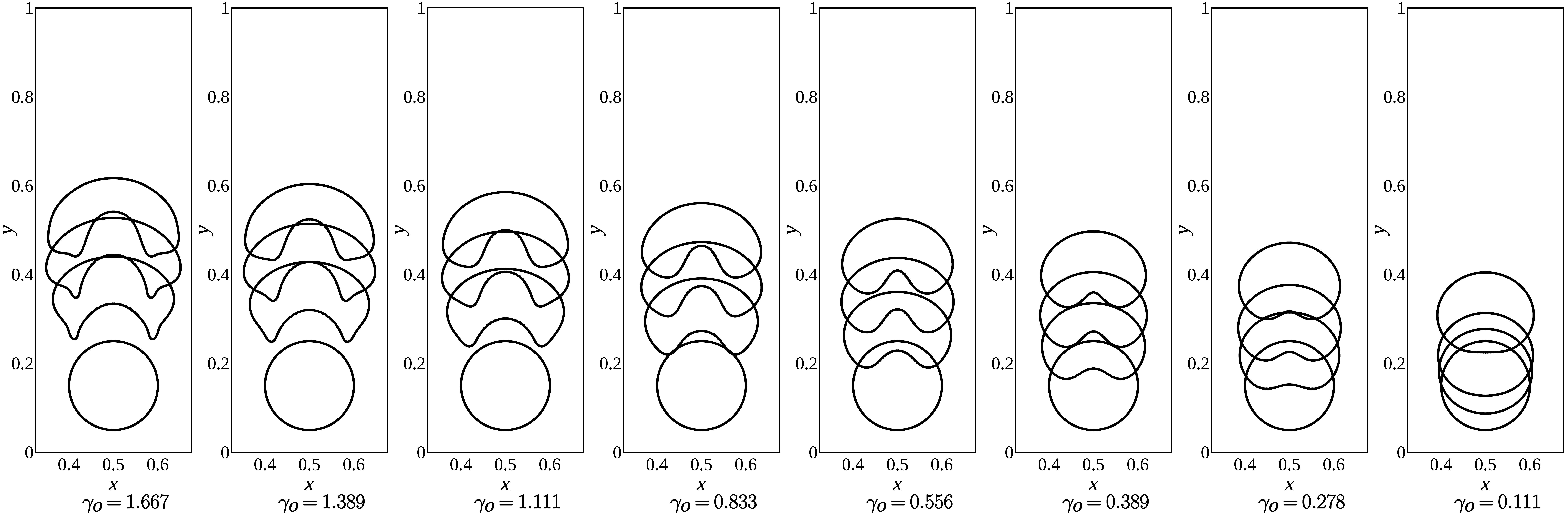}}
	\caption{The effect of parameter $\gamma_o$ on the levitating droplet - decreasing from left to right. The four time instants in each frame are $t=0.0,0.8,1.5$ and $5.0$, from bottom to top. The permeability in the expression of $La_m$ is now field dependent while the fixed parameters are - $Ga=0.1,La=0.1,\xi_o=0.185$.}
	\label{figEffectOfNonLinearMagnetization}
\end{figure}

The constant permeability assumption for the ferrofluid phase, so far, has helped to understand the very basic features of the levitation phenomenon. As we discussed in \S \ref{secGoverningEquationsDimensional}, this simplifying assumption of constant permeability of ferrofluid is used in the first two set of simulations for its computational effectiveness, and it has served as a reference for further refining of our simulations using a more realistic field-dependent permeability model. The assumption is very rational for the low applied magnetic field regimes. For high applied fields, at least a switch in the susceptibility magnitude is necessary after a certain strength of the magnetic field, around which the magnetization curve changes its slope considerably. If only a single constant value for the susceptibility is considered, then this assumption does not put any upper bound on the ferrofluid magnetization. In such a case, the solution might be unrealistic. To refine this, the effect of non-linear magnetization is incorporated in the numerical scheme by considering a field dependent magnetic susceptibility instead of a constant one, as described in \S\ref{secGoverningEquationsDimensional} (equation \ref{eqPhiBothPhases}) and \S\ref{secDescriptionOfAlgorithm}. The total number of influencing parameters has now increased since an additional parameter $\gamma_o=3\chi_o H_o/M_s=3\chi_o\xi_o$, which changes the characteristics of the magnetization curve, enters the simulations (equation \ref{equFunctionalFormV}). The parameter $\gamma_o$ is varied from $1.667$ to $0.111$ for given $Ga$ and $La$. The results are depicted in figure \ref{figEffectOfNonLinearMagnetization}.

\subsection{The effect of parameter $\gamma_o$}
Apparently, the parameter $\gamma_o$ has considerable effect on the deformation of the droplet, as well as on the levitation height. For $\gamma_o=0.111$, the droplet has a nearly round shape during its rise (figure \ref{figEffectOfNonLinearMagnetization}). A concavity starts appearing at the tail of the droplet for $\gamma_o=0.278$, and for a given time, the extent of the concavity increases as $\gamma_o$ increases. The droplet shapes at $\gamma_o=0.278$ and $0.389$ are similar to that of a \emph{cardioid} without a cusp. For $\gamma_o=0.556,0.833$ and $1.111$, the bottom feature of the droplet has widened and at $\gamma_o=1.389$ and $1.667$, the droplet is more flattened and skirted. 

\begin{figure}
	\centerline{\includegraphics[width=0.65\textwidth]{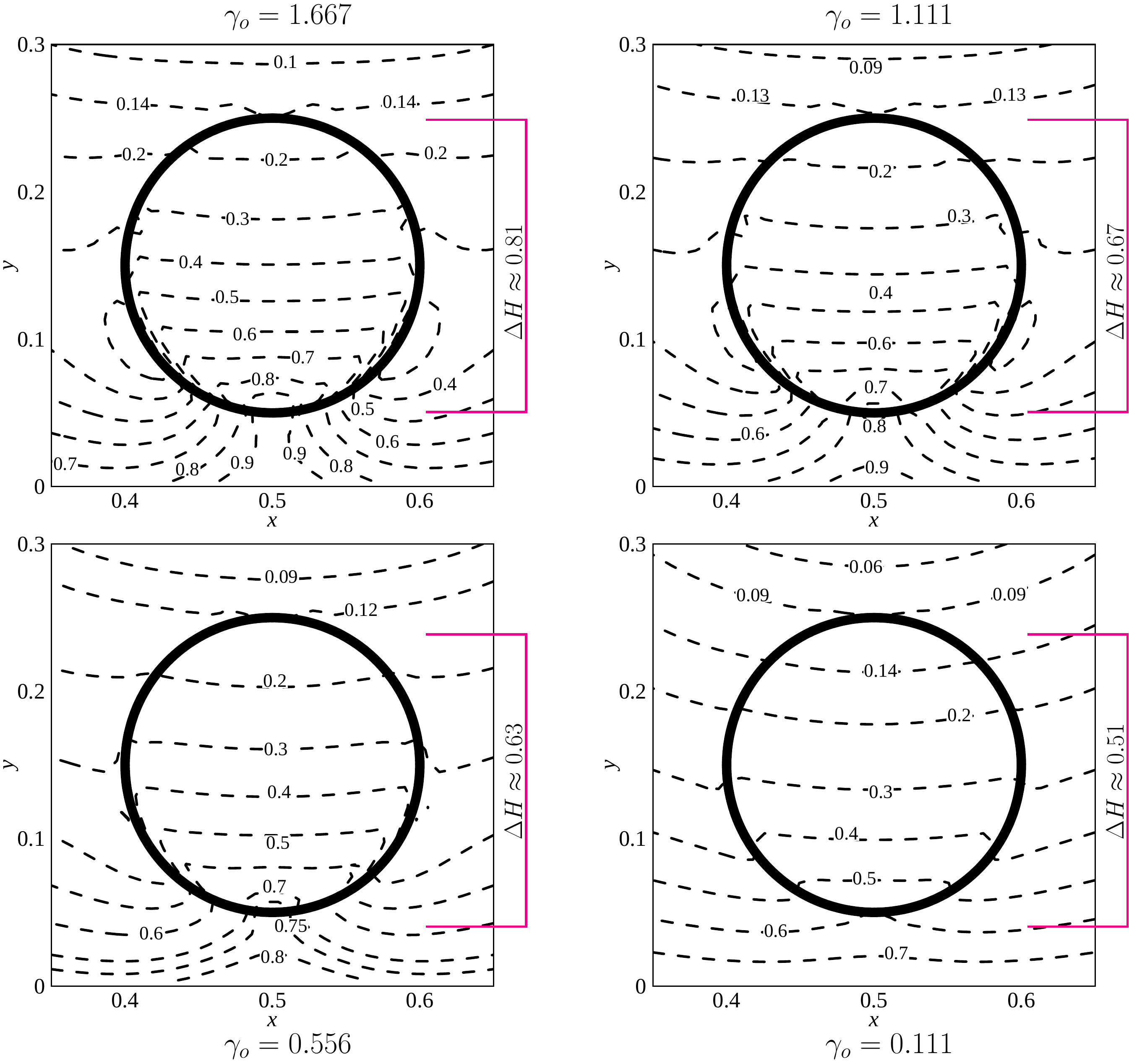}}
	\caption{The effect of parameter $\gamma_o$ on the field gradient across the top and bottom poles of the droplet (denoted by $\Delta H$ in the figure). The permeability in the expression of $La_m$ is now field dependent while the fixed parameters are - $Ga=0.1,La=0.1,\xi_o=0.185$. All the four frames are at initial time $t=0.0$. The field strength difference across the poles of the droplet ($\Delta H$) increases with increasing $\gamma_o$.}
	\label{figFieldGradientAcrossPoles}
\end{figure}

In this subset of simulations, the non-dimensional numbers $La,Ga$ and $\xi_o$ are all kept constant. Except the magnetic permeability, the other quantities in the expression of $La_m$ are also fixed. Thus, it essentially implies that the changes in the dynamics of the flow are coming due to the non-uniform spatial distribution of $\mu_f$. This non-uniformity makes it intuitive that $\mu_f$ along the interface of the droplet is no more an invariant. We seek the effects of the variability of $\mu_f$, due to different $\gamma_o$, by plotting the contours of $H/H_o$ around the initial state of the droplet. The contours are depicted in figure \ref{figFieldGradientAcrossPoles}. The four frames are for $\gamma_o=1.667,1.111,0.556$ and $0.111$. The spatial non-uniformity of $H/H_o$ is sensitive to the parameter $\gamma_o$. The difference in the field strength across the top and the bottom pole of the droplet ($\Delta H$), or in other words, the field gradient along the vertical axis of the droplet, in a global sense, increases with increasing $\gamma_o$. The local gradients across this axis are relatively gradual for lower values of $\gamma_o$, for example in the plot for $\gamma_o=0.111$. Thus, {a rational explanation is that the increase} in $\gamma_o$ increases the field gradient across the droplet, and thus the acceleration of the levitating droplet is higher for larger $\gamma_o$ (figures \ref{figEffectOfNonLinearMagnetization} and \ref{figFieldGradientAcrossPoles}).   

The fundamental change in the droplet response due to the non-linearity caused by the additional parameter $\gamma_o=(3\chi_o\xi_o)$, which basically controls the magnetization curve of the ferrofluid, is deducible by deriving asymptotic expressions for the order of magnetic force in the limits $\gamma_o\rightarrow 0$ and  $\gamma_o\rightarrow \infty$. By transforming the following expression for the field dependent permeability 
\begin{equation}
\displaystyle\mu_f=\mu_o\left(1+\chi(H)\right)=\mu_o\left(1+\frac{M_s}{|\nabla\phi|}\left[\mbox{coth}\;\gamma |\nabla\phi|-\frac{1}{\gamma |\nabla\phi|}\right]\right),
\end{equation}
to its non-dimensional form
\begin{equation}
\displaystyle\mu_f=\mu_o\left(1+\frac{1}{\xi_o|\nabla^*\phi^*|}\left[\mbox{coth}\;\gamma_o |\nabla^*\phi^*|-\frac{1}{\gamma_o |\nabla^*\phi^*|}\right]\right),
\end{equation}
that for bounded $|\nabla^*\phi^*|$, 
\begin{equation}
\left.
\begin{array}{cc}
\displaystyle \mu_f\sim\mu_o, \mbox{ as } \gamma_o\rightarrow 0, \mbox{ and, }\\[3pt]
\displaystyle \mu_f\sim\mu_o\left(1+\frac{1}{\xi_o}\right), \mbox{ as } \gamma_o\rightarrow\infty.
\end{array}
\right\}
\end{equation}
Using the above asymptotic relations, the order of the magnetic force, say $f_m$, becomes
\begin{equation}
\left.
\begin{array}{cc}
\displaystyle f_m\sim\left(\frac{\mu_f H_o^2}{R}\right)\sim\left(\frac{\mu_o H_o^2}{R}\right), \mbox{ as } \gamma_o\rightarrow 0,  \mbox{ and, }\\[3pt]
\displaystyle f_m\sim\left(\frac{\mu_f H_o^2}{R}\right)\sim\left(1+\frac{1}{\xi_o}\right)\left(\frac{\mu_o H_o^2}{R}\right), \mbox{ as } \gamma_o\rightarrow \infty.
\end{array}
\right\}
\end{equation}
Rewriting the quantity $\mu_o H_o^2/R$ as $\xi_o^2\mu_o M_s^2/R$ (as $H_o=\xi_o M_s$), the order of the magnetic force is rewritten to
\begin{equation}
\left.
\begin{array}{cc}
\displaystyle f_m\sim\xi_o^2\left(\frac{\mu_o M_s^2}{R}\right), \mbox{ as } \gamma_o\rightarrow 0,  \mbox{ and, }\\[3pt]
\displaystyle f_m\sim\xi_o^2\left(1+\frac{1}{\xi_o}\right)\left(\frac{\mu_o M_s^2}{R}\right), \mbox{ as } \gamma_o\rightarrow \infty.
\end{array}
\right\}
\end{equation}
The factor in the expression of the order of magnetic force is $\xi_o^2(1+1/\xi_o)$. The value of this function has a global minimum at $\xi_o=-1/2$. As by definition $H_o$ and $M_s$ are positive, $\xi_o$ is also positive in the interval of interest, and thus any increase in $\xi_o$ (or equivalently $\gamma_o$) will also increase $f_m$. However, as the behavior of $f_m$ near $\gamma_o=0$ and in the limit $\gamma_o\rightarrow \infty$ is different, its effect on the droplet response in two regimes is expected to be distinct. This argument supports the simulations in figure \ref{figEffectOfNonLinearMagnetization} that the droplet responds differently with increasing $\gamma_o$. 
\section{Stability of levitation}\label{secResultsStability}
\begin{figure}
	\centerline{\includegraphics[width=0.8\textwidth]{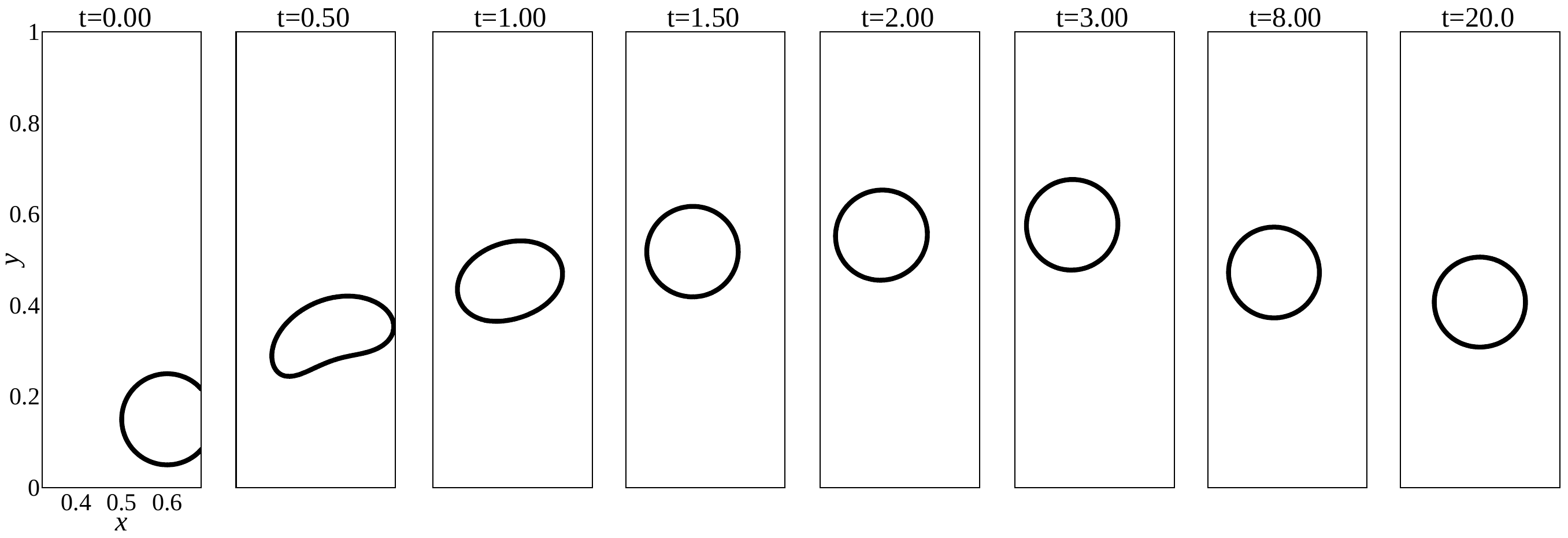}}
	\caption{The effect of eccentricity of the initial location of the droplet. The droplet stably approaches the same steady state location. In this case, all the four walls are covered with an alternate arrangement of magnets.}
	\label{figEffectOfEccentricity}
\end{figure}

The stability of levitation is highly desirable if the phenomenon is to replicated experimentally in a controlled fashion. We shift the initial location of the center of the droplet laterally from $(0.5,0.15)$ to $(0.6,0.15)$. As shown in figure \ref{figEffectOfEccentricity}, the droplet again approaches the stable horizontal location $x=0.5$. The levitation is stable under this Halbach configuration of the magnets where all the four walls are covered by an alternate arrangement of magnet poles. 
 
The special Halbach arrangement of magnets at all the four walls of the domain has proven to be an appropriate choice for ensuring the stability of the levitating droplet. The driving field gradient is due to the bottom two magnets. The magnets at the left and right side restrict sideway movement of the droplet. The role of the two top magnets is also crucial as their presence ensures that only one minimum field strength region is established, and that too near the center of the domain (topic of the subsection below). Under this arrangement, the droplet approaches a unique steady state location, even if any eccentricity in the initial location or any lateral movement of the levitating droplet is forced. This arrangement has also proven to be simpler in handling the field boundary conditions mathematically, as they are just sinusoidal function multiplying with the maximum strength on the magnet surface. In this section, we simulate the effects caused by altering the configuration of the Halbach array.

\subsection{Single and multiple stable and unstable states}
\begin{figure}
	\centerline{\includegraphics[width=0.999\textwidth]{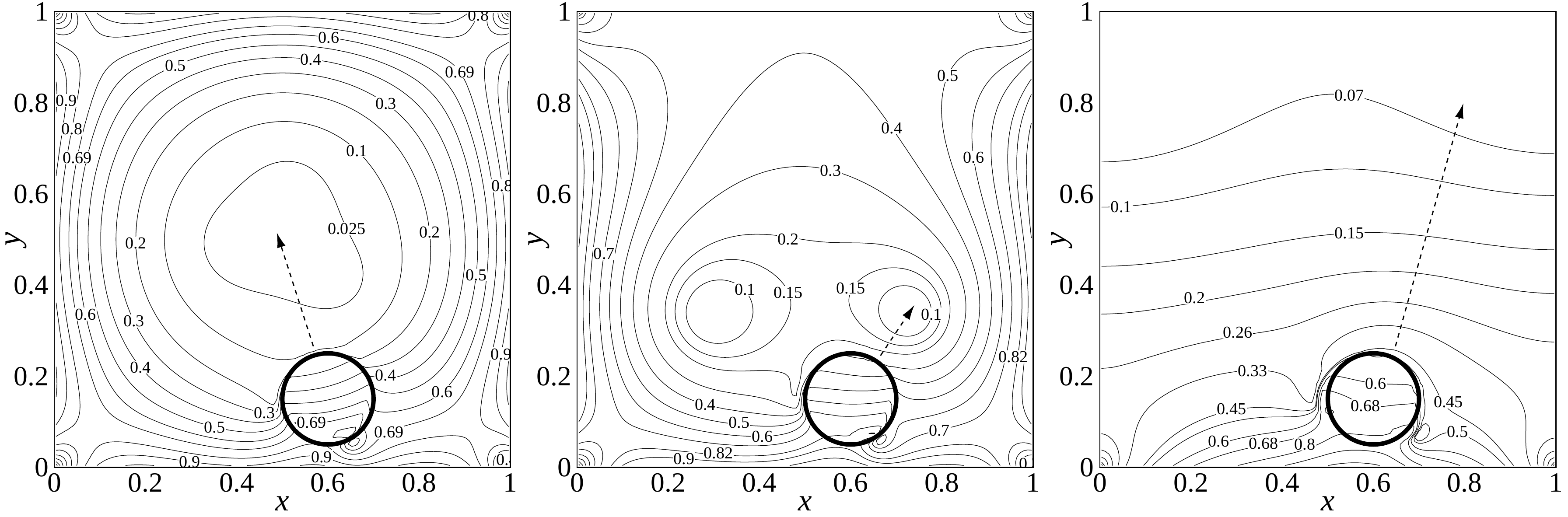}}
	\caption{The absolute magnetic field contours around the eccentrically placed droplet when - from left to right - (a) all the four walls are covered with alternate magnetic poles, (b) the bottom and both the side walls are covered, and (c) only bottom wall is covered.}
	\label{figEffectOfBcsInitialField}
\end{figure}

First, we study the field contours for three different arrangements of the magnetic poles at the boundaries. The three arrangements are - (a) all the four walls are covered with alternating magnetic poles, (b) the bottom and side walls are covered by alternating poles of magnets and only the top wall is free to the field, and (c) only the bottom wall is covered by the magnetic poles while the top and the side walls are free. Figure \ref{figEffectOfBcsInitialField} shows the absolute field contours around the initial state of eccentrically placed droplet for these three cases.

When all the four walls are covered with alternating magnetic poles, a minimum field region is created near the center of the domain (figure \ref{figEffectOfBcsInitialField}(a)). The field gradient is everywhere pointing inward to the domain and it is expected to be the equilibrium location for the droplet after levitation. 

In the second and particularly interesting case (figure \ref{figEffectOfBcsInitialField} (b)), where only the top wall is free from the magnetic poles, two minimum field regions are established. Which indicates the possibility of multiple stable states. The two low field regions give rise to a {\it cat-eye} like field configuration. As the droplet is more close to the right {\it eye}, there is an increased probability that it will be {\it trapped} in that region. 

In figure \ref{figEffectOfBcsInitialField} (c), the top and the side walls are now free from the magnetic poles while the bottom wall field boundary condition is the same as before. In this case, the droplet is expected to levitate, but as the field gradient is diverging or equivalently the contours are concave downward throughout the domain, the stability may not be ensured.

\begin{figure}
	\centerline{\includegraphics[width=0.999\textwidth]{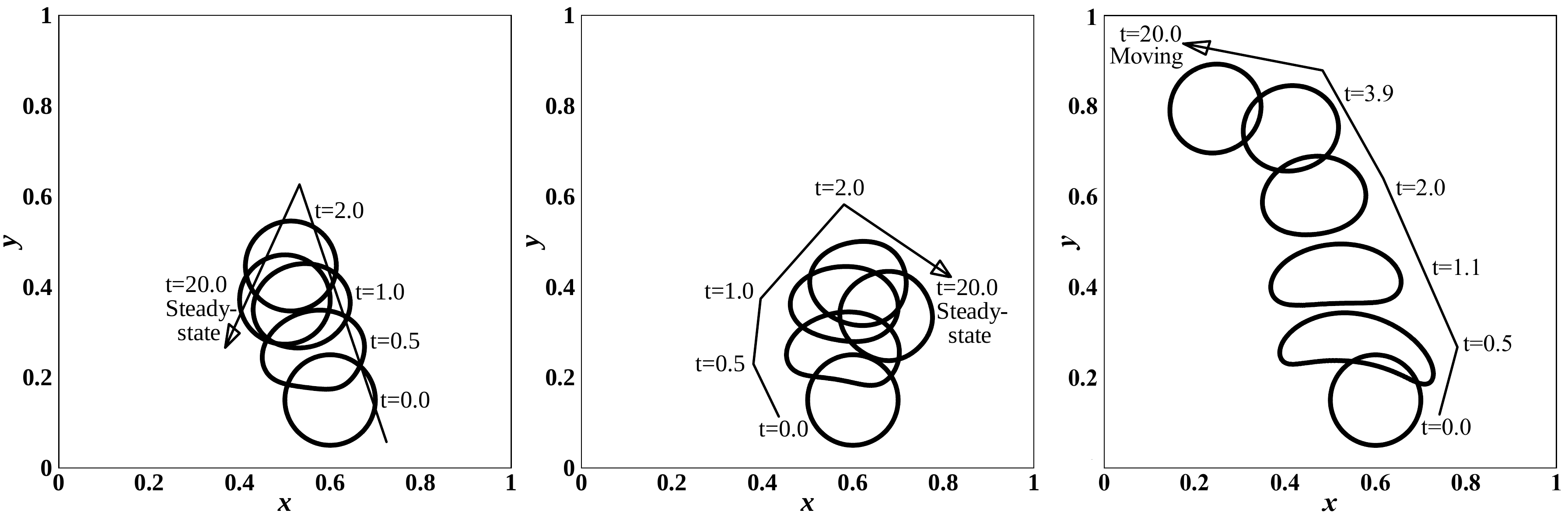}}
	\caption{The interface of the eccentrically placed droplet when - from left to right - (a) all the four walls are covered with alternate magnetic poles, (b) the bottom and both the side walls are covered, and (c) only bottom wall is covered.}
	\label{figEffectOfBcsInterface}
\end{figure}

The movement of the droplet is now simulated and the validity of the above intuitions, made by looking at the initial field contours alone, is tested. The movement of the droplet under the above stated three magnet arrangements is shown in figure \ref{figEffectOfBcsInterface}. In the first case the droplet seeks a stable horizontal location $x=0.5$. The vertical location is such that the magnetic forces are balancing the weight of the droplet. In the second case, though there are two minimum field locations, the droplet seeks the nearby one, as argued previously. In the third and the final case, the path of the droplet levitation is unstable. The equilibrium in this case is only possible when a physical support from the side walls is present; otherwise it can only be marginally stable if a perfect symmetry is maintained during the experiment. The mechanism is further explored below.

 
\subsection{Interplay between magnetic forces and vorticity: stability of levitation path} 
One non-trivial and counter intuitive observation is made in regard to the unstable path of the droplet levitation in figure \ref{figEffectOfBcsInterface} (c). At the first look, the field contours in figure \ref{figEffectOfBcsInitialField} (c) are such that the normal to them is pointing in the direction shown by the dashed line. As this direction is parallel to $\nabla H$, the droplet is expected to move along this direction. However, in figure \ref{figEffectOfBcsInterface} (c), the path of the droplet exhibits an opposite behavior. For some initial time, the droplet moves along $\nabla H$, alters its direction and then starts moving towards the north-west. This behavior of the droplet levitation directs that although the instability of the levitation is predictable from the configuration of the absolute magnetic field, the predictability of actual unstable path is more involved. Essentially, the information about the magnetic field patterns alone are not sufficient to predict the path of the droplet if the levitation becomes unstable. The magnetic field contours and the hydrodynamic flow field around the droplet should be analyzed in a coupled way. 

\begin{figure}
	\centerline{\includegraphics[width=0.999\textwidth]{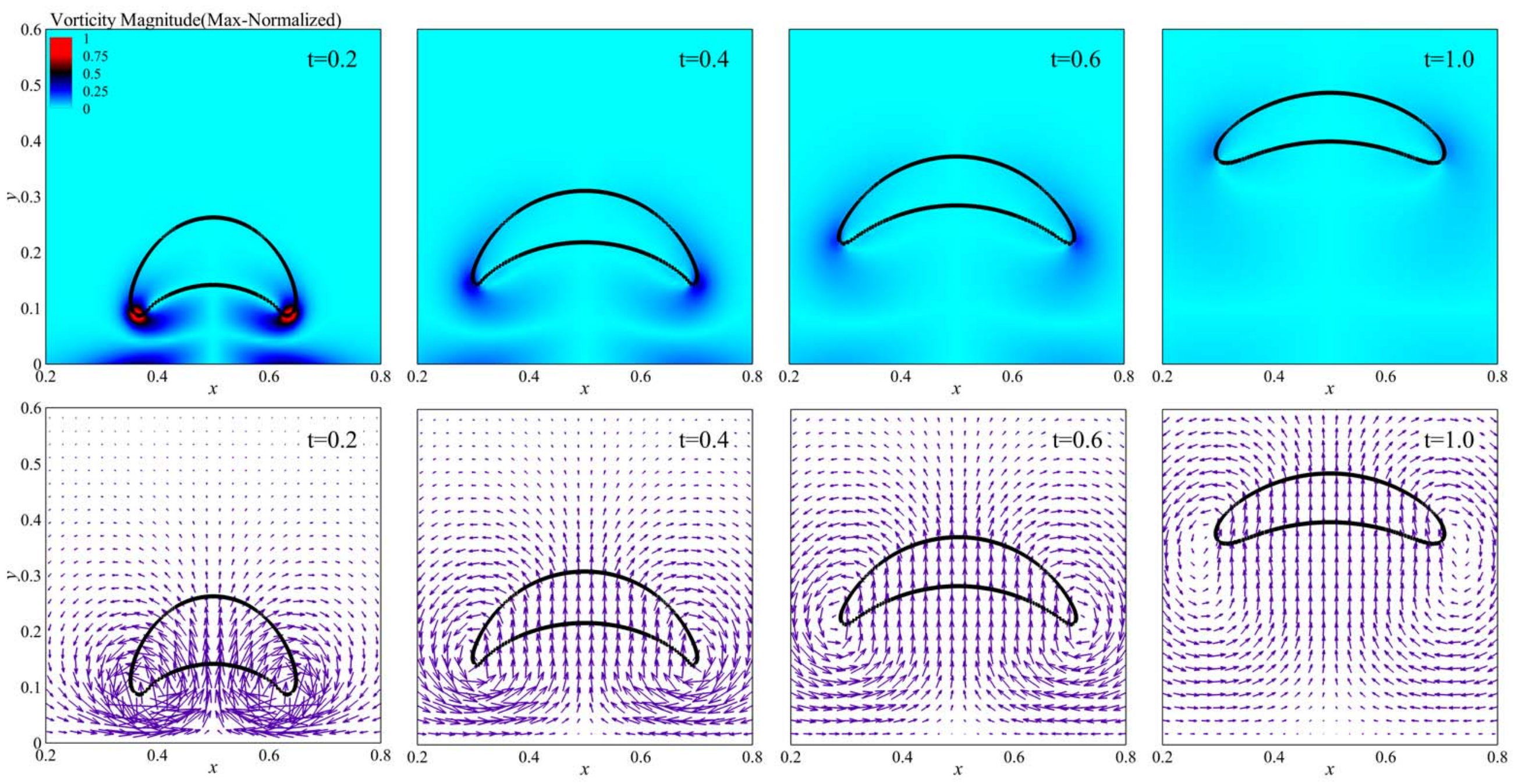}}
	\caption{The maximum value normalized absolute vorticity (top row), and the velocity vector field (bottom row) in and around the levitating droplet. The initial horizontal location of the droplet is centric in this case.}
	\label{figVelocityWithVorticityCentric}
\end{figure}

\begin{figure}
	\centerline{\includegraphics[width=0.999\textwidth]{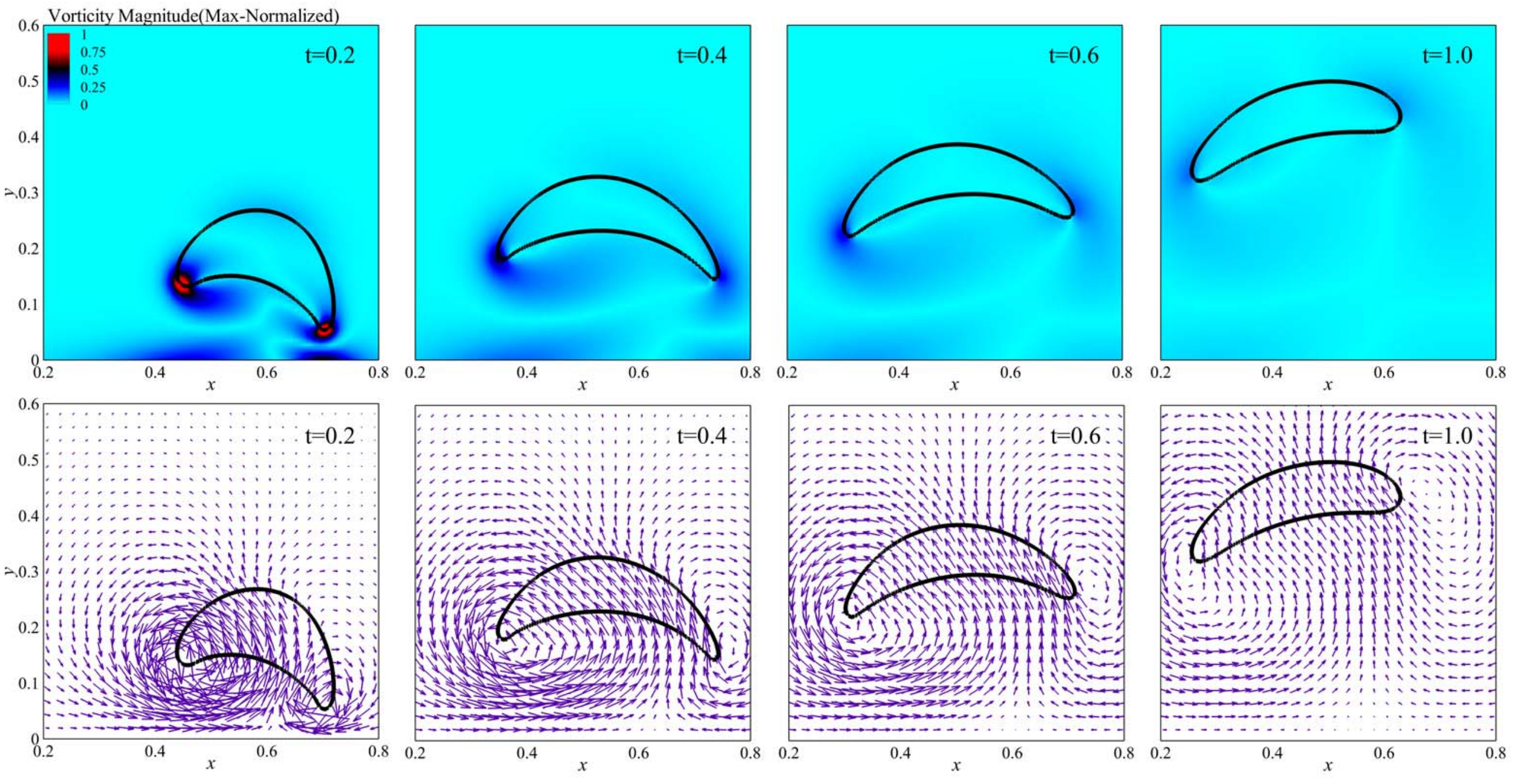}}
	\caption{The maximum value normalized absolute vorticity (top row), and the velocity vector field (bottom row) in and around the levitating droplet. The initial horizontal location of the droplet is eccentric in this case.}
	\label{figVelocityWithVorticityEccentric}
\end{figure}

We study the velocity field solution when the eccentricity in the horizontal location of the droplet is zero. The same is presented in figure \ref{figVelocityWithVorticityCentric}. In this case, there is no sideway movement of the droplet and the path of levitation is a straight vertical line. The magnetic field is due to only two magnets with alternate poles at the bottom wall, while the other three walls are not covered with the magnets. As can be observed, two counter rotating vortices originate at the tail of the droplet during the very initiation of the phenomenon. The situation is horizontally symmetric and there are no net lateral forces on the droplet due to the flow dynamics. The flow, however, do not show a similar character when the initial location of the droplet is not horizontally centric. The velocity field solution for this situation is shown in figure \ref{figVelocityWithVorticityEccentric}. The asymmetry in this situation causes one of the vortices, { which is closer to the horizontal center ($x=0.5$) to have more size and strength}. This is evident by looking the left counter-clock wise rotating vortex in figure \ref{figVelocityWithVorticityEccentric} for $t=0.2$, and the corresponding vorticity map. The magnetic forces has given rise to a vortex structure which actually is tilting the droplet laterally away from the expected path of the droplet. The levitation path eventually deviates from the direction along $\nabla H$ due to a coupled interplay between the magnetic and the hydrodynamic flow field.

The arrangement of the magnets has a vital role in the stability of levitation. In the absence of the magnets at the top and side walls, the lateral perturbations in the droplet path do not decay. The selection of the magnetic source arrangements, thus, should be the most carefully considered design feature and the stable levitation is certainly possible under appropriately magnetic field. 

\begin{figure}
	\centerline{\includegraphics[width=0.65\textwidth]{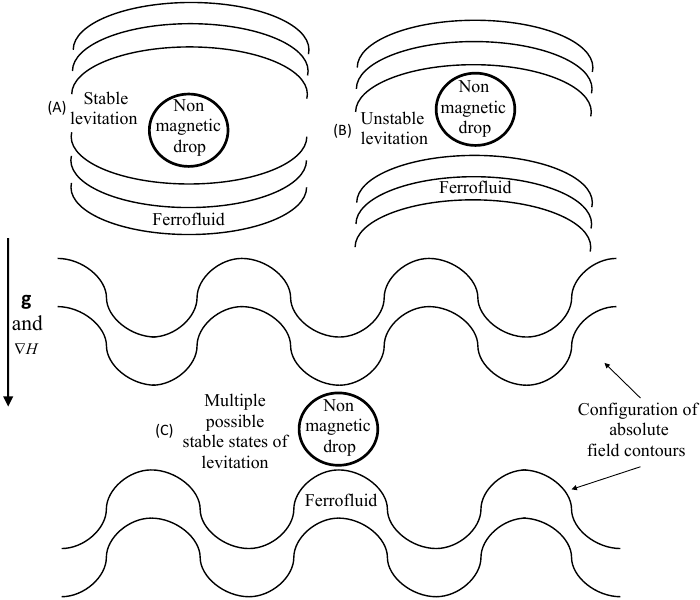}}
	\caption{The schematic for determining about the stability of levitation - (A) stable, (B) unstable, and (C) possibility of multiple stable-states.}
	\label{figStabilitySchematic}
\end{figure}

Based on above observations from the simulations a schematic can be proposed, in terms of the initial field contour configurations, which can help to depict the stability, instability or multiple stability of the droplet levitation. The schematic is presented in figure \ref{figStabilitySchematic}. The mechanism is analogous to the classical {\it ball-on-ramp} example which is often used to understand the concept of stability, and this analogy can be applied to study the stability of levitation by solving for the absolute field contours in the domain produced by a given arrangement of magnets. The stability of the droplet path, however, can be non-trivial and must be analyzed in conjunction with the hydrodynamic flow field near the droplet.

\section{Comparison with experiment}\label{secResultsExperiments}
\begin{table}
	\begin{center}
		\def~{\hphantom{1}}
		\scalebox{0.85}{
			\begin{tabular}{ll}
				
				Domain size  			&  $17.7\:\si{mm}\times17.7\:\si{mm}$\\[3pt]
				
				Initial location of droplet center& $(8.850\:\si{mm},1.725\:\si{mm})$ \\[3pt]
			
				Droplet shape and size$^*$  & Elliptic, $a=R+0.4125\:\si{mm},b=R-0.4125\:\si{mm},R=1.35\:\si{mm}$ \\[3pt]
				
				Grid resolution		    & $\Delta =R/17.39\:\si{mm}$ \\[3pt]
				
				Time step		    	& $\Delta t=0.5\times10^{-4}\:\si{s}$ \\[3pt]
				
				Solver absolute convergence tolerances & $1.0\times10^{-6}$\\[3pt]
				
			\end{tabular}}
			\caption{The simulation parameters for the comparison of numerical solution with the experimental results in \S\ref{secResultsExperiments}. The fluid properties are given in table \ref{tableExpVsSim}.\\
			$^*$Here $a$ and $b$ represent semi-major and semi-minor axis of the elliptical droplet respectively.}
			\label{tableExpVsSimDomain}
		\end{center}
	\end{table}
\begin{figure}
	\centering
	\begin{subfigure}[]{0.5\textwidth}
		\centerline{\includegraphics[width=1.0\textwidth]{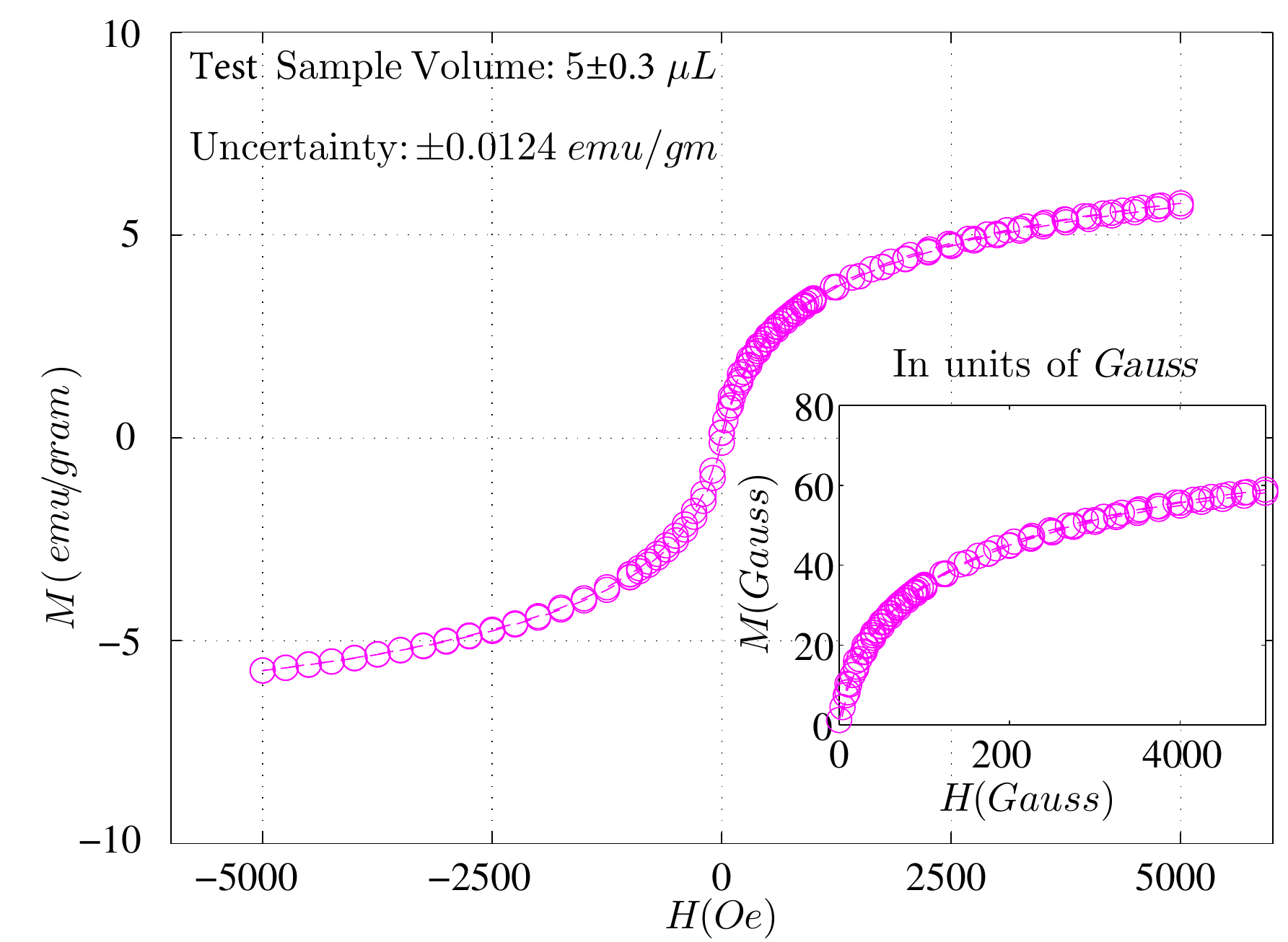}}
	\end{subfigure}
	\begin{subfigure}[]{0.3\textwidth}
		\centerline{\includegraphics[width=1.0\textwidth]{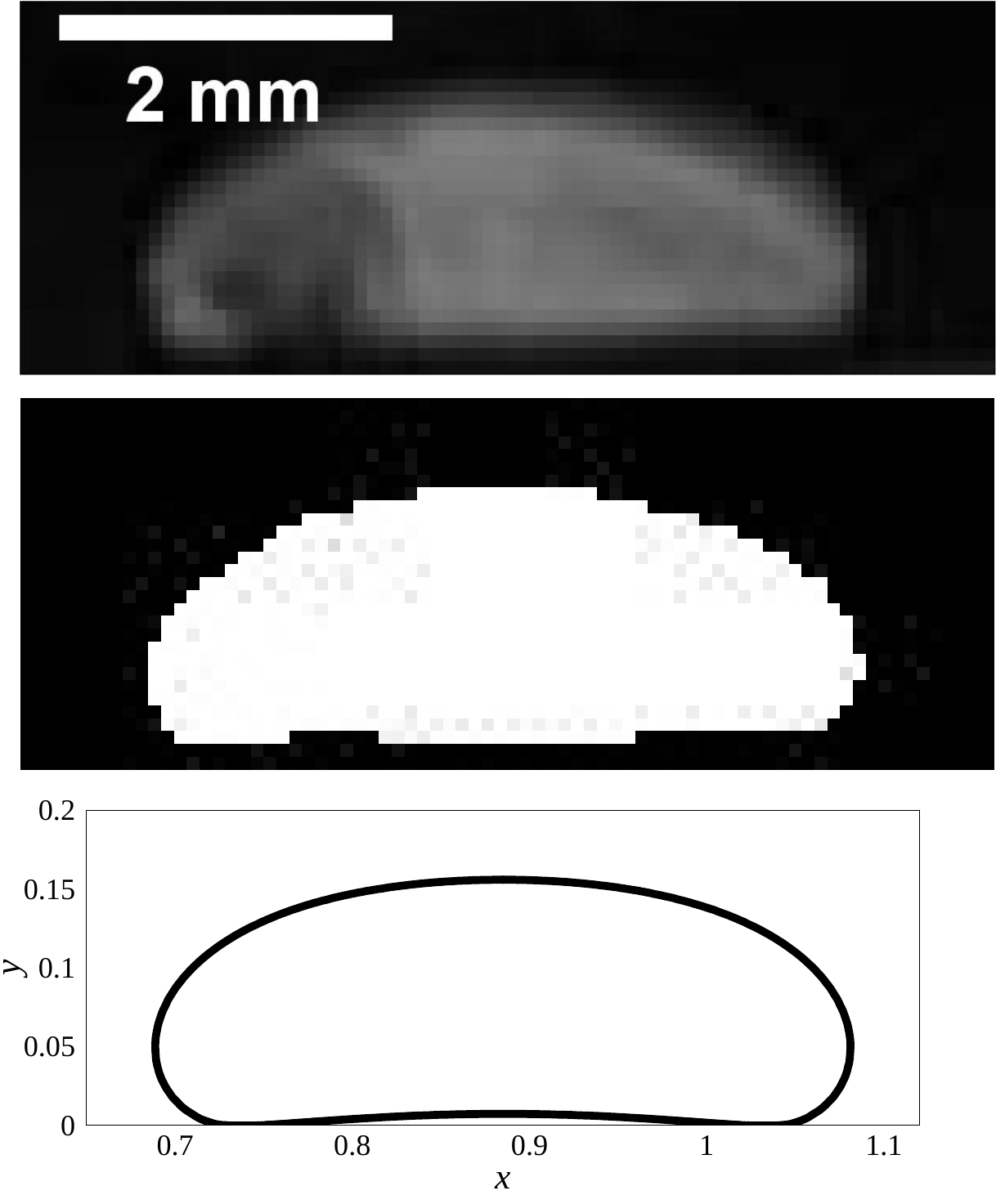}}		\label{figInitialInterfaceWithoutField}
	\end{subfigure}	
	\caption{(Left) The magnetization curve of the ferrofluid sample used in demonstrative experiments. (Right) The shape of the droplet when it is settled at the bottom of the cell in the absence of bottom pair of magnets. The image from the experiment (top), the binary counterpart of the same (middle) and the shape from the simulation (bottom).}
	\label{figCellAndMagnets}
\end{figure}
\begin{figure}
	\centerline{\includegraphics[width=0.999\textwidth]{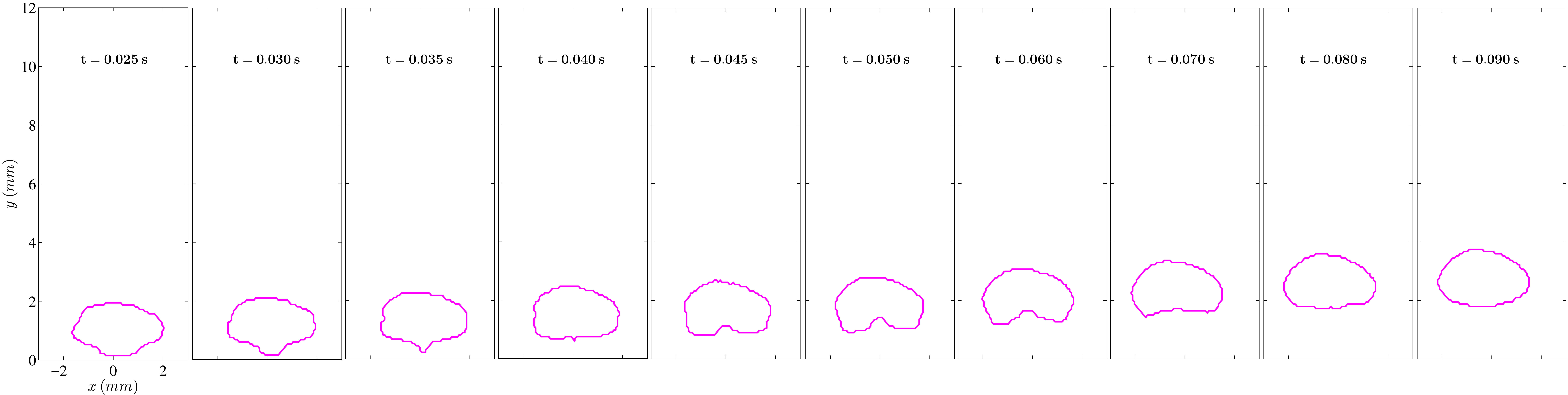}}
	\caption{The phase boundaries extracted from the binary image sequence from the experiment.}
	\label{figExpVsSimAll}
\end{figure}
From experiments, the typical time scale of the phenomenon is noted to be of the order of fraction of a second, and thus it is recorded at a high frame frequency of 2000 frames per second for a sufficient temporal resolution using a monochromatic camera (Phantom v7.0). Due to small size of the cell, the original gray scale image of the droplet is not very clear, as can be noticed in figure \ref{figExpTest3}. For this, we zoom near the relevant portion of the domain and then sharpen the images, converting them to binary form. Eventually the phase boundaries are then extracted from the binary image data, shown in figure \ref{figExpVsSimAll}. This sequence gives a comparatively clear description of the interface of the droplet. The image-processing sufficiently resolves the changes in the configuration of the interface and the dynamics of the tail of the droplet during its rise. After calibrating the frames for further measurements, an $xy$ co-ordinate system is chosen with its origin placed at the center of the bottom wall (figure \ref{figExpVsSimAll}). The co-ordinates are measured in $\si{mm}$. The time instant at which the droplet just starts experiencing the influence of the magnetic field is marked as $t=0.0\:\si{s}$. 
\begin{figure}
	\centerline{\includegraphics[width=0.77\textwidth]{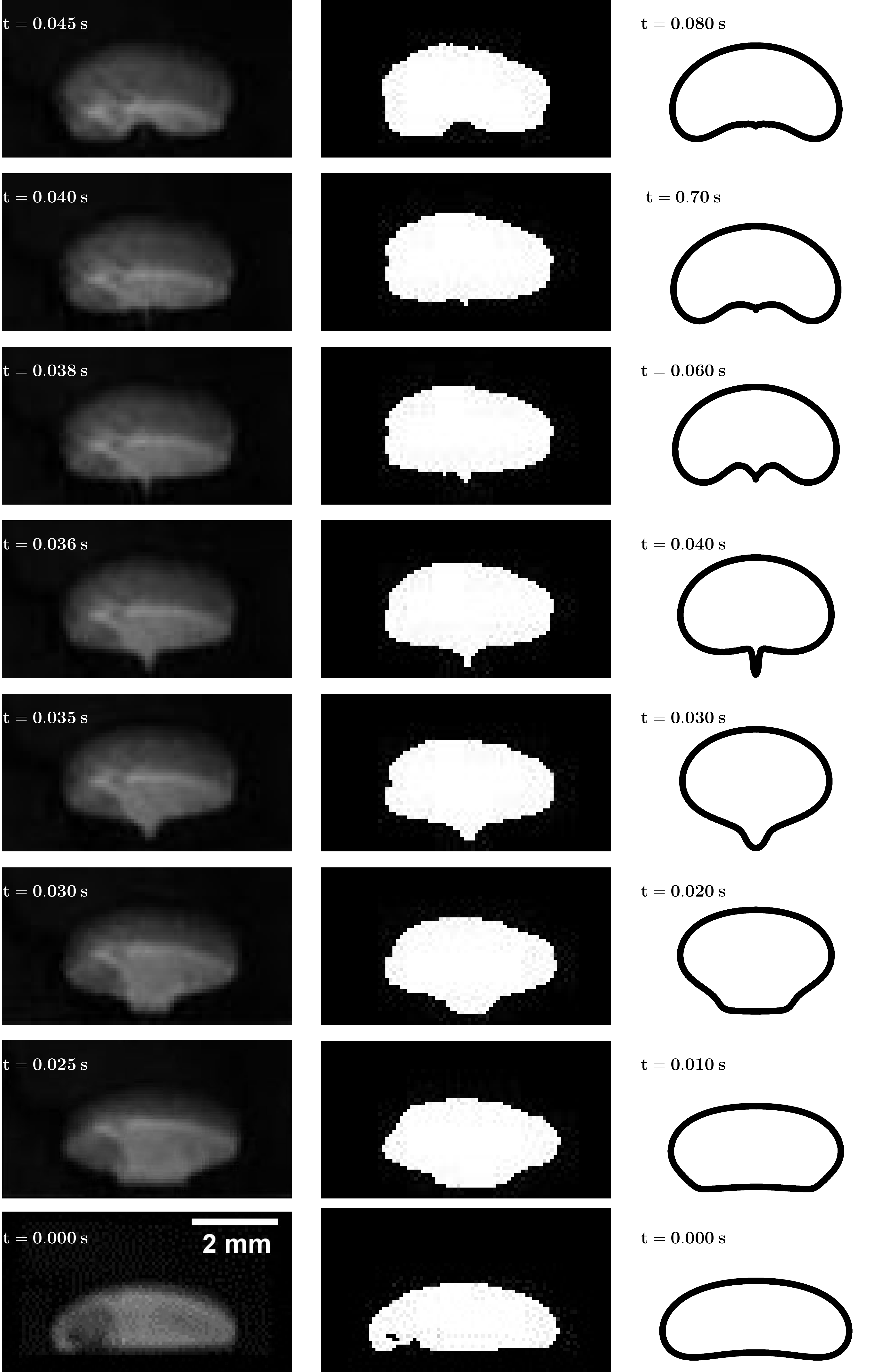}}
	\caption{Comparison between the experimental image sequence and the simulation. The tail of the droplet passes through a stage of singularity, captured both in experiment and simulation. The binary images are utilized to make the singularity more visible. The time increases from bottom to top.}
	\label{figExpVsSimFirst4Frames}
\end{figure}
The domain dimensions, the droplet size at the initial condition and the details about the pixel resolutions of the area covering the droplet and the domain, are given in table \ref{tableExpVsSimDomain}, while the fluid properties were given in table \ref{tableExpVsSim}. In this simulation set, we use the fluid properties and other relevant parameters in their dimensional form, so that the results can directly be compared with the experimental measurements. Standard properties of water are taken for $\Omega_d$. For properties of $\Omega_f$, the FF sample has been characterized, except for its interfacial tension with water. For this we use the method of \citet{zhu2011nonlinear} where the researchers compared the initial conditions of the experiment and simulation to determine the coefficient of interfacial tension. Using this approach we predict that $\sigma=0.0097\:\si{N.m^{-1}}$. The initial shape of the droplet in simulations compares well with the experimental initial shape. The comparison is shown in figure \ref{figCellAndMagnets} (right). The magnetization curve for the ferrofluid sample has been characterized using a EverCool SQUID VSM DC magnetometer and it is shown in figure \ref{figCellAndMagnets}. The initial susceptibility $\chi_o$ and saturation magnetization $M_s$, as depicted from the magnetization curve, are $0.0189$ and $57.7\:\si{G}$ respectively. Thus the FF sample is weakly magnetizable and a shorter levitation height of the water droplet is expected.

\subsection{Singularity at the tail of the droplet}
The shapes of the droplet at different instants of time from the experiment and simulation are compared in figure \ref{figExpVsSimFirst4Frames}. For above characterized samples of fluids, we notice that the levitation height is small, which is due to the weakly magnetizable nature of the ferrofluid sample ($M_s=57.7,\;{\chi_o}=0.0189$). Interestingly, a formation of singular projection at the tail of the levitating droplet is observed in both the experiment and the simulation. The central portion of the tail of the droplet remains adhered in the vicinity of the bottom wall for some initial time, although there is a layer of ferrofluid separating it from the bottom wall. The width of this adhered part of the interface reduces gradually and approximately at $t=0.035\:\si{s}$ in the experiment, it detaches completely. The singularity is formed after this instant at the tail of the droplet ($t=0.036\:\si{s}$). As the droplet levitates, the singular point appears to being pulled towards the bulk of the droplet. Eventually the singularity disappears and the droplet tail becomes smooth and concave downward. The simulations have predicted this phenomenon reasonably well. It also strongly support our argument in \S\ref{secResultsViscosityRatio} that the interfacial singularities or cusps at the interface of the levitating droplet in a ferrofluid can be ubiquitous, especially for low interfacial tension (low $La$). 

As initially a segment of the droplet interface is settled near the bottom wall, the field strength along this portion of the interface is approximated as $H(x)|_{y=0}=-H_o sin(2\pi x/L)$, according to the field boundary condition at the bottom wall. For a small initial time $\delta t$ we assume that the magnetic field at this segment of the interface is close to the field value at the wall itself. The order of the vertical magnetic force on this segment of the interface is then $f_m\sim \bnabla\bcdot\mathsfbi{S}_m\sim-(1/2)H^2\Delta\mu$, where $\Delta \mu$ is the permeability difference across the interface. Using the field strength along this interface segment, $f_m\sim-(1/2)H_o^2 \sin^2 (2\pi x/L) \Delta\mu$. For small initial period of time $\delta t$ the segment will rise faster if $f_m$ at the wall is higher and vice-versa. If the displaced vertical location of this segment after $\delta t$ is $\delta y$, then $\delta y(x) \sim f_m(x) \sim sin^2(x)$. The squared sine interface description at the tail of the droplet is closely resembled at $t=0.030\;\si{sec}$ in simulations and at $t=0.035\; \si{sec}$ in experiments in figure \ref{figExpVsSimFirst4Frames}, where the tip of the projection is still close to the wall and is not cusped (also indicated by smoothness of $\sin ^2(x)$). However, at relatively larger time the $\sin ^2(x)$ feature transits towards a cusp, and the approximation that the field value at the segment equal to the field value at the bottom wall is no more valid. After this regime, the hydrodynamic and the effect of higher outer liquid viscosity (see \S \ref{secResultsViscosityRatio}) becomes significant.  

{ The sharp singular tips are observed in droplet systems, for example in \cite{stone1999drops} under electric field and by \cite{rowghanian2016dynamics} for magnetic drop under uniform magnetic field. \cite{stone1999drops} have concluded that the conical tips appear for certain threshold electric field strength and the dielectric constant ratio. \cite{rowghanian2016dynamics} focused curved tip regimes, however, they clearly discuss that the sharp tips are possible due to destabilization of the droplet under high field strengths and high localization of the magnetic stresses at the droplet poles. In present study the situation is, however, dynamic and the field is non-uniform. A low interfacial tension (low Laplace number $La$) in combination with the hydrodynamic viscous stress difference (exhibited through viscosity ratio change) across the interface are two additional important factors which can help realizing such sharp conical tips. In practical situations, the former is reasonably modifiable using a surfactant solution and the latter by taking a different non-magnetic droplet medium in a ferrofluid (or a non-magnetic medium outside a ferrofluid droplet).}

{ Although the droplet is essentially moving in a quasi-two dimensional cell, and it might be more appropriate to use gap-averaged Navier-Stokes equations, it turns out that the current 2D mathematical description captures the droplet's shape reasonably well. Thus the \emph{spatial} solution is quite accurate. However during the later stages of the droplet rise, the solution is \emph{temporally} deviating (Fig. 24). As far as the \emph{spatial} shapes of the droplet are concerned, any 3D effects due to the out-of-plane curvature of the droplet in the Hele-Shaw gap are not dominant. We attribute the accurate capturing of the droplet interface shape to the following points --- (a) careful characterization of the ferrofluid sample’s non-linear magnetization curve, as well as some other properties, and implementing the same in the simulations using the Langevin’s non-linear magnetization model (Fig. 22), (b) implementing the initial interface condition precisely similar to one which we find for a resting droplet in the experiments, (c) convenience in closely realizing the actual field boundary-conditions due to the Halbach-array, (d) modeling similar flow boundary-conditions as that in experiments, and (e) using some state-of-the-art numerical techniques for computational accuracy. In addition, the ferrofluid sample in the experiments is characterized and found to be weakly magnetizable. The numerical solution is then expected to be more  accurate due to less sharper change of the magnetic properties and field (and thus a \emph{softer} jump in the magnetic force term in the NS equations) across the interface. The possible reasons why \emph{temporally} the 2D assumption is only reasonable might be due to the following unrelaxed assumptions --- (a) neglecting the out-of-plane curvature in simulations due to H-S geometry, or in other words, neglecting the quasi-2D nature of the problem and considering it as a 2D problem in simulations, (b) neglecting one very important aspect of the magnetization-relaxation dynamics in ferrofluids and considering instantly relaxing magnetization when the magnetic field changes. We believe that incorporation of the magnetization relaxation time effect coupled with a two-phase model for ferrofluid interface, together with corrections due to the third dimension (e.g. implementing the Darcy law \citep{gondret1997shear} together with inertial corrections to it \citep{ruyer2001inertial}), can closely resolve the temporal details also.}
\section{Dynamical model}\label{secOneDimensional}
A differential model for the vertical trajectory of the levitating droplet is constructed to obtain an analytical expressions for the onset condition of levitation, levitation height, and conditions for transitions in the droplet time response near the equilibrium. This was previously analyzed by the displacement and velocity plots from the simulations in \S \ref{secDisplacementAndVelocity}. { In this model, the levitation path is assumed to be stable in the lateral direction and thus the droplet is assumed to move only in the vertical dimension. A single magnetic source of strength $H_o$ is considered at the bottom wall while no source is assumed on the side or top. Additionally, we assume the droplet to be spherical. For simplicity of the formulation, the local magnetic field distortions due to the presence of the droplet are neglected, and the local filed is assumed equal to a local field which would be present if there is no droplet.} At any time $t$, if the vertical position of the droplet is denoted by $\zeta(t)$, then the Newton's equation of motion for the droplet can be written as
\begin{equation}
\begin{array}{ll}
\displaystyle \rho_d V_d \ddot{\zeta}=
\underbrace{\rho_f V_d g}_\text{buoyancy}
-\underbrace{\rho_d V_d g}_\text{gravity}
-\underbrace{6\pi \eta_f R \dot{\zeta}}_\text{Stoke's drag}
-\underbrace{\mu_o \chi_f H(\zeta) \frac{dH(\zeta)}{d\zeta}V_d}_\text{magnetic levitation force},\\[3pt]
\end{array}
\label{eq1dEquationOfMotion}
\end{equation}
where $H(\zeta)$ is the field strength along the vertical direction $\zeta$. We assume that the field strength decays as $H(\zeta)=H_o e^{-k\zeta}$ from the magnetic source of strength $H_o$ at $\zeta=0$ with a decay constant $k$ (in units of $1/{\mbox{length}}$). Notice that $\frac{dH}{d\zeta}=-kH_oe^{-k\zeta}$ is negative and thus provides a positive upward levitation force. {The above equation describing the decay of the magnetic field along the vertical direction ($+y$ or $\zeta$) is assumed to be valid only along the vertical direction but only away from the bottom wall, due to the fact that the droplet’s initial location in simulations was considered above the bottom wall (see for example figure \ref{figPhenomena}).} Using this in equation \ref{eq1dEquationOfMotion}, and rearranging, we get 
\begin{equation}
\begin{array}{ll}
\displaystyle \ddot{\zeta}
+{\left[\frac{6\pi \eta_f R}{\rho_d V_d}\right] \dot{\zeta}}
+{\left[\frac{\rho_d-\rho_f}{\rho_d}\right] g}
=\left[\frac{\mu_o k \chi_f H_o^2}{\rho_d}\right]e^{-2k\zeta}.\\[3pt]
\end{array}
\label{eq1dEquationOfMotionRearranged}
\end{equation}
To compare the vertical trajectories $\zeta(t)$ with simulations, the above equation is non-dimensionalized using the same reference scales used in equation \ref{eqScales}, which casts it in terms of $La_m$ and $Ga$ as 
\begin{equation}
\begin{array}{ll}
\displaystyle \ddot{\tilde{\zeta}}

+\underbrace{\frac{9}{2}\left[\frac{\eta_f}{\eta_d}\right] }_\text{$\mathcal{V}$}\dot{\tilde{\zeta}}

+\underbrace{Ga\left[1-\frac{\rho_f}{\rho_d}\right] }_\text{$\mathcal{G}$}

=\underbrace{La_m\left[1-\frac{\mu_o}{ \mu_f}\right] ({2 k R}) } _\text{$\mathcal{L}$} e^{-\overbrace{(2kR)}^\mathcal{K} \tilde{\zeta}}.

\\[3pt]
\end{array}
\label{eq1dEquationOfMotionNonDim}
\end{equation}
where $\tilde{\zeta}=\zeta/R$, and the right hand side is basically a non-linear forcing function due to the magnetic source. For simplicity, we write the above equation as
\begin{equation}
\begin{array}{cc}
\ddot{\tilde{\zeta}}+\mathcal{V}\dot{\tilde{\zeta}}+\mathcal{G}=\Psi(\tilde{\zeta}),\\[3pt]
\displaystyle \Psi(\tilde{\zeta})=\mathcal{L}e^{-\mathcal{K}\tilde{\zeta}}.\\[8pt]
\end{array}
\label{eq1dEquationOfMotionRearranged}
\end{equation}
The equation is sensitive to four parameters -- $\mathcal{V}=\mathcal{V}(\eta_f/\eta_d),\; \mathcal{G}=\mathcal{G}(Ga,\rho_f/\rho_d),\; \mathcal{L}=\mathcal{L}(La_m,\mu_o/\mu_f)$, and the non-dimensional constant $\mathcal{K}=2kR$ carrying information about the vertically decaying magnetic field strength. 
\subsection{Levitation height and onset condition}
The differential equation \ref{eq1dEquationOfMotionRearranged} is a second order ordinary but non-linear due to the form of $\Psi(\tilde{\zeta})$. It is reduced to a first order system of two differential equations by introducing a variable $\Pi=\dot{\tilde{\zeta}}$, written as
\begin{equation}
\begin{array}{cc}
\dot{\tilde{\zeta}}=\Pi,\\[3pt]
\dot{\Pi}=-\mathcal{V}\Pi-\mathcal{G}+\Psi(\tilde{\zeta}).\\[3pt]
\end{array}
\label{eq1dEquationOfMotionSystem}
\end{equation}
The long-time equilibrium or fixed point of the system is obtained by setting the droplet speed $\dot{\tilde{\zeta}}=\Pi$, and acceleration $\dot{\Pi}$ to $0$ in \ref{eq1dEquationOfMotionSystem}, which for non-zero $\Psi$, gives the steady state fixed point $(\Pi_*,\tilde{\zeta}_*)$
\begin{equation}
\left.
\begin{array}{ll}
\displaystyle \Pi_*=0, \mbox{ steady state velocity},\\[3pt]
\displaystyle \Psi(\tilde{\zeta}_*)=\mathcal{G},\mbox{ equation for steady state levitation height}.\\[3pt]
\end{array}
\right\}
\label{eqFixedPoint}
\end{equation}
As $\Psi(\tilde{\zeta})=\mathcal{L}e^{-\mathcal{K}\tilde{\zeta}}$, it turns out that
\begin{equation}
\left.
\begin{array}{ll}
\displaystyle \Pi_*=0, \mbox{ steady state velocity},\\[3pt]
\displaystyle \tilde{\zeta}_*=-\frac{1}{\mathcal{K}}\ln\frac{\mathcal{G}}{\mathcal{L}}=\frac{1}{\mathcal{K}}\ln\frac{\mathcal{L}}{\mathcal{G}}, \mbox{ steady state levitation height}.\\[3pt]
\end{array}
\right\}
\label{eqFixedPoint2}
\end{equation}
The steady state levitation height of the droplet depends on parameter $\mathcal{K}$ and the ratio $\mathcal{G}/\mathcal{L}$ (which is $\propto Ga/La_m$), and is independent of the parameter $\mathcal{V}$. This is expected as $\mathcal{V}$ is the strength of the viscous drag force opposing the droplet motion and it must vanish at steady state. If the levitation height $\tilde{\zeta}_*$ is to be a non-zero positive, then $\frac{1}{\mathcal{K}}\ln\frac{\mathcal{L}}{\mathcal{G}}\geq 0$, which provides the condition for the onset of levitation
\begin{equation}
\mathcal{L}\geq\mathcal{G},\;\;\mbox{or,}\;\;\frac{La_m}{Ga}\geq\alpha_1 \frac{(1-\rho_f/\rho_d)}{(1-\mu_o/\mu_f)},
\end{equation}
where $\alpha_1=1/\mathcal{K}$ is constant. If the onset condition is satisfied, the levitation height is
\begin{equation}
\tilde{\zeta}_*=\alpha_1\ln\frac{La_m(1-\mu_o/\mu_f)}{Ga(1-\rho_f/\rho_d)}.
\end{equation}
The above expression is physically is in agreement with the facts -- the rise of the droplet will increase in a ferrofluid sample of higher permeability, with lower droplet mass densities, with the increase in $La_m$ or decrease in $Ga$.  
\subsection{Transition in the nature of the fixed point}
The transitions in the nature of the fixed point are reveled by evaluating the eigenvalues of the Jacobian $\mathsfbi{J}$ for the system \ref{eq1dEquationOfMotionSystem}, which is
\begin{equation}
\setlength{\arraycolsep}{0pt}
\renewcommand{\arraystretch}{1.4}
\mathsfbi{J} = \left[
\begin{array}{cc}
\displaystyle 0 \quad& \quad 1\\
\displaystyle \frac{d\Psi}{d\tilde{\zeta}} &\quad  -\mathcal{V} \\
\end{array}  \right].
\label{equJacobian}
\end{equation}
If $\lambda$ are the eigenvalues of $\mathsfbi{J}$, then the characteristic equation $|\mathsfbi{J}-\lambda\mathsfbi{I}|=0$ is $\lambda^2+\mathcal{V}\lambda-{d\Psi}/{d\tilde{\zeta}}=0$
and the eigenvalues or the characteristic values of the fixed point, $\lambda_{1,2}$, are $\displaystyle \lambda_{1,2}=-({\mathcal{V}}/{2})\pm({1}/{2})(\mathcal{V}^2-4(-{d\Psi}/{d\tilde{\zeta}}))^{1/2}$.
This gives the possibilities
$\mathcal{V}^2>-4\frac{d\Psi}{d\tilde{\zeta}}$ (distinct real eigenvalues),
$\mathcal{V}^2=-4\frac{d\Psi}{d\tilde{\zeta}}$ (identical real eigenvalues), and
$\mathcal{V}^2<-4\frac{d\Psi}{d\tilde{\zeta}}$ (complex conjugate eigenvalues).
\begin{figure}
	\centerline{\includegraphics[width=0.95\textwidth]{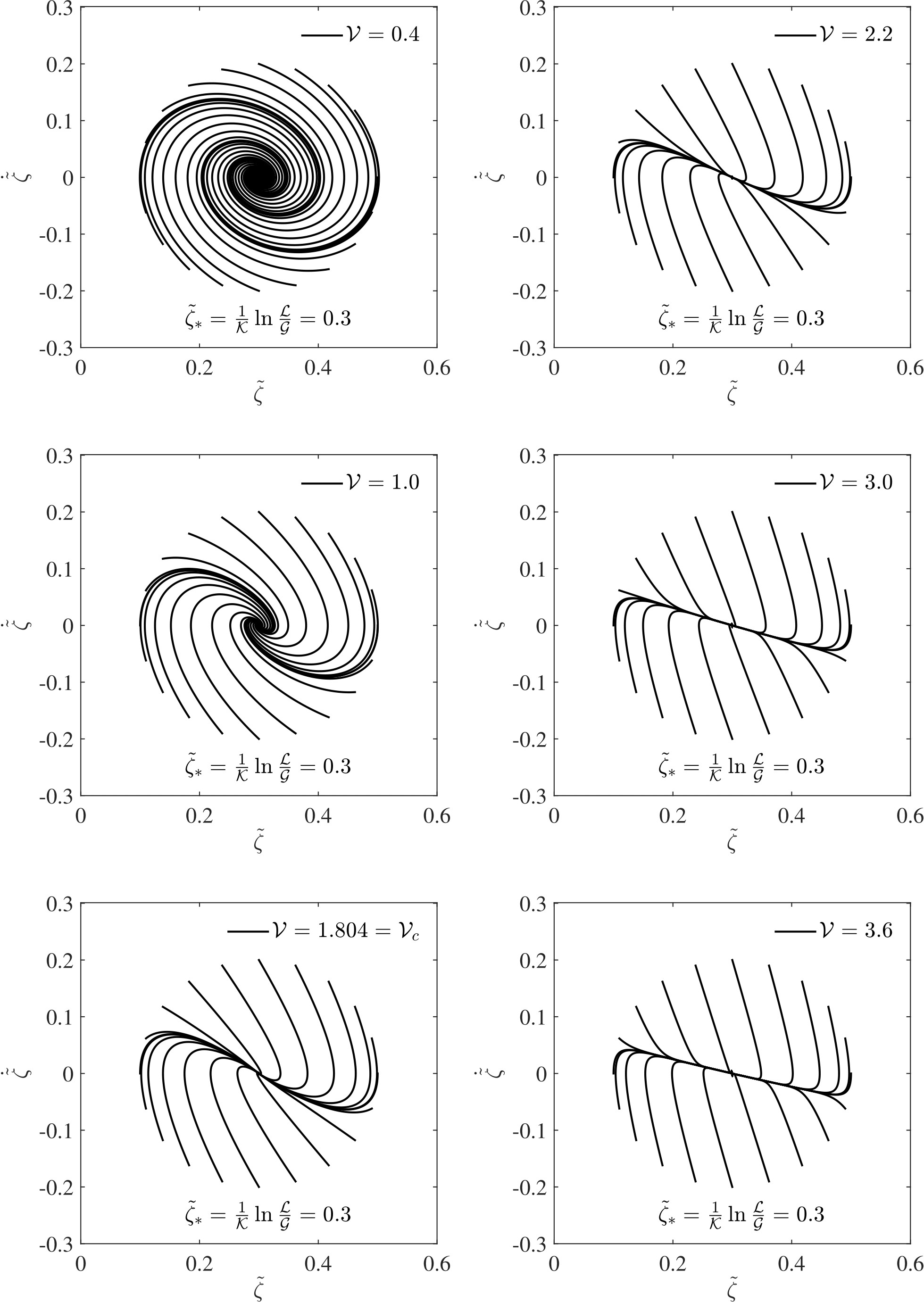}}
	\caption{The behavior of the solution of equation \ref{eq1dEquationOfMotionNonDim} near the equilibrium/fixed point, depicted by the phase portraits on displacement-velocity $(\tilde{\zeta},\dot{\tilde{\zeta}})$ plane with varying $\mathcal{V}$ ($\mathcal{V}$ is proportional to the viscosity ratio). The nature of the fixed point $(\tilde{\zeta}_*,\dot{\tilde{\zeta}}_*)=((1/\mathcal{K})\ln{(\mathcal{L}/\mathcal{G}),0)}$ changes at critical $\mathcal{V}_c=1.804$ from a stable \emph{spiral} to a stable \emph{node}. Here, $\mathcal{L}/\mathcal{G}=1.21$ and $\mathcal{K}=\pi/5$.}
	\label{figPortraits}
\end{figure}
The first case physically suggests that the fixed point is of type \emph{node} and the droplet will approach the steady state monotonically. The third case of complex eigenvalues, however, suggest that the fixed point is of type \emph{spiral} and the droplet will oscillate around the fixed point before reaching the steady state. The second case gives the critical value of the parameter $\mathcal{V}$, at which the stable fixed point $\zeta_*$ changes its type from \emph{spiral} to \emph{node} or vice-versa, according to
\begin{figure}
	\centerline{\includegraphics[width=0.61\textwidth]{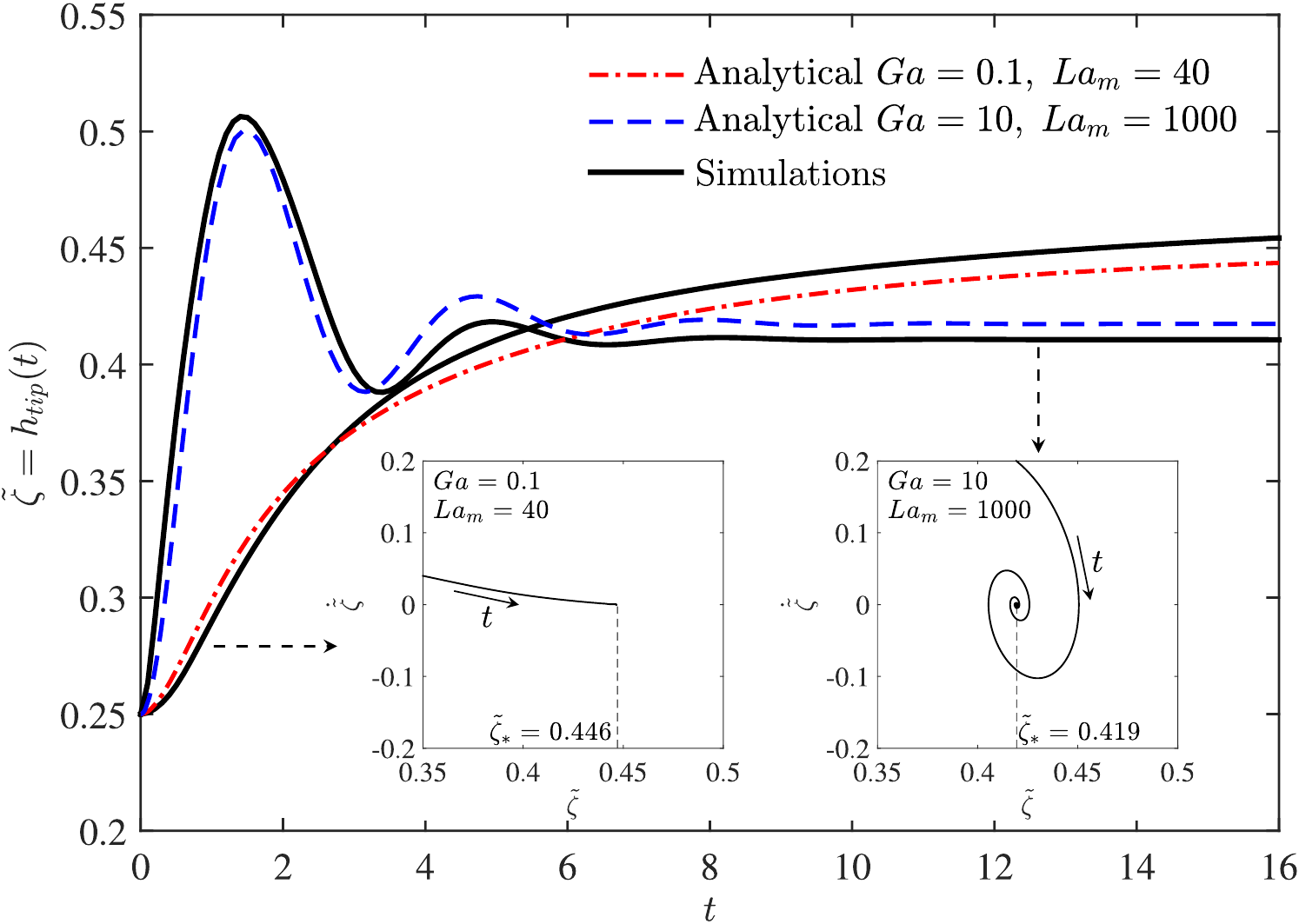}}
	\caption{The comparison of the time-displacement of the droplet between the simulations and the analytical model for two extreme cases of $Ga\;(=0.1$ and $10)$. The $La_m$ for the two cases is chosen such that levitation height $\tilde{\zeta}_*\;(\propto La_m/Ga)$ remains in the same range. Both approaches have predicted that there is a transition between a \emph{node} and a \emph{spiral}.}
	\label{figComparisonSimVsModel}
\end{figure}

\begin{equation}
\displaystyle \mathcal{V}_c^2=-4\left[\frac{d\Psi}{d\tilde{\zeta}}\right]_{\tilde{\zeta}_*}=-4\left[\frac{d}{d\tilde{\zeta}}\mathcal{L}e^{-\mathcal{K}\tilde{\zeta}}\right]_{\tilde{\zeta}_*}=4\mathcal{K}\mathcal{G}.
\label{equDpsiByDzeta}
\end{equation}
Thus from the condition of complex conjugate eigenvalues, $\mathcal{V}^2<-4\frac{d\Psi}{d\tilde{\zeta}}$, the oscillations around the equilibrium point do occur if
\begin{equation}
\displaystyle \mathcal{V}^2<4\mathcal{K}\mathcal{G},\;\;\mbox{or,}\;\;\frac{\eta_f}{\eta_d}<\alpha_2\sqrt{\left(1-\frac{\rho_f}{\rho_d}\right)Ga},
\label{equConditionForOscillations}
\end{equation}
where $\alpha_2=(4/9)\mathcal{K}^{1/2}$ is constant. It is important to note that the parameter $\mathcal{L}$ do not influence the condition for the oscillations around the equilibrium, it only appears in the expression of the steady state levitation height.

The nature of the stable fixed point $\zeta_*$ and transitions in its behavior is further studied through phase portraits on $(\zeta,\dot{\zeta})$ plane. The differential equation for $(\zeta,\dot{\zeta})$ from \ref{eq1dEquationOfMotionSystem} is
\begin{equation}
\displaystyle \frac{d\Pi}{d\tilde{\zeta}}=\frac{d\dot{\tilde{\zeta}}}{d\tilde{\zeta}}=\frac{\dot{\Pi}}{\dot{\tilde{\zeta}}}=\frac{-\mathcal{V}\Pi-\mathcal{G}+\Psi(\tilde{\zeta})}{\Pi}.
\label{eq1dEquationOfPhasePortrait}
\end{equation}
Using this, first we show the effect of variation of $\mathcal{V}$ ($\propto$ viscosity ratio) at fixed $\mathcal{G}$ ($\propto Ga$) and $\mathcal{L}$ ($\propto La_m$). This is depicted in figure \ref{figPortraits} for different initial conditions $(\zeta(0),\dot{\zeta}(0))$ on a circle around the fixed point. For $\eta_f/\eta_d=0.5$, $\mathcal{V}=2.25$, and thus we vary $\mathcal{V}$ around $\mathcal{O}(1)$. The behavior of the linearized solution around the fixed point $\zeta_*$ is clearly sensitive to $\mathcal{V}$, although the actual magnitude of $\zeta_*$ depends on $\mathcal{L/G}$. The variation in $\mathcal{V}$ changes the nature of the fixed point from being a \emph{spiral} to a \emph{monotonic attractor} at $\mathcal{V}_c$. This observation is in well agreement with the simulations in figure \ref{figDisplacement} and \ref{figVelocity}, where the transition between monotonicity and oscillations of the curves were noted. 

We conclude by explicitly comparing the solution of the analytical model \ref{eq1dEquationOfMotionNonDim} to the simulations of figure \ref{figDisplacement}. In simulations we noted that the response changes from monotonic to undulatory when $Ga$ is increased (figure \ref{figDisplacement}). Here we take the extreme cases of $Ga=0.1$ and $10$ considered there. The comparison is shown in figure \ref{figComparisonSimVsModel}. The corresponding trajectories on ($\tilde{\zeta},\dot{\tilde{\zeta}}$) plane are also depicted in the inset. The $La_m$ value is chosen from the results of figure \ref{figDisplacement} such that the ratio $\mathcal{L/G}$ is more or less the same in the dynamical model. This results in nearly same final levitation height according to $\tilde{\zeta}_*={\mathcal{K}^{-1}}\ln(\mathcal{L/G})$, and thus only the nature of response can be closely compared. We have obtained a reasonable agreement between the simulations and the dynamical model about the statement of transition between monotonic to undulatory droplet levitation.

Thus the model captures the essential features of the levitation dynamics of the droplet, and at the same time, it provides a simpler alternative in place of full fledged simulations to predict the condition for the onset of levitation, the levitation height, the condition for oscillations around the equilibrium and the nature of the solution around the equilibrium. The model, however, do not care about the lateral stability of the levitation path. Those conditions for the lateral stability needed a coupled analysis of the magnetic field and the flow solutions (\S \ref{secResultsStability}).  
\section{Conclusions and discussion}\label{secConclusions}
In this study, we show that the {\it stable} levitation of a non-magnetizable droplet immersed inside a ferrofluid is possible with the help of an appropriately generated spatially non-uniform magnetic field; the levitation can be {\it unstable}, or can have multiple-stable states, essentially depending upon the spatially inhomogeneous magnetic field strength. The dynamics of the levitating droplet is analyzed primarily through computations based on a conservative finite-volume based pressure projection algorithm coupled with a multigrid solver for the magnetic field solution and the front-tracking algorithm for the interfacial advections. Physical demonstrations to support the simulations are presented. A dynamical model is proposed for the prediction of the onset of levitation, the steady state levitation height, and the time evolution of the droplet in the vertical direction. 
In this inverse system, where the droplet is non-magnetizable and the outer liquid is ferrofluid, the droplet is forced in the direction opposite to the field gradient.

The conclusions from the study are focused in terms of the following three fundamental curiosities - (a) the shape of the levitating droplet and the appearance of interfacial singularity, (b) the stability of the levitation path and the existence of multiple possible final states, and (c) the nature of the droplet time response around the steady state levitation point. The conclusions about these three aspects from our study are discussed below one by one.
 
\paragraph{\it a. The shape of the levitating droplet and the appearance of interfacial singularity.}
The non-magnetizable droplet levitation in a ferrofluid is simulated inside a bounded square domain under the magnetic field generated by a Halbach array of magnets. The system in general is sensitive to the Laplace number $La$, the magnetic Laplace number $La_m$ and the Galilei number $Ga$. The shape of the levitating droplet primarily depends on the magnitudes of $La$ and $La_m$; the $La_m$ dependence being apparent only at low $La$. The shape is weakly influenced by the changes in $Ga$. For high $La$, the shape of the droplet remains nearly circular, or at most, attains only a slight deformation. On the other hand, the deformation can be more pronounced at low $La$ due to spatially complex magnetic field. Also, this minor deviation from the circular shape in case of higher $La$ occurs only during the initial transience. As the $La$ number decreases, the deformation of the interface at the tail of the droplet increases. The deformation becomes more pronounced for $La$ of order $10^{-1}$. In this regime, unique shapes of the levitating droplet are observed, {\it e.g}, the {\it segmented-ring, crescent} and {\it tooth-like}; the shape transitions from one type to another occur over time or with change of the above control parameters.    

The above observations are for the droplet having viscosity higher than that of the ferrofluid. If the viscosity of the droplet is lower than that of the ferrofluid, it is noticed that the local dynamics of the interface alters. Under these circumstances, the tail of the droplet might show cusped features at sufficiently high $La_m$. The previously discussed {\it crescent-like} shapes now approach exact {\it crescent} shapes. It is shown that there is a possibility of the appearance of singularities at the surface of the non-magnetizable droplet during its field guided motion inside ferrofluid; such singular projection at the tail of the droplet is physically demonstrated and also predicted by the simulation.

The deformation of the droplet is also sensitive to the degree of non-linearity in the magnetization curve of the ferrofluid sample. A change of even one order in the magnitude of the parameter $\gamma_o=3\chi_o H_o/M_s$ can significantly alter the droplet shape, nature of interfacial projections and the levitation height. 
 
\paragraph{\it b. The stability of the levitation path.}
The path of the levitating droplet has shown quite a sensitivity to the spatial description of the magnetic field; not every arrangement of magnets can provide a stable levitation mechanism. The motion of the non-magnetic droplet, and the flow resulting due to this motion, is primarily due to the application of the magnetic field gradient and is parallel to $\nabla H$. But interestingly if the horizontal symmetry of the field around the initial location of the droplet is not maintained, the flow vortices generated near the tail projections of the droplet can be capable of deviating the droplet from its path parallel to $\nabla H$. The magnetic forces and the resulting flow show mutually competing influences on the trajectory of the levitating droplet and this interaction can cause the droplet trajectory to go away from what is expected. We show that appropriate magnetic arrangement can constrain this instability of the droplet levitation path.

\paragraph{\it c. The nature of the droplet time response and the existence of multiple possible final states.}
Besides the levitation path, the stability of the final equilibrium location of the droplet is also investigated. It is found that there may exist magnetic fields which can give rise to multiple stable states of the levitated droplet. The regions of minima of the magnetic field strength, local or global, act as attractors to the droplet. If the magnet arrangement generates multiple such local minima, multiple stable states can exist. The final equilibrium of the droplet is then affected by the initial location of the droplet and its relative distance to various regions of field minima. 

The onset condition for the levitation, the temporal evolution of the droplet, its steady state levitation height and the nature of the stationary point are predicted by a dynamical model, provided that the droplet levitation path is laterally stable/constrained. The model has verified the statement, which was initially based on the simulations, that the response of the droplet can be either monotonic, or it can oscillate about the equilibrium location before reaching to the steady state depending upon the viscosity ratio, density ratio and $Ga$.  Specifically a transition between a stable \emph{spiral} and a stable \emph{node} is identified; the transition between the two occur at $\mathcal{V}_c^2={4\mathcal{KG}}$, and the steady state levitation height can be quickly predicted by $\tilde{\zeta}_*=\mathcal{K}^{-1}\ln({\mathcal{L}}/{\mathcal{G}})$. 
\section{Acknowledgements}
Authors gratefully thank fruitful discussions with Prof. Sandipan Ghosh Moulic. Authors also thank the Council for Scientific and Industrial Research, New-Delhi, for partial funding. A preliminary part of the work was presented at the $9^{th}$ International Conference on Multiphase Flow, 2016, Florence, Italy. 
\appendix
\section{Grid and time step independence}\label{appA}
To study the grid and time step independency of the simulations, a combination of non-dimensional groups is first selected, from the range simulated in the present study, for which the droplet goes under maximum deformation and levitation height. It is intuited that a grid resolution which is sufficient for this extreme case will serve the purpose for the rest of the simulation sets. For constant permeability assumption, the droplet deformation is maximum for $La=0.1$ and its levitation is maximum when $Ga=0.1$ and $La_m=1000$. For variable permeability the same is true for $La=0.1, Ga=0.1, La_m=1000^*$ and $\gamma_o=1.666$.  
\begin{figure}
	\centerline{\includegraphics[width=0.9\textwidth]{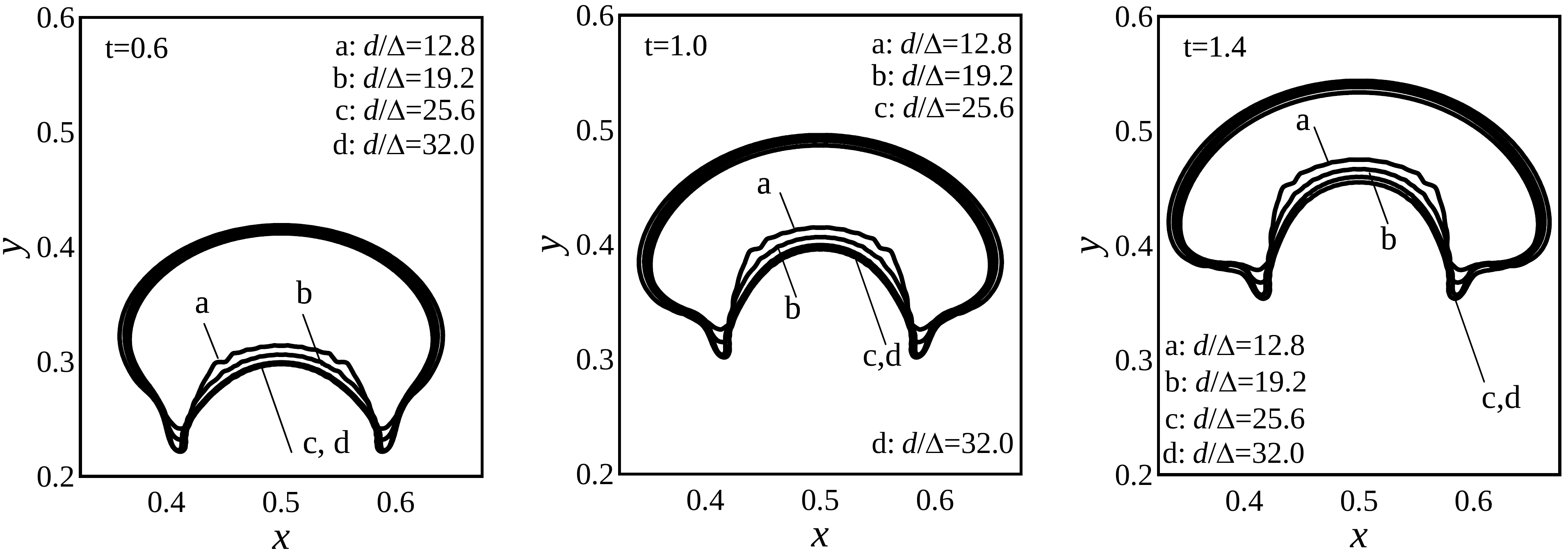}}
	\caption{The interface of the levitating droplet compared for different grid resolutions under constant permeability assumption. The non dimensional parameters are $La=0.1, Ga=0.1$ and $La_m=1000$.}
	\label{figGridIndependenceConstantMum}
\end{figure}
\begin{table}
	\begin{center}
		\def~{\hphantom{1}}
		\scalebox{1.0}{
			\begin{tabular}{cccccc}
				$d/\Delta$&$9.6$& $12.8$ &$19.2$&$25.6$&$32.0$ \\[8pt]
				\begin{tabular}{@{}c@{}}Constant permeability\\ assumption\end{tabular} &$25.76\%$& $9.11\%$ & $3.32\%$  & $0.71\%$ &$0.00\%$\\[8pt]
				\begin{tabular}{@{}c@{}}Variable permeability\end{tabular} & $15.36\%$& $6.35\%$ & $2.76\%$  & $0.67\%$ & $0.00\%$\\
			\end{tabular}}
			\caption{The relative mean percentage error (RMPE) in the droplet deformation $(\mathcal{D}(t))$ curve with respect to the grid resolution, both for constant permeability assumption and the variable permeability formulation.}
			\label{tableGridAndTime}
		\end{center}
\end{table}
\begin{figure}
	\centerline{\includegraphics[width=0.9\textwidth]{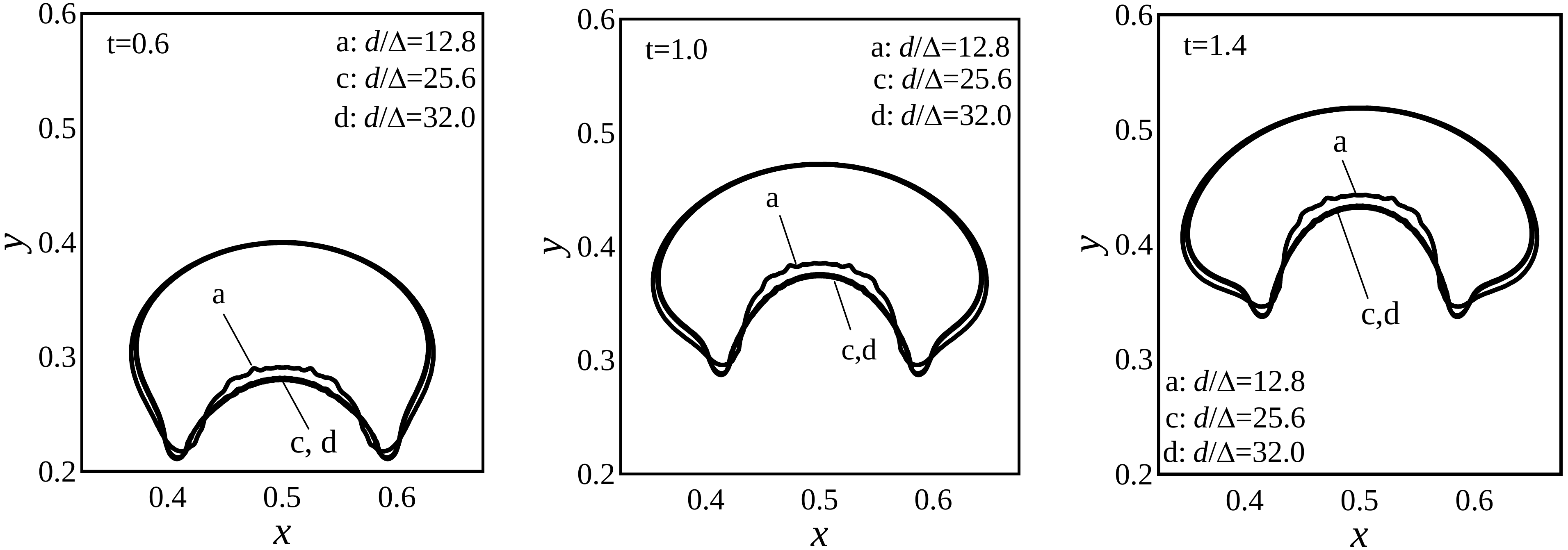}}
	\caption{The interface of the levitating droplet compared for different grid resolutions under variable permeability formulation using Langevin's relation. The non dimensional parameters are $La=0.1, Ga=0.1$ and $La_m=1000^*$.}
	\label{figGridIndependenceVariableMum}
\end{figure}
Thus for these set of parameters, we resolve the grid starting from $d/\Delta=9.6$ to $32.0$ and look for the saturation in the outcomes at some grid resolution. Here $d$ is the diameter of the round droplet and $\Delta$ is the computational cell size. The outcomes judged are the interface shapes (figure \ref{figGridIndependenceConstantMum} under constant permeability assumption and figure \ref{figGridIndependenceVariableMum} for variable permeability case) and the relative mean percentage error in the droplet deformation parameter curve $\mathcal{D}(t)$ (table \ref{tableGridAndTime} and figure \ref{figGridIndependenceAspectRatioWrtTime}). The grid resolution of $d/\Delta=25.6$ has proven to be a reasonable choice.
\begin{figure}
	\centerline{\includegraphics[width=0.9\textwidth]{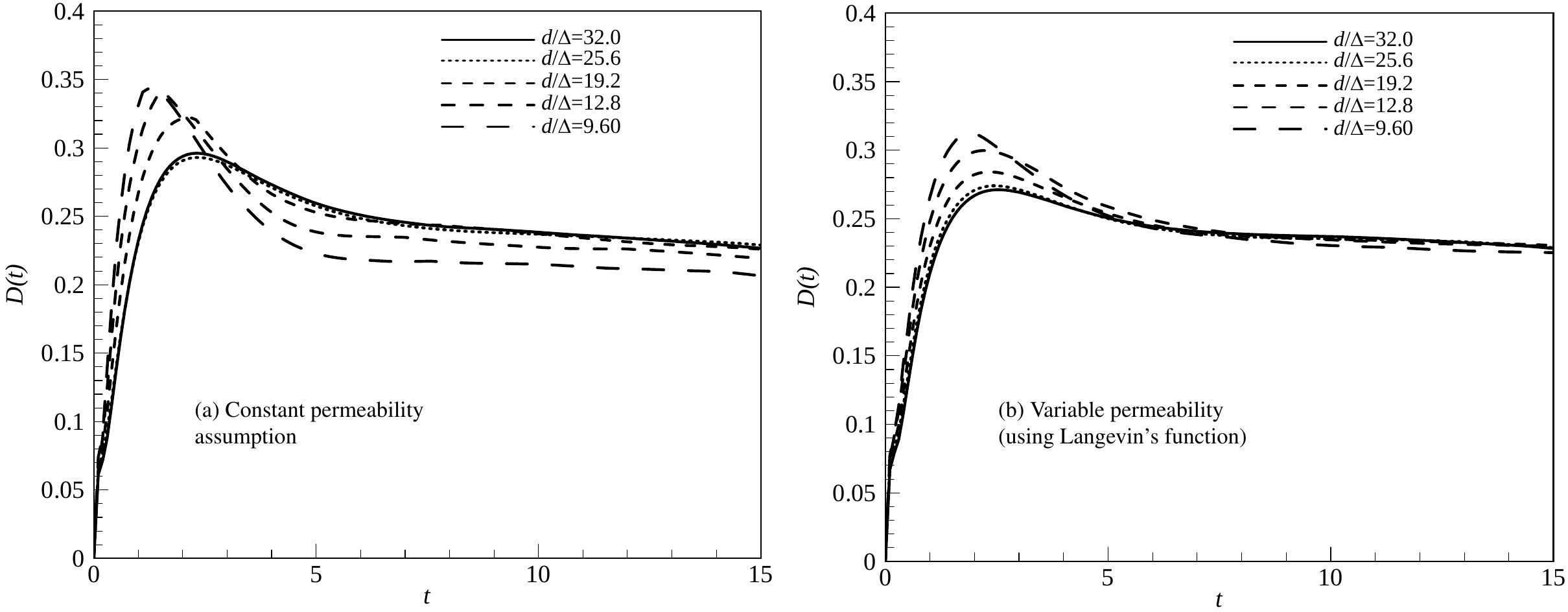}}
	\caption{The droplet deformation $(\mathcal{D}(t))$ curve for different grid resolution, both for constant permeability and variable permeability. The non dimensional parameters are $La=0.1, Ga=0.1$ and $La_m=1000^*$.}
	\label{figGridIndependenceAspectRatioWrtTime}
\end{figure}

\begin{figure}
\centerline{\includegraphics[width=0.9\textwidth]{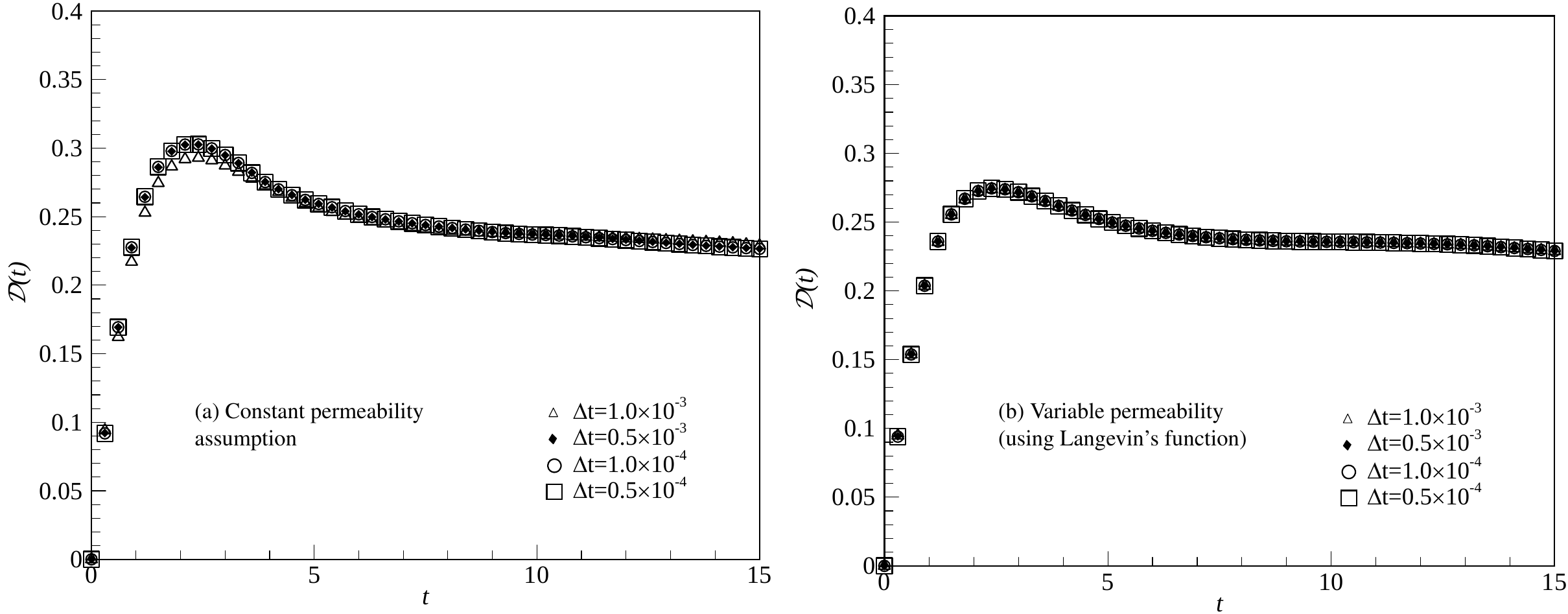}}
	\caption{The droplet deformation $(\mathcal{D}(t))$ curve for different time steps, both for constant permeability and variable permeability. The non dimensional parameters are $La=0.1, Ga=0.1$ and $La_m=1000^*$.}
	\label{figTimeStepIndependenceAspectRatioWrtTime}
\end{figure}

Likewise, the time independence of the simulations is also assured by comparing the signature of the droplet deformation parameter with respect to time for different time steps (figure \ref{figTimeStepIndependenceAspectRatioWrtTime}). The simulations have shown time step independence at $\Delta t=1.0 \times 10^{-4}$. The droplet shapes are also found to be independent of the time step below $\Delta t=1.0 \times 10^{-4}$, and thus this value is adopted. 
\bibliographystyle{jfm}

\end{document}